\documentclass[11pt,a4paper]{article}
\usepackage{amsmath, amsfonts,amssymb,mathtools,nicefrac,nccmath,cases}
\usepackage{mathrsfs}
\usepackage{microtype,booktabs,multirow}
\usepackage[usenames,dvipsnames,table]{xcolor}
\usepackage{colortbl}
\usepackage{tikz}
\usetikzlibrary{shapes,arrows}
\usetikzlibrary{calc,shapes.callouts,shapes.arrows,bending,intersections}
\usetikzlibrary{shapes.geometric}
\usetikzlibrary{arrows.meta,arrows}
\usetikzlibrary{decorations.pathmorphing}
\usetikzlibrary{decorations.markings}
\usetikzlibrary{shapes.multipart}
\usetikzlibrary{tikzmark}
\usepackage{arydshln}
\usepackage{caption}
\usepackage{subcaption}

\usepackage[nosort]{cite}
\usepackage{colortbl}

\usepackage{xcolor}
\usepackage{bm}
\usepackage{lipsum}

\usepackage[linktocpage=true]{hyperref}
\hypersetup{colorlinks=true,linkcolor=nicecolor,citecolor=nicecolor,urlcolor=nicecolor}
\definecolor{nicecolor}{rgb}{0.1, 0.3, 0.4}

\definecolor{blue}{rgb}{0.06, 0.3, 0.57}
\definecolor{Gray}{gray}{0.4}

\definecolor{nicecolor}{rgb}{0.1, 0.3, 0.4}
\definecolor{blue}{rgb}{0.06, 0.3, 0.57}
\definecolor{Gray}{gray}{0.4}
\colorlet{tableheadcolor}{gray!15} 
\colorlet{tablerowcolor}{gray!7} 

\def\hybrid{\topmargin 5pt    \oddsidemargin 0pt
	\headheight 0pt \headsep 0pt
	\textwidth 6.25in       
	\textheight 9.5 in       
	\marginparwidth .875in
	\parskip 5pt plus 1pt 
	\jot = 1.5ex
}
\hybrid
\numberwithin{equation}{section}
\numberwithin{table}{section}
\usepackage{float}
\usepackage{tabularx}
\newcolumntype{D}{>{\centering\arraybackslash}X}
\newcolumntype{L}{>{$}l<{$}}
\newcolumntype{R}{>{$}r<{$}}
\newcolumntype{C}{>{$}c<{$}}
\usepackage{IEEEtrantools}
\usepackage{makecell}

\newcommand{\beq}{\begin{equation}}  \newcommand{\eeq}{\end{equation}}
\newcommand{\bal}{\begin{aligned}}   \newcommand{\eal}{\end{aligned}}
\newcommand{\bea}{\begin{eqnarray}}  \newcommand{\eea}{\end{eqnarray}}
\def\beqa{\begin{eqnarray}}
\def\eeqa{\end{eqnarray}}

\newcommand{\bmat}{\left(\begin{array}}
\newcommand{\emat}{\end{array}\right)}

\newcommand{\bbC}{\mathbb{C}}
\newcommand{\bbR}{\mathbb{R}}

\def\d{{\rm d}}
\def\del{\partial}

\newcommand{\cO}{\mathcal{O}}
\newcommand{\cT}{\mathcal{T}}

\newcommand{\cP}{\mathcal{P}}
\newcommand{\cC}{\mathcal{C}}
\newcommand{\cD}{\mathcal{D}}

\newcommand{\cS}{\mathcal{S}}

\newcommand{\cN}{\mathcal{N}}
\newcommand{\cW}{\mathcal{W}}
\newcommand{\cG}{\mathcal{G}}
\newcommand{\cA}{\mathcal{A}}
\newcommand{\cH}{\mathcal{H}}
\newcommand{\cB}{\mathcal{B}}

\newcommand{\MM}{\mathcal{M}}

\newcommand{\cQ}{\mathcal Q}

\def\calb         {{\cal B}}
\def\calc         {{\cal C}}
\def\cald         {{\cal D}}

\def\calg         {{\cal G}}

\def\calh         {{\cal H}}

\def\calk         {{\cal K}}

\def\calm         {{\cal M}}
\def\caln         {{\cal N}}
\def\calo         {{\cal O}}
\def\calp         {{\cal P}}
\def\calq         {{\cal Q}}
\def\calr         {{\cal R}}

\def\calt         {{\cal T}}

\def\calz         {{\cal Z}}

\usepackage{cancel}

\newcommand{\be}{\begin{equation}}
\newcommand{\ee}{\end{equation}}

\newcommand{\half}{\frac{1}{2}}

\newcommand{\bbZ}{\mathbb{Z}}

\newcommand{\nmod}{{\hat n}}

\definecolor{Gray}{gray}{0.95}

\def\d {{\rm d}}
\def\del          {\partial}

\def\ii           {{\rm i}}

\def\Re           {{\rm Re\hskip0.1em}}
\def\Im           {{\rm Im\hskip0.1em}}

\usepackage{xcolor}
\definecolor{colorloc1}{RGB}{0,0,102}  
\definecolor{colorloc2}{RGB}{0,125,253} 
\usepackage[framemethod=default]{mdframed}
\newmdenv[skipabove=10pt,
skipbelow=7pt,
rightline=true,
leftline=true,
topline=true,
bottomline=true,
linecolor=colorloc1,
backgroundcolor=colorloc2!5,
innerleftmargin=4pt,
innerrightmargin=0pt,
innertopmargin=0pt,
leftmargin=2pt,
rightmargin=0pt,
linewidth=2pt,
innerbottommargin=0pt]{lbBox}
\newenvironment{importantbox}{\begin{lbBox}\vspace{1 mm}
	} {\vspace{1.5 mm}\end{lbBox}}

\begin{document}

\baselineskip=14pt
\parskip 5pt plus 1pt

\vspace*{-1.5cm}
\begin{flushright}    
  {\small 
  IFT-UAM/CSIC-20-93
  }
\end{flushright}

\vspace{2cm}
\begin{center}        

  {\huge Swampland Conjectures for Strings and Membranes \\
   [.3cm]  }
\end{center}

\vspace{0.5cm}
\begin{center}        
{\large  Stefano Lanza,$^{1}$ Fernando Marchesano,$^2$ Luca Martucci,$^3$ and Irene Valenzuela$^1$}
\end{center}

\vspace{0.15cm}
\begin{center}        
 \emph{$^1$Jefferson Physical Laboratory, Harvard University, 
  Cambridge, MA 02138, USA}\\[2mm] 
$^2$ \emph{Instituto de F\'{\i}sica Te\'orica UAM-CSIC, Cantoblanco, 28049 Madrid, Spain} 
\\[2mm] 
${}^3$ \emph{Dipartimento di Fisica e Astronomia ``Galileo Galilei",  Universit\`a degli Studi di Padova} \\ 
\emph{\& I.N.F.N. Sezione di Padova, Via F. Marzolo 8, 35131 Padova, Italy} 
 \\[.3cm]

\end{center}

\vspace{2cm}


\begin{abstract}
\noindent
Swampland criteria like the Weak Gravity Conjecture should not only apply to particles, but also to other lower-codimension charged objects in 4d EFTs like strings and membranes. However, the description of the latter is in general subtle due to their large backreaction effects. In the context of  4d $\mathcal{N}=1$ EFTs, we consider  $\frac12$BPS strings and membranes which are fundamental, in the sense that they cannot be resolved within the EFT regime. We argue that, if interpreted from the EFT viewpoint, the 4d backreaction of these objects translates into a classical RG flow of their couplings. Constraints on the UV charges and tensions get then translated to constraints on the axionic kinetic terms and scalar potential of the EFT. This uncovers new relations among the Swampland Conjectures, which become interconnected by the physical properties of low-codimension objects. In particular, using that string RG flows describe infinite field distance limits, we show that the WGC for strings implies the Swampland Distance Conjecture. Similarly, WGC-saturating membranes generate a scalar potential satisfying the de Sitter Conjecture.

\end{abstract}

\thispagestyle{empty}
\clearpage

\setcounter{page}{1}


\newpage

  \tableofcontents

\newpage

\section{Introduction}
\label{sec:intro}

The Swampland program \cite{Vafa:2005ui,Brennan:2017rbf,Palti:2019pca} aims at identifying universal criteria that any effective field theory (EFT) should satisfy to admit a UV embedding in a consistent theory of quantum gravity. One of the most studied swampland criteria is the Weak Gravity Conjecture (WGC) \cite{ArkaniHamed:2006dz}: given a EFT with a $p$-form gauge field weakly coupled to Einstein gravity, there must exist at least an electrically charged $(p-1)$-brane with a charge-to-tension ratio bigger than the one of an extremal black brane in that theory. This conjecture has been extensively checked in the past years (see \cite{Palti:2019pca} and references therein). However, these studies have always been focused on codimension $>2$ charged objects.\footnote{Although  precisely the generalisation to codimension 1 objects plays a crucial role in the conjecture that any non-supersymmetric vacuum is unstable \cite{Ooguri:2016pdq,Freivogel:2016qwc}.} The reason is that applying the WGC to low codimension objects is much trickier, as these objects induce a strong backreaction that modifies the asymptotic structure of the vacuum and cannot be neglected. Even the definition of tensions $\cT$ and charges $\cQ$ of these objects appears problematic, as they cannot be defined asymptotically. Up to now, any generalisation of the WGC to low codimension objects \cite{Ooguri:2016pdq,Ibanez:2015fcv,Hebecker:2015zss,Craig:2018yvw,Font:2019cxq} usually involves dealing with probe branes, but a proper treatment taking into account their backreaction is missing. In this paper, we take the first steps in this direction.

Understanding how to deal with low codimension objects can also provide valuable information about the low-energy EFT actions that are consistent with quantum gravity. Interestingly, the gauge kinetic function of $p$-form (with $p\geq d-2$) gauge fields not only fixes the physical charges $\cQ$ and the strength of the gauge interaction between the objects, but also parametrises the axionic kinetic terms and scalar potentials of the low energy effective action. In 4d effective actions, scalars with approximate continuous shift symmetries can be dualised to 2-form gauge fields, with the gauge kinetic function given by the axionic field metric. Similarly, dynamical discrete parameters in the potential (usually corresponding to internal fluxes in string compactifications) can be dualised to 3-form gauge fields such that their gauge kinetic function parametrises the F-term potential. Hence, constraints on the EFT kinetic terms and potentials can be translated to properties of the strings and membranes charged under the 2-form and 3-form gauge fields respectively, and vice versa.

Notice that some swampland conjectures (like the WGC) imply constraints on the properties of the charged objects, while other swampland conjectures constrain the behaviour of the field metrics and scalar potentials, like e.g. the de Sitter conjecture of  \cite{Obied:2018sgi}. But in the context of strings and membranes, they all get connected for the reasons explained above. In fact, our work will uncover new connections among the Swampland Conjectures, which supports the perspective that all these conjectures should not be regarded as independent statements but just as different faces of the same underlying quantum gravity principles. In particular, we will show that the Distance \cite{Ooguri:2006in} and the asymptotic de Sitter  \cite{Grimm:2019ixq} conjectures result as a consequence of applying the WGC for strings and membranes.

We will focus on describing the physics of BPS strings and membranes in 4d $\cN=1$ EFTs, where these objects become particularly significant. Indeed, controlled settings to check the Swampland conjectures usually require to deal with BPS objects in a supersymmetric setup. However, in 4d $\cN=1$ EFTs, only strings and membranes can be BPS. In this sense, a proper analysis of these objects seems essential in order to extend the tests of the conjectures -- up to now mostly performed in settings with 8 supercharges or more -- to more phenomenologically attractive theories like 4d $\cN=1$ EFTs.

In particular, we will consider {\em fundamental} strings and membranes, namely those satisfying $\Lambda^p<\cT_{p-1}< M_{\rm P}^2\Lambda^{p-2}$ for an EFT with cut-off $\Lambda$. We will argue that the aforementioned  difficulties associated with low-codimension objects  can be avoided by treating them  as localised operators entering the EFT, rather than as states in a given vacuum. One can then use the bulk-plus-brane EFT to study the low energy dynamics of the complete system. In the spirit of \cite{Goldberger:2001tn,Michel:2014lva,Polchinski:2015bea}, the classical backreaction can then be understood as a classical RG flow of the brane couplings. This induces a flow of the light scalar fields which makes the brane tension already scale dependent at the classical level. Hence, the tension entering in the EFT action should be regarded as the tension $\cT$ of the object  evaluated at the cut-off scale $\Lambda$.  For codimension $\geq 2$, the brane couplings correspond to irrelevant operators, while for codimension 2 and 1 they become marginally relevant and relevant operators respectively. This implies that we can only study the properties of these objects at sufficiently high cut-off scale in an appropriate perturbative regime, which is equivalent to be located close enough to the localised object.

The RG flow induced by the backreaction of the strings forces some scalars to travel an infinite field distance when approaching the string core. In this way it is possible to probe a particular asymptotic limit in moduli space selected by the string charge. This hints to a one-to-one correspondence between strings and infinite field distance limits that leads us to propose the \emph{Distant Axionic String Conjecture}, which will be analysed in detail in a companion paper \cite{Lanza:2021qsu}. Essentially, the conjecture states that all infinite distance limits of a 4d EFT can be realised as an RG flow endpoint of a fundamental axionic string. When turning on a potential for the scalars, some strings become anomalous and membranes need to end on them. We show that, if the mass for the scalars is below the EFT cut-off $\Lambda$, then there is a sublattice of fundamental membranes such that the anomaly can be cured at the EFT level, which is a non-trivial consistency check. Indeed, the EFT lattice of fundamental membranes $\Gamma_{\rm EFT}$ guarantees that the energy scales of the induced potential are compatible with the EFT regime, and it depends on the perturbative/asymptotic regime under consideration, which is in turn selected by the string flow. We discuss how to identify this EFT lattice and analyse the scalar flow induced by the membrane backreaction when moving away from it.

Simply using supersymmetry, we derive some off-shell identities relating the charge and the tension that any BPS string and membrane in 4d $\cN=1$ theories need to satisfy. They can be interpreted as  conditions for the balance of forces between two identical strings or membranes, although this interpretation only makes sense at a high enough cut-off, i.e. close enough to the object. In particular, the F-term potential that can be dualised to 3-form gauge fields can always be interpreted as a no-force condition for a membrane with a charge $\cQ^2= 2 V$. A Repulsive Force Condition \cite{Palti:2017elp,Heidenreich:2019zkl} could then potentially constrain the supersymmetry breaking mechanisms in the EFT. 

Taking the EFT viewpoint, we revisit the interpretation and connections among the Swampland conjectures for low codimension objects. We study a possible interpretation of the WGC for strings and membranes and point out that, unlike the no-force identity, saturating the WGC implies a specific behaviour of the K\"ahler potential which is indeed characteristic at the asymptotic limits in string compactifications. We then show that satisfying the WGC for strings implies that the string tension goes to zero exponentially in terms of the proper field distance as approaching the core of the string. Since the string tension acts as a cut-off of the EFT, this is precisely a realisation of the Swampland Distance Conjecture (SDC) \cite{Ooguri:2006in} in $\cN=1$ settings, namely that there is always an infinite tower of states becoming exponentially light at the infinite field distance limits. In this case, the tower would correspond to the string excitation modes, as the Emergent String Conjecture \cite{Lee:2019wij} implies in some instances, and the exponential rate is fixed in terms of the extremality bound for the string.  If our Distant Axionic String Conjecture indeed holds in any setting, this would imply that the SDC is always just a consequence of a string satisfying the WGC! And the string extremality factor would always provide a universal bound for the exponential rate of the tower. Furthermore, the RG flow interpretation provides evidence for the emergence proposal \cite{Grimm:2018ohb,Heidenreich:2017sim} as the infinite distance emerges indeed from integrating out high energy modes as $\Lambda\rightarrow 0$. Similarly, we show that membranes saturating the WGC imply a contribution to the scalar potential that always satisfies the initial de Sitter conjecture  \cite{Obied:2018sgi}. More precisely, since we can only identify WGC-saturating membranes at the perturbative regimes, this provides a concrete realisation of the asymptotic version of the de Sitter conjecture \cite{Grimm:2019ixq}, yielding runaway potentials towards infinite field distance.

The outline of the paper goes as follows. In Section~\ref{sec:N=1EFT} we summarise how to include fundamental BPS strings and membranes in $\mathcal{N}=1$ supergravity theories and show that they satisfy some off-shell identities, which can be na\"ively interpreted as no-force conditions between them. In Section~\ref{sec:EFTRG} we illustrate how to properly treat these low-codimension objects within the EFT, regarding them as localised operators of the EFT and understanding their backreaction as an RG-flow thereof. In Section~\ref{s:asymptotic} we study the flow induced by elementary strings and membranes by considering specific asymptotic limits in field space and motivate the Distant Axionic String Conjecture, to be discussed in more detail in \cite{Lanza:2021qsu}. Finally, in Section~\ref{sec:SC} we show how elementary strings and membranes contain crucial information that allows us to revisit several Swampland Conjectures and to thread previously undisclosed links among them. Specifically, after showing how to properly formulate a WGC and a Repulsive Force Conjecture for strings and membranes, we will exhibit how elementary strings provide tools to explore infinite distance limits in relation to the SDC, and how membranes can be used to infer constraints on the potential, such as those required by the de Sitter conjecture.  

Technical details have been relegated to the appendices. Appendix \ref{app:RFC_Id} shows and interprets perturbatively the no-force identities for strings and membranes. Appendix \ref{app:multimoduli} describes the flux lattice structure in perturbative limits involving several fields. 

\emph{Note added:} While this paper was being prepared for submission we received \cite{Herraez:2020tih} which also points out the interpretation of the $\cN= 1$ F-term potential as a no-force condition for membranes. It therefore has partial overlap with section \ref{sec:NFI} and appendix \ref{app:RFC_Idm}.

\section{Extended objects in \texorpdfstring{\hbox{$\mathcal N=1$}}{N=1} EFTs}
\label{sec:N=1EFT}

Low-codimension extended objects are ubiquitous in EFTs. In particular, BPS strings and membranes arise naturally in 4d ${\cal N}=1$ theories with approximate shift symmetries and F-term potentials. Particularly interesting for our purposes are \emph{fundamental} strings and membranes, namely those whose tensions satisfy
\be\label{EFTregime}
\Lambda^{2} < \cT_{\rm str} < M^{2} _{\rm P} \, ,  \qquad \text{and}\qquad  \Lambda^3 < \cT_{\rm mem} < M^{2} _{\rm P}\Lambda\, ,
\ee 
respectively, where $\Lambda$ is the effective cut-off scale. Such objects must be included as localised operators in the theory, in the sense that they cannot be resolved within the EFT regime of validity.  In this section we review the description of these objects from the EFT viewpoint, and in particular the field-dependent expressions for their tensions and physical charges obtained in terms of dual effective actions. Such expressions directly lead to identities which, at least naively, can be interpreted as no-force conditions among extended objects of equal charge. They also allow one to characterise fundamental objects as those with mild backreaction effects compared to $\Lambda$. In fact, for low-codimension objects the notions of tension and mild backreaction are subtle. Therefore, some of the statements made in this section can only be made precise after our discussion in section \ref{sec:EFTRG}.

\subsection{Strings and membranes in four dimensions}
\label{sec:strmemgen}

The standard 4d $\cN=1$ bosonic effective action for a set of chiral multiplets $\{\phi^\alpha \}$ reads
\be
S=\int \left(\frac{M^2_{\rm P}}{2}R*1-M^2_{\rm P} K_{\alpha \bar \beta}\,\d \phi^\alpha\wedge*\d\bar\phi^{\bar \beta} -  V * 1 \right)\, ,
\label{effaction}
\ee
with $R$ the Ricci scalar, $K_{\alpha\bar{\beta}}\equiv \partial_\alpha\partial_{\bar \beta}K$ the K\"ahler metrics and $V$ the scalar potential. If the latter is an F-term potential coming from a superpotential of the form
\be
 W (\phi)=M_{\rm P}^3\, f_A  \Pi^A (\phi) \, , \qquad f_A \in \bbZ\, ,
 \label{supoflux}
\ee
then the potential can be expressed via the Cremmer et al. formula as
\be
V = \half T^{AB} f_A f_B\, ,
\label{Vflux}
\ee
with
\be
T^{AB} \equiv 2M^4_{\rm P}\, e^{K}\Re\left(K^{\alpha\bar\beta}D_\alpha\Pi^A \bar D_{\bar\beta} \bar\Pi^B- 3\Pi^A \bar\Pi^B\right)\,, 
\label{TAB}
\ee
where we employ the standard notation $D_\alpha\Pi^A \equiv \Pi^A_\alpha+K_\alpha \Pi^A$, with  $\Pi^A_\alpha\equiv\partial_\alpha\Pi^A$,  $K_\alpha\equiv\partial_\alpha K$, and with $K^{\alpha\bar\beta}$ the inverse of $K_{\alpha\bar{\beta}}$. The superpotential \eqref{supoflux} is quite common in 4d string compactifications with fluxes, with the so-called periods $\Pi^A (\phi)$ being regular holomorphic functions of the chiral fields, and $f_A$ the flux quanta. One may of course generalise the above expressions to include non-perturbative corrections to the superpotential, as we do below. 

In certain regimes, it is useful to resort to a dual formulation of the above effective action. A typical case is when there are periodic directions in moduli space that are promoted to approximate axionic shift symmetries. That is, for some fields $\{t^i\} \subset \{\phi^\alpha\} = \{t^i, \chi^\kappa\}$ with periodic directions
\be
\Re t^i\, \simeq \,\Re t^i + e^i\, , \qquad e^i \in \bbZ\, ,
\ee
an approximate continuous isometry $\Re t^i\, \rightarrow \Re t^i + \lambda\, e^i$, $\lambda \in \bbR$ is developed for the field space metric. Then each chiral field $t^i$  can be dualised to a linear multiplet, containing a two-form potential $\cB_{2\,i}$ and a dual saxion $\ell_i$. The axionic shift symmetries on $a^i = \Re t^i$ are made manifest by a K\"ahler potential $K$ that only depends on the saxionic components $s^i \equiv \Im t^i$. We then have that the dual variables are defined as
\be\label{dualfields}
\ell_i=-\frac12 \frac{\del K}{\del s^i}\, , \qquad \cH_{3\, i} = \d  \cB_{2\, i} =-M^2_{\rm P}\, \cG_{i j}*_4\d\Re t^j\, ,
\ee
where 
\be\label{GGmetric}
 \cG_{i j}\equiv \frac12 \frac{\del^2 K}{\del s^i\del s^j}\ .
\ee 
The kinetic terms in \eqref{effaction} which are quadratic in  the derivatives of the fields $t^i$ are dualised to 
\be\label{dualaction}
-\frac12\int \cG^{ij}\left( M^2_{\rm P}\,\d\ell_i\wedge *\d\ell_j+\frac{1}{M^2_{\rm P}}\cH_{3\, i}\wedge * \cH_{3\, j} \right)\, ,
\ee
to which we may easily add the action of a 4d string, coupling to the two-forms with charges $e^i$
\be\label{stringS}
S_{\rm string}=-M^2_{\rm P}\,\int_\cS  |e^i\ell_i|\sqrt{-h}+e^i \int_\cS \cB_{2\, i}\, ,
\ee
that is, implementing the monodromy $a^i \rightarrow a^i + e^i$. From here, it is straightforward to read the string tension and physical charge
\be
\label{strten}
\cT_{\bf e} = M^2_{\rm P} |e^i \ell_i|\, , \qquad \cQ_{\bf e}= M_{\rm P} \sqrt{\cG_{ij}e^i e^j}\, .
\ee
In fact, it is in this dual formalism that the definition of tension and physical charge of strings can be made precise, as well as the different interactions among them. 

Similarly, one can give a dual description of flux quanta in \eqref{Vflux} in terms of three-form potentials $A_3$ to which 4d membranes couple \cite{Bielleman:2015ina,Carta:2016ynn,Herraez:2018vae}. Doing it supersymmetrically is a subtle procedure, but it can be implemented following the approach in \cite{Farakos:2017jme,Farakos:2017ocw,Bandos:2018gjp,Lanza:2019xxg,Lanza:2019nfa}. Remarkably, it was found in \cite{Lanza:2019xxg} that in general not all flux quanta can be dualised to three-forms. This obstruction originates from different requirements, the most relevant for this work being that only a subset of membranes mediate transitions consistent with the EFT UV cut-off $\Lambda$. The membranes whose transitions are consistent with the cut-off form a lattice, dual to the lattice of {\it dynamical} fluxes: 
\be\label{GammaEFT}
\Gamma_{\rm EFT} = \left\{ \text{Fluxes dualisable to 3-forms \& coupled to membranes satisfying } \frac{\cT_{\rm mem} }{M^2_{\rm P}} <  \Lambda  \right\} .
\ee
We will come back to the above definition in section \ref{sec:EFTmemlat}.

Once that $\Gamma_{\rm EFT}$ has been identified, one can rewrite the superpotential \eqref{supoflux} as 
\be
\label{Wfa}
 W (\phi)=M_{\rm P}^3\, f_a  \Pi^a(\phi) + \hat{W}(\phi)\, ,
\ee
where $f_a$ belongs to $\Gamma_{\rm EFT}$ and $\hat{W}$ can now  include, in addition to the contribution  of non-dynamical fluxes, other  contributions like those associated with non-perturbative effects. The scalar potential \eqref{Vflux} then splits as 
\be\label{potsplit}
V=\frac12 T^{ab}f_a f_b +f_a \Upsilon^a + \hat V\, ,
\ee
where $T^{ab}$ is the restriction of \eqref{TAB} to $\Gamma_{\rm EFT}$, thanks to which it has an inverse $T_{ab}$, and
\begin{subequations}\label{hatV}
\begin{align}
\Upsilon^a &=2 M_{\rm P}\,e^{K}\Re\left(K^{i\bar\jmath}D_i\hat W\bar D_{\bar\jmath}\bar\Pi^a-3\hat W\bar\Pi^a\right)\, ,\label{hatV1}\\
\hat V&=\frac{e^K}{M^2_{\rm P}}\left(K^{i\bar\jmath}D_i \hat W\bar D_{\bar\jmath}\overline{ \hat W}-3|\hat W|^2\right)\, .\label{hatV2}
\end{align}
\end{subequations}
Assuming that $T^{ab}$ is independent of the $f_a$,\footnote{A flux-independent $T^{AB}$ follows from a superpotential \eqref{supoflux}  linear on the fluxes, and the latter is equivalent to a  canonical 3-form kinetic term $F_4 *F_4$ (i.e.\ quadratic on its field strength) typical of a perturbative regime.} we may dualise the potential term in \eqref{effaction} to
\be\label{dualF4lagr}
\begin{aligned}
S_{\text{3-forms}}=&-\int_{X_4} \Big[\frac12 T_{ab}\,F^a_4*\!F^b_4+ T_{ab}\Upsilon^aF^b_4+\Big(
\hat V-\frac12 T_{ab}\Upsilon^a \Upsilon^b\Big)*\!1\Big] \\
&+\int_{\del X_4}T_{ab}\left(* F^a_4+\Upsilon^a\right)A^b_3\, ,
\end{aligned}
\ee
where $A^a_3$ are three-form potentials and  $F_4^a = {\rm d} A_3^a$ the associated four-form field strengths. One recovers the previous expression for the potential after solving their equations of motion
\be\label{F4vevb}
 T_{ab}(*F^b_4+\Upsilon^b)= -f_a\, ,
 \ee
and inserting the solution into \eqref{dualF4lagr}. In this formalism one can add terms of the form $\cQ_I\int C^I_4$ implementing  tadpole conditions, with $C^I_4$ some four-form potentials gauging some three-forms $A_3^a$ \cite{Lanza:2019xxg}. In standard string theory examples these tadpole conditions read
\beq
\label{tadpole}
\cQ_I = \langle f , f \rangle_I +  {Q}_I^{\rm bg} = 0\,, 
\eeq
with ${Q}_I^{\rm bg}$ some fixed flux-independent quantity, for instance depending on the O-plane content, and $\langle \cdot , \cdot \rangle_I$ quadratic symmetric pairings on the fluxes. 
As shown in \cite{Lanza:2019xxg}, $\Gamma_{\rm EFT}$ must  be such that these pairings become linear after fixing the non-dynamical fluxes, a condition that can be made compatible with the definition \eqref{GammaEFT} in typical string theory examples.

\begin{center}
	\begin{figure}[thb]
		\centering
		\includegraphics[width=7cm]{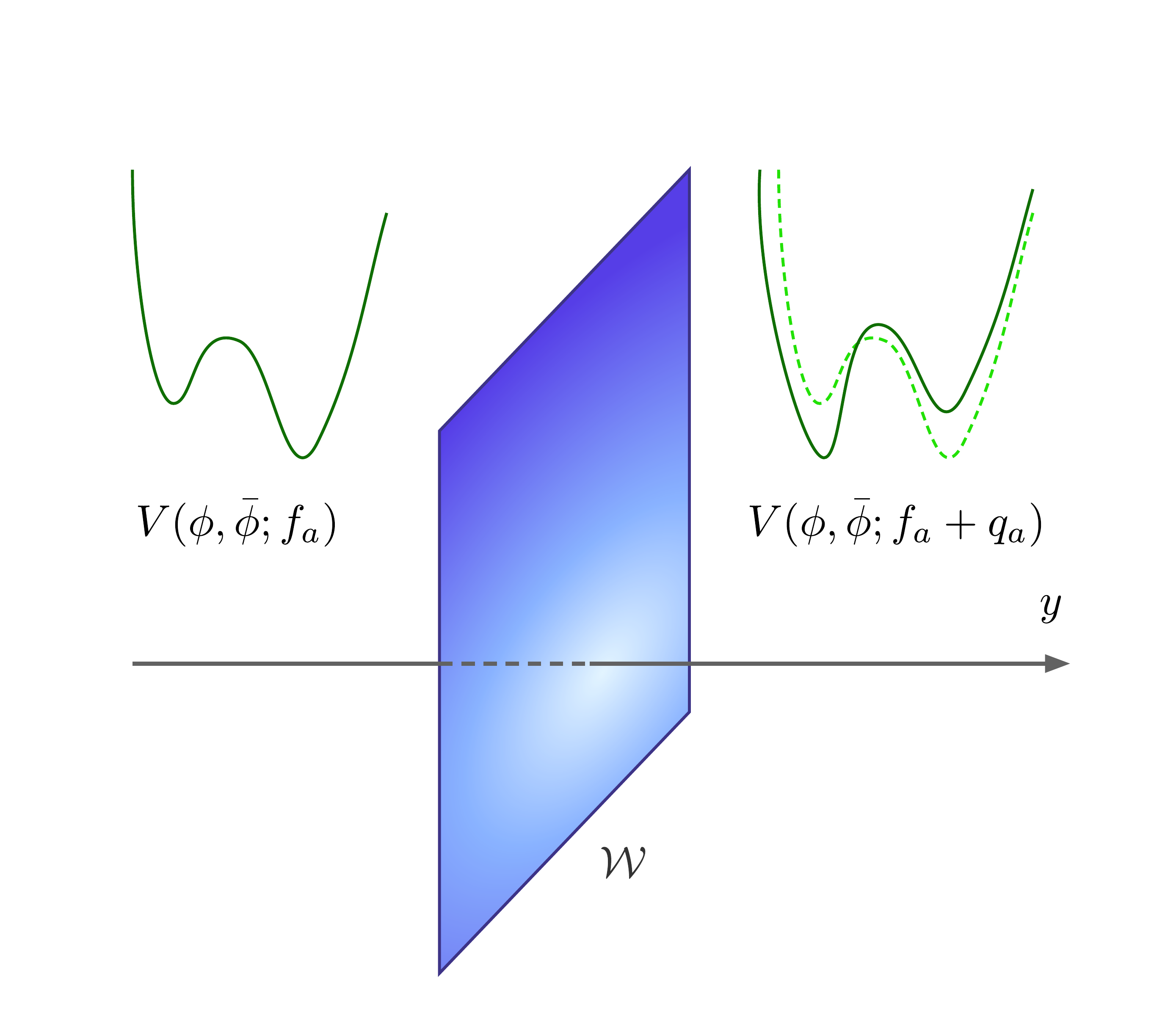}
		\caption{A membrane inducing a jump on the fluxes, thus modifying the potential in the different regions separated by the membrane. \label{Fig:MemStep}}
	\end{figure}
\end{center}

In this dual framework one can also describe the action of a membrane with charges $q_a$ as
\be\label{bosmem}
S_{\rm mem}=-2 M^3_{\rm P}\int_{\mathcal{W}} \d^3\zeta\, e^{\frac{1}{2}K}\left|q_a \Pi^a \right|\sqrt{-\det h}+q_a\int_{\mathcal{W}} A^a_3\, .
\ee
A membrane implements the flux jump $f_a \rightarrow f_a + q_a$, with $q_a \in \Gamma_{\rm EFT}$, directly at the level of effective description, resulting in a potential which is differently defined on its two sides, as depicted in Fig.~\ref{Fig:MemStep}. The membrane tension and physical charge are defined as follows
\be\label{memT}
\cT_{\bf q}=2 M^3_{\rm P}\,e^{\frac12K}\left|q_a \Pi^a \right| \, , \qquad  \cQ^2_{\bf q}=  T^{ab}q_aq_b\, .
\ee
One can further extend the above supersymmetric formulation by  gauging some two-form potentials under the gauge symmetries of the three-form potentials, and correspondingly add to the EFT bound-states of open membranes ending on strings  \cite{Lanza:2019xxg}.   

\vspace{-1cm}

As expected from string compactifications, the physical charge and tension of both membranes and strings depend on the fields $\{\phi^\alpha\}$ describing the EFT field space $\MM$. From the field theory viewpoint, however, we are used to think of these values as dependent on the EFT cut-off $\Lambda$ as well. This observation will be central for the discussion of the next section and crucial for the rest of the paper, as it will provide a vantage point to analyse swampland criteria.

\subsection{No-force identities}
\label{sec:NFI}

Preserving part of the bulk $\mathcal{N}=1$ supersymmetry over the worldvolume of strings and membranes leads to stringent constraints over their tensions and charges. As explained above,  these   ought to depend on the bulk fields as in \eqref{strten} and \eqref{memT}. 
We now show that this dependence is related to some important off-shell identities.
First, let us notice that \eqref{strten} and \eqref{memT}  are general expressions  derived just from assuming $\caln=1$ supersymmetry and shift symmetries. Consequently, the following identities hold as long as the underlying EFT is supersymmetric.

Indeed, one can easily check that the string tension \eqref{strten} obeys the identity
\be
\label{No-fo_Ids}
 \|\del\cT_{\rm str}\|^2  = M_{\rm P}^2\cQ^2_{\bf e} \, ,
\ee
where $\|\del\cT_{\rm str}\|^2  \equiv \cG_{ij} \partial_{\ell_i} \mathcal{T}_{\rm str} \partial_{\ell_j} \mathcal{T}_{\rm str}$, 
and the physical charge $\cQ^2_{\bf e}$ is defined as in \eqref{strten}.
The tension \eqref{memT} of a supersymmetric membrane instead satisfies the equality
\be
\begin{split}
	\label{No-fo_Idm}
	&\|\del\cT_{\rm mem}\|^2 - \frac32 \cT_{\rm mem}^2  = M_{\rm P}^2  \cQ_{\bf q}^2\,,
\end{split}
\ee
where we have denoted $\|\del\cT_{\rm mem}\|^2  \equiv 2 K^{\alpha\bar \beta} \del_{\alpha} \cT_{\rm mem} \bar\del_{\bar \beta} \cT_{\rm mem}$ and introduced the physical charge as in \eqref{memT}. The detailed proof of the identity \eqref{No-fo_Idm} can be found in Appendix~\ref{app:RFC_Idm}. Intuitively, it can be understood as the Cremmer et al. F-term scalar potential generated by dynamical fluxes, and then identifying such fluxes with fundamental membrane charges.

It is important to remark that the identities \eqref{No-fo_Ids} and \eqref{No-fo_Idm} directly rely on the bulk $\cN=1$ supersymmetry. On the one hand, SUSY constrains the kinetic matrices for the gauge two- and three-forms as they appear in \eqref{dualaction} and \eqref{dualF4lagr}. On the other hand, it forces the string and membrane tensions to acquire the precise form \eqref{strten} and \eqref{memT}. As a result, the only information about the extended object that appears in the relations \eqref{No-fo_Ids} and \eqref{No-fo_Idm}  is just their charge: the string charges $e^i$ in the former and the membrane charges $q_a$ in the latter.
Furthermore, both the identities \eqref{No-fo_Ids} and \eqref{No-fo_Idm} are realised \emph{off-shell}, that is independently of the point of field space, which may or may not preserve supersymmetry and not even be a vacuum of the theory.  In this sense, \eqref{No-fo_Ids} and \eqref{No-fo_Idm} can be regarded as tautological, as long as the underlying bulk theory and the strings or membranes are $\cN=1$ supersymmetric.

From the EFT viewpoint, the identities \eqref{No-fo_Ids} and \eqref{No-fo_Idm} entail interesting phenomenological implications. More precisely, they may be interpreted as conditions for the balance of mutual forces between two identical strings or membranes. 

In fact, in a $d$-dimensional theory, given $(p+1)$-dimensional objects with codimension  greater than two, that is $d-p-3>0$, it does make sense to define states asymptotically, where their backreaction is negligible. We may then regard the states as classical sources, interacting among each other. Consider, for instance, two identical $p$-branes, charged under the gauge fields $A_{p+1}^a$ with charge $q_a$ and whose tensions $\cT(\phi)$ depends on some real bulk scalar fields $\varphi^n$. They exert to each other the following mutual forces per unit area \cite{Polchinski:1996fm,Heidenreich:2019zkl}
\begin{equation}
\label{No-fo_Ftot}
    F_{\rm tot} = \frac{f_{\rm tot}}{r^{d-p-1}}\,,
\end{equation}
with $r$ their mutual distance and
\begin{equation}
\label{No-fo_RFCeq}
	f_{\rm tot} = F^{ab} q_a q_b - G^{nm} \partial_n \cT \partial_m \cT  - \frac{(p+1)(d-p-3)}{d-2} \cT^2\, ,
\end{equation}
where $F^{ab}$ and $G^{nm}$ are the inverse, respectively, of the gauge kinetic functions $F_{ab}(\varphi)$ and scalar kinetic matrix $G_{nm}(\varphi)$. In writing \eqref{No-fo_Ftot} we have neglected possible contributions which are subleading in the distance. A positive (negative) contribution to $f_{\rm tot}$ means a repulsive (attractive) force, and each contribution in \eqref{No-fo_RFCeq} has a clear interpretation. The first one comes from the electric repulsion between two branes, while the second contribution is attractive and due to the interactions of the branes with the bulk scalar fields $\varphi^n$. The last contribution in \eqref{No-fo_RFCeq} is the gravitational one: for branes of codimension greater than two, gravity is always attractive. Two identical objects satisfy the \emph{no-force condition} $f_{\rm tot} = 0$, whenever the electric repulsion is exactly balanced by their scalar and gravitational attractions.

The promotion of $f_{\rm tot} = 0$ to a no-force identity for codimension-two and codimension-one objects, like 4d strings and membranes, requires particular care. As will be explored in the next section, these have a strong effect on the IR physics, resulting in the impossibility to define the notion of asymptotic, free state. Still, one may include membranes and strings within effective descriptions. In particular, as explained in Appendix~\ref{app:RFC_Id}, under certain assumptions it makes sense to compute an effective potential between two identical strings or membranes, as long as the IR physics is not involved. Then, from the effective potential, one can extrapolate the mutual forces between these objects.

In the case of strings, given the action \eqref{dualaction}, one can show that the total force between two identical strings is given by
\begin{equation}
\label{No-fo_RFCq}
F_{\rm tot} = \frac{f_{\rm tot}}{r}\,, \qquad {\rm with}\quad f_{\rm tot} =  M_{\rm P}^2\cQ^2_{\bf e} - \|\del\cT_{\rm str}\|^2\, .
\end{equation}
Thus, requiring the no-force condition coincides with \eqref{No-fo_Ids}. Comparing this with \eqref{No-fo_RFCeq}, we recognise that two strings do not exert \emph{any} gravitational force between each other \cite{Vilenkin:1981zs,Dabholkar:1990yf,Buonanno:1998kx}. The balance of forces between them can then be achieved if and only if  the electric repulsion, which is mediated by the gauge two-forms $\cB_{2\, i}$, is compensated by the scalar interaction, mediated by the dual saxions $\ell_{i}$. The BPS-strings \eqref{strten} do satisfy the no-force condition trivially.

Performing a field theory computation as in Appendix~\ref{app:RFC_Idm} (see also \cite{Herraez:2020tih}), we find out that the net force between two membranes is constant and given by
\begin{equation}
\label{No-fo_RFCm}
F_{\rm tot} =  M_{\rm P}^2  \cQ_{\bf q}^2 - \|\del\cT_{\rm mem}\|^2 + \frac32 \cT_{\rm mem}^2  \,,
\end{equation}
and thus \eqref{No-fo_Idm} can be interpreted as a no-force condition between two identical membranes. In this case, some comments are in order. First, in \eqref{No-fo_RFCm} the gravitational force is repulsive \cite{Vilenkin:1981zs,Garriga:2003gv}. Moreover, the gauge kinetic matrix $T_{ab}$ for the three-form in \eqref{dualF4lagr} may not be positive definite. Therefore, the electric force contribution  $M_{\rm P}^2  \cQ_{\bf q}^2$ in \eqref{No-fo_RFCm} may be either positive or negative. The no-force condition that is read from \eqref{No-fo_RFCm} then tells that two identical membranes do not exert forces between one another whenever the scalar attraction and the gravitational repulsion are balanced by the electric forces, which can be either attractive or repulsive. Recall that \eqref{No-fo_Idm} simply amounts to the F-term potential generated by fundamental membranes, and that is why the no-force condition for them is valid off-shell.

While quite suggestive, eqs.\eqref{No-fo_RFCq} and \eqref{No-fo_RFCm} rely on the assumption that interactions among strings and membranes occur in an approximately flat background, see Appendix \ref{app:RFC_Id}. However, the mere presence of these low-codimension objects substantially modifies the background asymptotics through their backreaction. As we will discuss in section \ref{sec:EFTRG}, to properly interpret the identities \eqref{No-fo_Ids} and \eqref{No-fo_Idm}, it is necessary to develop a precise EFT understanding of the tensions and physical charges of strings and membranes.

\subsection{EFT lattice of membranes}
\label{sec:EFTmemlat}

A direct consequence of \eqref{No-fo_Idm} is that it allows one to draw a more precise understanding of the lattice of dynamical fluxes $\Gamma_{\rm EFT}$, as defined in \eqref{GammaEFT}. Indeed, on general grounds, within an EFT with cut-off scale $\Lambda$ a membrane will be considered \emph{fundamental} if its thickness is below $\Lambda^{-1}$. Describing such a membrane at the EFT level will only be possible if the typical bulk energy scales characterising the membrane backreaction are below $\Lambda$. Needless to say, this must be combined with the condition 
\be\label{TTT}
\calt_{\bf q} \geq \Lambda^3\, ,
\ee
or else the vibrational modes of the membrane will enter the EFT regime. 

A natural energy scale regulating the gravitational membrane backreaction may be identified with  $\calt_{\bf q}/{M^2_{\rm P}}$. In particular, the jump in the trace $\calk$ of the extrinsic curvature induced by the membrane is proportional to  $\calt_{\bf q}/M^2_{\rm P}$ \cite{Brown:1988kg}.
The trace of the extrinsic curvature can be considered as an estimate of the  energy scale associated with the slope change of the metric around the membrane. By imposing the EFT requirement $\Delta\calk$ is small with respect to $\Lambda$, we get the condition $\calt_{\bf q}/(M_{\rm P}^2\Lambda) < 1$ should be small enough, in agreement with definition \eqref{GammaEFT}.

Similarly, a membrane induces  a jump proportional to $\del_{\alpha}\calt_{\bf q}(\phi)$  
in the gradient of the scalars as one crosses the membrane along the normal direction. We can then take $\|\del\calt_{\bf q}\|/M^2_{\rm P}$, where $\|\del\calt_{\bf q}\|$
was introduced in \eqref{No-fo_Idm}, as an estimate of the energy scale associated with such a transition. The validity of the EFT regime imposes that this energy scale should be small compared to $\Lambda$:
\be
\label{condm2}
\frac{\|\del\calt_{\bf q}\|}{M^2_{\rm P}}< \Lambda\, .
\ee
Finally, a similar argument applied to the energy scale associated with the jump $q_a$ in the flux quanta $f_a$ induced by the membrane leads to the EFT condition that $\calq^2_{\bf q}/(M^2_{\rm P}\Lambda^2)$ should be small too:
\be\label{condm3}
\frac{\calq^2_{\bf q}}{M^2_{\rm P}}< \Lambda^2\, .
\ee 

To have a consistent EFT description of fundamental membranes,  \eqref{condm2} and \eqref{condm3} should be imposed on top of  $\calt_{\bf q}/M^2_{\rm P} < \Lambda$. However, in an $\caln=1$ EFT the off-shell identity \eqref{No-fo_Idm} shows that these three conditions are not independent of each other: by imposing two of them automatically guarantees the third one.  Furthermore, as will be discussed in section \ref{sec:WGCmem}, under certain circumstances which are typically realised in string/M-theory setups, a membrane counterpart of the WGC relation  $\calt^2_{\bf q}\sim M_{\rm P}^2\calq_{\bf q}^2$ holds. Then one may obtain a single independent EFT condition, say $\calt_{\bf q}/M^2_{\rm P} <  \Lambda$, which is the one appearing in the definition of the EFT flux lattice. This explains why the definition \eqref{GammaEFT} captures the necessary conditions for membranes to describe dynamical fluxes, at least in the string/M-theory examples analysed in \cite{Lanza:2019xxg}. In fact, the discussion of \cite{Lanza:2019nfa} clearly indicates that, at least in string/M-theory models, these EFT conditions can be satisfied by considering large distance regimes in field space and  by restricting to corresponding sublattices of membranes. In the next sections we will develop this idea and relate it to the existence of  asymptotic field space limits associated with string RG flows.

\section{Low codimension branes and EFT RG flows}
\label{sec:EFTRG}

In this section, we discuss in more detail the EFT interpretation of ${\cal N}=1$ supergravity theories coupled to strings and membranes introduced in section \ref{sec:N=1EFT}. This will provide an appropriate conceptual framework to formulate possible swampland criteria for such extended objects, as well as other codimension $\leq 2$ branes in more general theories.

Indeed, as follows from our previous discussion, a straightforward extension of the usual swampland arguments to electrically charged localised objects of codimension $\leq 2$  is not obvious. The difficulty is related to the fact that their presence strongly affects the asymptotic vacuum structure of the theory. For such objects, one cannot naively identify tension, charge, and  associated extremality bounds in terms of some asymptotic fluxes, as  usually done for higher codimension black branes. 

These difficulties, however, are overcome if one considers these codimension $\leq 2$ branes as localised operators entering the EFT, rather than as states of a given vacuum sector of the theory, and then use the complete bulk-plus-brane EFT to study the low-energy dynamics of the complete system. In fact, as long as the corresponding gauge field is part of the elementary degrees of freedom, the completeness criterion \cite{Polchinski:2003bq} requires the introduction of the corresponding charged brane operators. 
From this perspective, any swampland criterion for codimension $\leq 2$ branes should be {\em directly} formulated in terms of the corresponding contribution to the EFT. Needless to say, they may have important physical implications.

Let us consider a $d$-dimensional bulk EFT that contains a set of scalars $\phi^\alpha$ and their supersymmetric partners. Then, just by diffeomorphism invariance, the leading bosonic contribution of a charged $p$-brane to the EFT takes the form\footnote{We neglect the presence of possible world-volume fields, besides the geometric ones describing the embedding.}
\be\label{genbrane}
-\int\d^{p+1}\xi\, \calt(\phi)\sqrt{-h} +e\int \calb_{p+1}\, ,
\ee
where $\sqrt{-h}$ is the induced volume density, $\calt(\phi)$ represents the   tension of the brane, and $e$ is its quantised charge. In fact, we will consider the dependence of the tension $\calt(\phi)$ on the scalars as completely fixed by supersymmetry, as in the four-dimensional models described in section \ref{sec:N=1EFT}.  

Of course, we are assuming that \eqref{genbrane} can be treated in the semiclassical EFT regime. So, the relevant time and length scales describing the dynamics of the brane should be large compared to $\Lambda^{-1}$.  Furthermore, consider for instance a(n unstable) configuration in which a $p$-brane is rolled to form a $p$-sphere $S^p$ of radius which is of the same order of the UV length-scale, $L\sim \Lambda^{-1}$. The corresponding mass $M_*$ would be of order $\calt \Lambda^{-p}$.  This would be compatible with the semiclassical EFT regime only for $M_*\geq \Lambda$, from which we obtain the EFT lower bound  
\be \label{Tlambda}
\calt \geq  \Lambda^{p+1}\, ,
\ee
on the $p$-brane tension. For fundamental strings  this reduces to the  bound $ T_{\rm F1}\sim \frac{1}{\alpha'}\geq \Lambda^2$, familiar for the UV cut-off scale of ten-dimensional supergravity seen as an EFT of string theory. 

Now, it is important to realise that a proper interpretation of the brane contribution \eqref{genbrane} requires some care, in particular for codimension $\leq 2$ branes. Indeed, in general, a brane  (classically) backreacts on the underlying geometry and, furthermore, this backreaction typically diverges on the brane. So, at first sight it seems that, once one tries to take into account the brane's backreaction, the localised terms \eqref{genbrane} do not make sense anymore. However, as emphasised for instance in \cite{Michel:2014lva,Polchinski:2015bea},  terms of the form \eqref{genbrane} make perfect sense even after their backreaction is taken into account, as long as we appropriately interpret them in the general framework of EFTs \cite{Weinberg:1978kz}: the brane couplings should be regarded as defined at the EFT cut-off scale $\Lambda$ and the classical brane backreaction can be interpreted in terms of a {\em classical} RG-flow of the brane couplings, in the spirit of \cite{Goldberger:2001tn}. In particular, the brane backreaction induces a flow of the light scalar fields and then the brane tension $\calt(\phi)$ typically becomes scale dependent already at the classical level.  This observation will be of crucial importance in this paper.

For codimension $>2$ branes, the background deformation induced by the brane  typically goes to zero at large distances from the brane. For instance, the scalar fields $\phi^\alpha$ should flow, at large distances, to their background values $\phi^\alpha_0$  in absence of the brane.  
From the EFT viewpoint of \cite{Goldberger:2001tn,Michel:2014lva,Polchinski:2015bea}, in which one may identify the minimal accessible  distance from the brane with $\Lambda^{-1}$, such a back-reaction implies that the brane couplings should be considered as {\em irrelevant} and that $\calt_0\equiv \calt(\phi_0)$ can be interpreted as the IR fixed point value of the brane tension, corresponding to $\Lambda\rightarrow 0$. In other words, for low enough cutoff scale $\Lambda$, one can  reliably  use the brane action \eqref{genbrane} in the probe approximation. In order to better estimate the maximal  $\Lambda$ at which the probe-approximation breaks down and the  RG-flow becomes relevant, one should look at the backreaction of the brane, which generically depends on the details of the EFT. If for instance one uses the Schwarzschild radius $\sim r_{\rm S}\equiv  \left(\calt/M^{d-2}_{\rm P}\right)^{1/(d-p-3)}$ to estimate the distance at which the gravitational backreaction becomes relevant, then the RG-flow corrections are negligible and the probe approximation is reliable  as long as $r_{\rm S} <  \Lambda^{-1}$, which is equivalent to  
\be\label{RSbound} 
\calt < \frac{M^{d-2}_{\rm P}}{\Lambda^{d-p-3}} \, .
\ee
More directly, the condition $r_{\rm S} <e  \Lambda^{-1}$ is necessary for a  black brane with finite horizon to be described by a localised action of the form \eqref{genbrane}.

For codimension $\leq 2$ branes, the story is different. While the above arguments imply that the codimension $>2$ branes are typically associated with irrelevant operators, codimension $\leq 2$ operators can be associated to relevant or marginally relevant operators. This suggests that their tensions and charges should  be defined, in an appropriate perturbative regime, at a sufficiently high cut-off scale $\Lambda$.   
The typical situation is schematically summarised in the following table:
\begin{equation}
	\label{Tab:Branec}
	\begin{tabular}{|c|c|}
		\hline
		 Codimension & Brane coupling 
		\\
		\hline \hline
	 ${\rm codim}>2$ &  irrelevant  
		\\
	${\rm codim}=2$ &	marginally relevant
		\\
	${\rm codim}=1$ &	relevant 
		\\
		\hline
	\end{tabular}
\end{equation}
To better elaborate on this point, let us go  back to the strings and membranes of the  $\caln=1$ four-dimensional EFTs described in section \ref{sec:N=1EFT}.

\subsection{String  flows}\label{sec:stringflow}

We first focus on strings in four-dimensional $\caln=1$ theories. Supersymmetry then forces their EFT contribution  to take the form \eqref{stringS} (excluding possible additional ``non-universal" world-sheet fields).   For concreteness, let us consider the simple Lagrangian associated with a single axionic chiral field $t$, with $t\simeq t+1$, and K\"ahler potential 
\be\label{simpleK}
K=- n \log \Im t\, ,
\ee
which is ubiquitous in string theory models. 
Let us split $t=a+\ii s$, with $a\simeq a+1$ the axion and $s$ the saxion, assumed to be positive.  From \eqref{dualfields} one obtains the corresponding dual saxion
\be
\label{ells_nd}
\ell=\frac{{n}}{2s}\, ,
\ee
and from \eqref{stringS} we see that the contribution to the EFT of a string of (integral) charge $e>0$ under the two-form $\calb_2$ dual to the axion $a$, is
\be\label{simplestring}
-\int\d^2\xi\, \sqrt{-h}\, \calt(\ell) +e\int \calb_{2}\, ,
\ee
with
\be
\calt(\ell)\equiv M^2_{\rm P}\, e \ell\, .
\ee
This localised source produces a flow of the scalars fields which, in the neighbourhood of a straight enough piece of string, looks like \cite{cstring}
\be\label{tsol3}
t(z)=t_0+\frac{e}{2\pi\ii} \log\frac{z}{z_0}\, ,
\ee
where the complex coordinate $z$ parametrises the transverse string directions, $z=0$ represents the location of the string and $t_0$ is the value of $t$ at an arbitrary point $z=z_0$. By setting $z=r e^{\ii\theta}$, we then see that $a=a_0+\frac{e\theta}{2\pi}$ and then the axion undergoes a monodromy $a\rightarrow a+e$ around $z=0$, while the  saxion flows in the radial direction as follows
\be\label{simplell}
s(r)=s_0-\frac{e}{2\pi}\log\frac{r}{r_0}\quad~~~~~\Leftrightarrow\quad~~~~~\ell(r)=\frac{{n}}{2s_0-\frac{e}{\pi}\log\frac{r}{r_0}}\, .
\ee
Since the string is located at $r=0$ and $\ell(r=0)=0$, by naively plugging the profile \eqref{simplell} back into   \eqref{simplestring}, one would get a string action with vanishing tension, violating the EFT bound \eqref{Tlambda}.

However, one should remember that a Wilsonian EFT, and then also the string tension, is associated with a given cut-off energy scale $\Lambda$. Hence, by identifying $\Lambda$ with a minimal distance\footnote{There are in fact logarithmic corrections to this relation \cite{Lanza:2021qsu}, which do not modify the discussion substantially.} $r_\Lambda=\Lambda^{-1}$, in the spirit of \cite{Goldberger:2001tn,Michel:2014lva,Polchinski:2015bea} it is natural to propose the following formula for the EFT tension at a cut-off scale $\Lambda$:
\be\label{Tsimple}
\frac{\calt(\Lambda)}{M^2_{\rm P}}=e\ell(r_\Lambda)=\frac{{n}e}{2s_0+\frac{e}{\pi}\log(\Lambda r_0)}\, .
\ee
The logarithmic behaviour is typical of a marginal coupling, as anticipated. The corresponding  RG-flow differential equation 
\be\label{Tdiff}
\Lambda \frac{\d }{\d\Lambda}\left(\frac{\calt}{M^2_{\rm P}}\right)=-\frac1{{n}\pi}\left(\frac{\calt}{M^2_{\rm P}}\right)^2\, ,
\ee
can be immediately integrated into
\be
\frac{\calt(\Lambda')}{M^2_{\rm P}}=\frac{1}{\frac{M^2_{\rm P}}{\calt(\Lambda)}+\frac{1}{{n}\pi}\log\frac{\Lambda'}{\Lambda}}\, ,
\ee
which can of course be obtained directly from \eqref{Tsimple}. Reversely, given a certain cut-off scale $\Lambda$, we may choose the arbitrary $r_0$ to be the minimal possible one, by identifying $r_0=r_\Lambda\equiv\Lambda^{-1}$, and write the saxionic profile \eqref{tsol3} for $r\geq \Lambda^{-1}$ in the form
\be\label{simplesr}
s(r)=\frac{{n}e\,M^2_{\rm P}}{2\calt(\Lambda)}-\frac{e}{2\pi}\log(\Lambda r)\, ,
\ee
which is clearly $\Lambda$-independent, if   $\calt(\Lambda)$ changes according to \eqref{Tdiff}. As discussed in more detail in \cite{Lanza:2021qsu}, the tension renormalisation can be also  identified with the renormalisation due to the contribution of the energy density of the flowing scalar. So, by lowering the EFT cut-off $\Lambda$ one integrates out bulk high energy modes whose contribution to  the linear energy density is then reabsorbed into a renormalisation of the effective string  tension.     

The above interpretation of the string backreaction in terms of the classical logarithmic  RG flow of the string tension has important implications. First of all, given a certain $\calt(\Lambda)$, we see that $s(r)$ starts from a certain initial value $s(r_\Lambda)$ at the minimal EFT distance $r_\Lambda=\Lambda^{-1}$ and then flows to smaller values for $r>r_\Lambda$. Notice that in quantum gravity models the axionic symmetries  should be only approximately valid in the large field distance regime $s\gg 1$ and actually broken by exponentially suppressed non-perturbative corrections of the form $e^{2\pi\ii n t}$, for $n\in \mathbb{Z}_+$.
Only in this perturbative regime we can trust the K\"ahler potential \eqref{simpleK}. So, the flow \eqref{simplesr} drives the theory to a strong coupling regime $s\simeq 0$ in which the above EFT description breaks down. This fact implies that the  string localised operators can be interpreted as   marginally relevant.

However, if $s(r_\Lambda)$ is large enough, and then $\calt(\Lambda)$ is small enough, the  breakdown should happen at an IR  strong-coupling energy scale of order
\be\label{Lstrongstring}
 \Lambda_{\rm strong}\equiv\frac{1}{r_{\rm strong}}=\Lambda\exp\left[-\frac{{n}\pi\, M_P^2}{\calt(\Lambda)}\right]\, .
\ee
Hence, for small  
\be\label{stensionbound}
\frac{\calt(\Lambda)}{M^2_{\rm P}}< 1\, ,
\ee
the strong coupling scale $\Lambda_{\rm strong}$ is exponentially smaller than $\Lambda$  and the above flows remain in the weak-coupling regime for large distances. 
So, if for instance  the string forms a circle of radius $R\gtrsim \Lambda^{-1}$, the backreaction is  not expected to exit the perturbative regime as long as $R\ll r_{\rm strong}$. Notice that \eqref{stensionbound} is the four-dimensional string counterpart  of the bound \eqref{RSbound} identified for higher-codimension branes.

According to the above discussion, given a certain  EFT cut-off $\Lambda=r_\Lambda^{-1}$ we can associate $s(r_\Lambda)$ with a background saxionic vev. If we increase the cut-off $\Lambda$,  the string tension decreases and the corresponding background saxionic vev $s(r_\Lambda)$ increases. This guarantees that the EFT description in terms of a fundamental charged localised string is self-consistent, since this would be possible only in the  large $s(r_\Lambda)$ regime, in which the approximate axionic symmetry allows for the dual  formulation in terms of the two-form $\calb_2$.\footnote{In contrast, at large length scales the string backreaction enters a strong coupling region,  $\calb_2$ is no longer an appropriate elementary degree of freedom and the description in terms of a fundamental string breaks down.}    
Furthermore, we see that the bulk saxion $s(r_\Lambda)$  is driven to infinite distance as we increase $\Lambda$. This RG flow should certainly stop at a maximal break-down scale  $\Lambda_{\rm max}\simeq \sqrt{\calt(\Lambda_{\rm max})}$
at which the string  EFT condition \eqref{Tlambda} (with $p=1$) is violated. $\Lambda_{\rm max}$ provides a pure EFT  upper-bound on the EFT break-down scale, which may be in fact smaller because of some UV degrees of freedom invisible in the original EFT. 

So far we have considered just the simple one-field model \eqref{simpleK}. As we will argue in section \ref{s:asymptotic}, these features are common to a large set of fundamental strings, in settings with a richer set of string charges. A more complete discussion of all these aspects will also be presented in \cite{Lanza:2021qsu}.    

\subsection{Membranes}
\label{sec:EFTmem}

Let us now turn to the BPS electrically charged membranes described by \eqref{bosmem}. As in the string case, supersymmetry completely fixes the form of $\calt_{\bf q}(\phi)$ in terms of  the bulk scalar sector.  In the EFT perspective of \cite{Goldberger:2001tn,Michel:2014lva,Polchinski:2015bea}, the membranes  can be considered as {\em relevant} EFT operators. In particular, differently from the strings and from more generic higher codimension objects, no singularities appear in their backreaction near their core. As a result, no dependence on the UV cutoff $\Lambda$ is needed in order to regularise it, at least at tree level. Correspondingly, one does not need to regularise the bare tension $\calt_{\bf q}$ appearing in \eqref{memT} by a $\Lambda$-dependent term.

The fact that a membrane acts as a relevant operator implies also that its effects typically become strong in the IR. A natural  coupling controlling a corresponding perturbative expansion is given by the dimensionless combination
\be\label{memlambda}
\lambda(\Lambda)\equiv \frac{\calt_{\bf q}}{M^2_{\rm P}\Lambda}\, ,
\ee
which is clearly classically relevant, in the sense that it becomes stronger if one decreases $\Lambda$. As argued in section \ref{sec:EFTmemlat}, $\lambda(\Lambda)$ controls the different aspects of a membrane backreaction. Therefore, it should be a good EFT perturbative parameter. From this, one obtains the following estimate of the IR strong-coupling scale
\be\label{memLstrong}
\Lambda_{\rm strong}\equiv \frac{\calt_{\bf q}}{M^2_{\rm P}}\, . 
\ee
It is interesting to compare \eqref{memLstrong} and \eqref{Lstrongstring}. In the string case, we can write $\Lambda_{\rm strong}=e^{-\frac{\pi}{\hat\lambda(\Lambda)}}\Lambda $, where $\hat\lambda(\Lambda)=\frac{\calt_{\rm st}(\Lambda)}{M^2_{\rm P}}$ is the natural perturbative parameter entering \eqref{stensionbound}. In the membrane case, instead,   $\Lambda_{\rm strong}=\lambda(\Lambda)\Lambda$. We see that, with respect to the string case in which $\Lambda_{\rm strong}$ is exponentially suppressed with respect to $\Lambda$,  the EFT membrane description breaks down  quite quickly as one lowers $\Lambda$.

The effective  coupling $\lambda(\Lambda)$ is defined by using the `bare' membrane tension $\calt_{\bf q}$, given by the value of the bulk fields at the membrane location. Therefore, given our discussion in section \ref{sec:stringflow}, one may wonder whether one could arrive to a different conclusion by starting from an effective tension which includes  some additional running effect due to possible flow of the bulk scalars. 

Locally, we may consider a membrane as flat, and located at the origin of a transverse coordinate $y$. Since we are assuming that supersymmetry is preserved at the cut-off scale $\Lambda$, we may then identify an effective scale dependent membrane tension  $\calt_{\bf q}^{\rm eff}(\Lambda)$ as the jump of $|\calz|$ along an interval $\Delta y=2\Lambda^{-1}$ around the membrane:\footnote{We have chosen the orientation of $y$ so that the jump of $|\calz|$ is positive as we cross the membrane in the positive $y$ direction and we have assumed for simplicity that $\calz$ does not change phase across the membrane. The generalisation to other orientations and phases is straightforward and does not change the following discussion.} 
\be\label{effmemT}
\calt^{\rm eff}_{\bf q}(\Lambda)=2\left(|\calz|_{y=\Lambda^{-1}}-
|\calz|_{y=-\Lambda^{-1}}\right)\, ,
\ee
where $\calz\equiv e^{\frac{K}2}W$ is the normalised superpotential, including also the flux-dependent part generated by integrating out the three-form potentials.
$\calt_{\bf q}^{\rm eff}(\Lambda)$ reduces to the bare tension $\calt_{\rm q}$ used above if $\phi^\alpha$ can be considered constant along the interval $\Delta y=2\Lambda^{-1}$, since in this case the variation of $|\calz|$ in \eqref{effmemT} is only due to the jump $f_a\rightarrow f_a+q_a$ of the background flux quanta. This identification is stable if we increase $\Lambda$ (without violating \eqref{TTT}) and then  we can identify $\calt_{\bf q}=\lim_{\Lambda\rightarrow\infty}\calt^{\rm eff}_{\bf q}(\Lambda)$.

On the other hand, one may try to reduce $\Lambda$ enough to see a non-negligible deviation of \eqref{effmemT} from the bare value of the tension,  due to the running of the bulk fields induced by the membrane. In particular,
one may identify an effective coupling $\lambda^{\rm eff}(\Lambda)$ defined as in \eqref{memlambda} but with $\calt_{\bf q}$ replaced by $\calt^{\rm eff}_{\bf q}(\Lambda)$. 
Now, as we lower $\Lambda$, $\calt^{\rm eff}_{\bf q}(\Lambda)$ changes in a way dictated by the flow of the scalars along the $y$ direction. Around a straight enough membrane, one can assume such flow to be governed by the BPS flow equations discussed in \cite{Cvetic:1992bf,Cvetic:1996vr,Ceresole:2006iq,Bandos:2018gjp}. As we will show in Section~\ref{sec:memscflow}, these imply that the effective membrane tension flows as 
\be
\calt^{\rm eff}_{\bf q}(\Lambda)=\frac{\calt_{\bf q}}{1-{\alpha^2}\frac{|\mathcal{T}_{\bf q}|}{4M_{\rm P}^{2}\Lambda }}\equiv \frac{\calt_{\bf q}}{1-\frac14 \alpha^2\lambda(\Lambda)}\, ,
\label{Tefflambda}
\ee
where we have introduced the (bare) dimensionless perturbative coupling $\lambda(\Lambda)$ defined in \eqref{memlambda}, and $\alpha^2>0$  is related to the growth of the central charge $|\mathcal{Z}|$ in the given asymptotic regime. Alternatively, we may write down a corresponding flow for the effective coupling as
\be
\label{memleff}
\lambda^{\rm eff}(\Lambda)\equiv \frac{\calt^{\rm eff}_{\bf q}(\Lambda)}{M^2_{\rm P}\Lambda}=\frac{\lambda(\Lambda)}{1-\frac14 \alpha^2\lambda(\Lambda)}\,.
\ee
Hence, it is clear that, along the flow\footnote{Note that \eqref{memTeff} holds for general BPS flows activated by membranes \cite{Bandos:2018gjp}, since it follows from the monotonic behaviour  of $|\calz|$ \cite{Ceresole:2006iq,Bandos:2018gjp} along the flow.}
\be
\label{memTeff}
\calt^{\rm eff}_{\bf q}(\Lambda)\geq \calt_{\bf q}\, \qquad \Rightarrow \qquad \lambda^{\rm eff}(\Lambda)\geq \lambda(\Lambda)
\ee
and, in particular, the effective coupling {\em diverges} at $\Lambda=\frac14 \alpha^2\Lambda_{\rm strong}$, rather then being just of order $\calo(1)$ as the bare coupling $\lambda(\Lambda)$. Furthermore, we explicitly see that all the profiles \eqref{memleff} and \eqref{memTeff} are governed by the combination $\Lambda_{\rm strong} y$, which implies that all bulk modes excited by the membrane are characterised by energy scales of  order $\Lambda_{\rm strong}$. Thus, as anticipated, we cannot see any important deviation of the effective tension $\calt^{\rm eff}_{\bf q}(\Lambda)$ from the bare one $\calt_{\bf q}$ without entering the strong coupling region, in agreement with the above general arguments.

\subsection{Interpreting the no-force identities}
\label{sec:INFI}

From our discussion on 4d string and membrane backreaction, one can infer some intrinsic limitations of our previous interpretation of the no-force identities \eqref{No-fo_Ids} and \eqref{No-fo_Idm}. Indeed, the standard no-force condition  for higher codimension objects sees the different terms in \eqref{No-fo_RFCeq} as the force induced by the exchange of gauge potentials, scalars and gravitons between objects. The computations of Appendix \ref{app:RFC_Id} leading to \eqref{No-fo_RFCq} and  \eqref{No-fo_RFCm} take the same viewpoint, with the assumption that the exchange of particles is made on top of a Minkowski background metric.  This latter assumption, completely justified for long-range forces between higher codimension objects, is very restrictive for strings and membranes. 

As we have seen, the backreaction of strings grows as we separate from their core, and becomes singular when we reach the distance $r_{\rm strong} = \Lambda^{-1}_{\rm strong}$ given by \eqref{Lstrongstring}. The same occurs for membranes, with now the much shorter distance given by \eqref{memLstrong}. Therefore, unless two strings or membranes are located at distances much smaller than $r_{\rm strong}$, their backreaction will be significant, and the assumption made in Appendix \ref{app:RFC_Id} will not hold. Note that the obstruction to apply such a field theory computation not only amounts to implement the exchange of modes depicted in figures \ref{Fig:FT_intstrings} and \ref{Fig:FT_intmem} in a curved background, but it is more dramatic. Indeed, let us consider the case of the string backreaction \eqref{tsol3}. As we proceed away from the string core, the saxion $s$ flows to smaller and smaller values, such that at some point the axionic shift symmetry of its axionic partner is badly broken by non-perturbative effects of the form $e^{2\pi \ii n t}$, $n \in \mathbb{N}$. At this point, it does not make sense to dualise the saxion as in \eqref{dualfields}, or to consider a two-form $\cB_2$ dual to the axion $a$. Therefore, the expressions for the string tension and the physical charge given by \eqref{strten} lose meaning, just as the whole perturbative approach behind figure \ref{Fig:FT_intstrings}.

These difficulties are partially overcome if one interprets \eqref{No-fo_Ids} and \eqref{No-fo_Idm} from the EFT perspective discussed in this section. With such a prescription, $\cT_{\rm str}$ must be understood as the tension evaluated at the cut-off scale $\Lambda$. This is well-defined independently of the separation between strings, and in fact its computation depends on the neighbourhood of each string core. In this neighbourhood, the axionic shift symmetry is recovered, and so it makes sense to perform the dualisation \eqref{dualfields} and to  describe the electric coupling of the string to  $\cB_2$. As such one may define the physical charge $\cQ_{\bf e}$ at the cut-off scale by evaluating \eqref{strten} at a distance $\Lambda^{-1}$ from the string core. Finally, because the identity \eqref{No-fo_Ids} holds off-shell it will be valid along the whole string flow, up to the cut-off scale. 
In particular, it will be independent of where we locate the string, and the same applies if we have a more complicated multi-string solution (see e.g. \cite{Lanza:2021qsu}), as long as we remain within the perturbative regime. In this sense, one can interpret \eqref{No-fo_Ids} as a balance between string forces at distances  which are larger than $\Lambda^{-1}$ but smaller than $\Lambda_{\rm strong}^{-1}$, so that the perturbative description can be applied.  
Similar comments hold for $\cT_{\rm mem}$ and $\cQ_{\bf q}$.

\section{Strings and Membranes in asymptotic limits}
\label{s:asymptotic}

Understanding the classical RG flow of strings and membranes in terms of their backreaction brings us to the question of which different kinds of backreaction there are, and what do they imply in terms of the physics of extended objects. In this section we will address this question for both strings and membranes, finding that the different backreaction behaviours are classified in terms of the lattices of string and membrane charges. 

In the case of strings, a subcone of $\half$BPS charges dubbed $\cC^{\text{\tiny EFT}}_{\rm S}$ corresponds to the fundamental axionic strings of the theory. Through their backreaction, the elements of $\cC^{\text{\tiny EFT}}_{\rm S}$ can also be identified with the different infinite-distance, weak-coupling limits of the EFT. Along each of these limits, the lattice of membrane charges $\Gamma_{\rm F}$ splits into two: those that asymptote to super-Planckian ($\Gamma_{\rm heavy}$) and sub-Planckian ($\Gamma_{\rm light}$) tensions. The latter feature a well-defined non-vanishing tension in the UV, and contain the lattice of dynamical fluxes $\Gamma_{\rm EFT}$ defined in section \ref{sec:strmemgen}. As a byproduct of our analysis, we give explicit expressions for the different kinds of string and membrane flows in terms of the discrete data of each perturbative limit. This will be used in section \ref{sec:SC} to link the physics of strings and membranes to different swampland criteria.

\subsection{Physical strings and infinite field distance limits}
\label{sec:infinite}

As we have emphasised, the EFT description of fundamental 4d strings and membranes in terms of their backreaction is necessary to fully understand the physics of these extended objects. Remarkably, this backreaction provides very valuable information on the field space of the unperturbed EFT, and even on its microscopic origin. In this subsection we will summarise the results of \cite{Lanza:2021qsu} that are important for our subsequent analysis.

A clear example of this phenomenon is given by the backreaction of 4d $\half$BPS strings, which generalises the one discussed in section \ref{sec:stringflow}. Consider a string transverse to the plane $(z, \bar{z})$, located at $z=0$. Imposing 2d Poincar\'e invariance along the directions $(t,x)$ parallel to the string leads to the following metric Ansatz 
\be\label{strmetr}
\d s^2= -\d t^2+\d x^2+e^{2D}\d z\d\bar z\, ,
\ee
where $D$ depends only on the dimensionless coordinates $(z,\bar z)$. For certain $\half$BPS strings, it can be shown that the backreaction on the fields $t^i$ can be approximated by
\be\label{tsol}
t^i=t_0^i+\frac{1}{2\pi\ii}e^i\log \left(\frac{z}{z_0}\right)\, ,
\ee
with $t_0^i, z_0 \in \bbC$ integration constants. Here the fields $t^i$ couple magnetically to the string, which is manifest by the fact that around it they undergo a monodromy
\be\label{tmon}
t^i\rightarrow t^i +e^i\, , \qquad \qquad e^i \in \bbZ\, ,
\ee
with ${\bf e} = \{e^i\}$ the string charges. 
To derive \eqref{tsol} one assumes that $\{t^i\}$ belong to the moduli space $\MM$ of the EFT \cite{cstring}, and that any correction of the form $e^{2\pi \ii m_i t^i}$, $m_i \in \mathbb{Z}$ compatible with the monodromy \eqref{tmon} are sufficiently suppressed. This selects a region $\MM_{\rm pert}$ in which $\half$BPS instanton effects do not induce a significant dependence of the superpotential on the fields $\{t^i\}$, as otherwise we would expect to occur \cite{Palti:2020qlc}. Additionally, in such a region one may set the warp factor in \eqref{strmetr} as $2D = K(t_0) - K(t)$, namely in terms of the K\"ahler potential evaluated on the backreacted fields \cite{Lanza:2021qsu}. If such fields probe a region of  $\MM_{\rm pert}$ in which the K\"ahler potential is invariant under continuous axionic shifts, the solution for the backreacted metric will display a radial symmetry. This will in turn allow to give a dual description of the solution in terms of two-form potentials $\calb_{2\, i}$, as reviewed in section \ref{sec:N=1EFT}. In this dual picture, $\half$BPS strings are described by the localised electrically charged couplings \eqref{stringS}, evaluated at the cut-off scale $\Lambda$ of their flow.

In general, one may assume that $t_0^i$ appearing in \eqref{tsol} belongs to $\MM_{\rm pert}$, and see how far the string solution  can be extended. For this, it is useful to decompose $t^i$ into  axionic and saxionic pieces, namely as $t^i = a^i+ i  s^i$, and write $z=re^{\ii\theta}$ so that \eqref{tsol} becomes
\begin{subequations}
\label{solsplit}
\bea\label{imt}
s^i & = & s_{0}^i-\frac1{2\pi}e^i\log \left(\frac{r}{r_0}\right)\, ,\\
\label{axionmon}
a^i &= &\frac{\theta}{2\pi}\,e^i+\text{const}\, .
\eea
\end{subequations}
In particular, one may analyse how the saxionic coordinates $s^i$ evolve as $r$ changes, by either approaching or moving away from the string. Notice that we can interpret \eqref{tsol} as a map from $\bbC$ to the moduli space $\MM$, and so \eqref{imt} as a one-parameter flow in the saxionic vevs. Reasoning as in the previous section, one can relate the physics of unperturbed vacua along this trajectory in $\MM$ with a change in the EFT energy cut-off.

As in section \ref{sec:N=1EFT} the BPS-ness of the solution implies that the string tension is given by\footnote{We dub $\half$BPS strings as those whose tension \eqref{strten} satisfies \eqref{strtenb}, and preserve a particular half of the bulk supersymmetry \cite{Lanza:2019xxg,Lanza:2019nfa}.  Those preserving the opposite half satisfy $\cT_{\bf e} =-M_{\rm P}^2\,e^i\ell_i(r_\Lambda) > 0$ and are dubbed $\half$BPS anti-strings. For the latter the solution \eqref{tsol} must be replaced by an  antiholomorphic profile.}
\be\label{strtenb}
\cT_{\bf e} =M_{\rm P}^2\,e^i\ell_i(r_\Lambda)\, ,
\ee 
where $\ell_i$ stand for the dual saxions that appear in the linear multiplet description of the string, and $r_\Lambda\simeq \Lambda^{-1}$ provides an estimate of  the minimal distance accessible by the EFT.

Suppose that the saxionic domain is defined by $s^i> 0$ and $\ell_i>0$.  Because $\cT_{\bf e} \geq 0$, one can assume that at least one charge entry $e^i$  is positive. Then, as we take $r$ large the corresponding entry of $s^i$ tends to zero, reaching the deep interior of $\MM$. There, instanton effects take us away from the perturbative regime $\MM_{\rm pert}$, so that the classical string solution breaks down at large distances from the string, similarly to what happens along  transverse directions to D7-branes in type IIB string theory, which can be described by the 8d counterpart of the model of section \ref{sec:stringflow}. 

If instead we extend the solution towards $r \rightarrow 0$ and assume that $e^i \geq 0$, $\forall i$, we obtain 
\be\label{infs}
s^i\rightarrow s^i_\infty = e^i \cdot \infty\, .
\ee
When mapped into a trajectory in the moduli space of vacua, this behaviour drives the scalar fields $t^i$ asymptotically towards the boundary of $\MM$. If the string charge {\bf e} is such that $e^i m_i \geq 0$ for all allowed  instanton charges {\bf m}, then all corrections to \eqref{tsol} of the form $e^{2\pi i m_it^i}$ will die off in the limit \eqref{infs}, and we will remain within $\MM_{\rm pert}$ as we approach the string core. In fact, as these terms measure the strength of $\half$BPS instantons and those are the leading non-perturbative corrections, all effects breaking the continuous shift symmetry $a^i\rightarrow a^i+e^i \times$constant should be suppressed in such a limit. We dub these strings as fundamental axionic strings, because the metric solution describing them is governed by a K\"ahler potential that develops an exact axionic symmetry as we approach their core. As such, the solution admits a dual description in terms of a two-form potential ${\cal B}_2$, to which the string couples electrically.

By standard quantum gravity arguments \cite{Banks:1988yz,Banks:2010zn}, we only expect to realise global continuous symmetries at points of infinite distance in moduli space. As a result, EFTs that are consistent with quantum gravity should map axionic string locations to points $s_\infty^i$ at infinite distance  in their moduli space, saxionic trajectory of the backreaction \eqref{imt} to infinite distance paths in $\MM$. In fact, in light of the results in \cite{Garcia-Etxebarria:2014wla,Grimm:2018ohb,Grimm:2018cpv,Corvilain:2018lgw,Lee:2018urn,Lee:2019xtm,Lee:2019wij} it is tempting to speculate that all infinite distance limits correspond to axionic strings. For instance, in  refs.\cite{Grimm:2018ohb,Grimm:2018cpv,Corvilain:2018lgw}  each infinite distance limit is characterised in terms of a monodromy, whose physical realisation in our 4d EFT language is nothing but the discrete shift \eqref{tmon}. In our setup, instead of acting on BPS particles as in \cite{Grimm:2018ohb,Grimm:2018cpv,Corvilain:2018lgw}, this monodromy will act on BPS membranes, see \cite{Font:2019cxq} and below.

One can summarise this proposal as follows:

\begin{importantbox}
\begin{center}

\vspace{1em}

{\bf Distant Axionic String Conjecture (DASC): \cite{Lanza:2021qsu}} \\ {\em All infinite distance limits of a 4d EFT can be realised  as  \\ an RG flow endpoint of a fundamental axionic string.}

\vspace{1em}

\includegraphics[width=10cm]{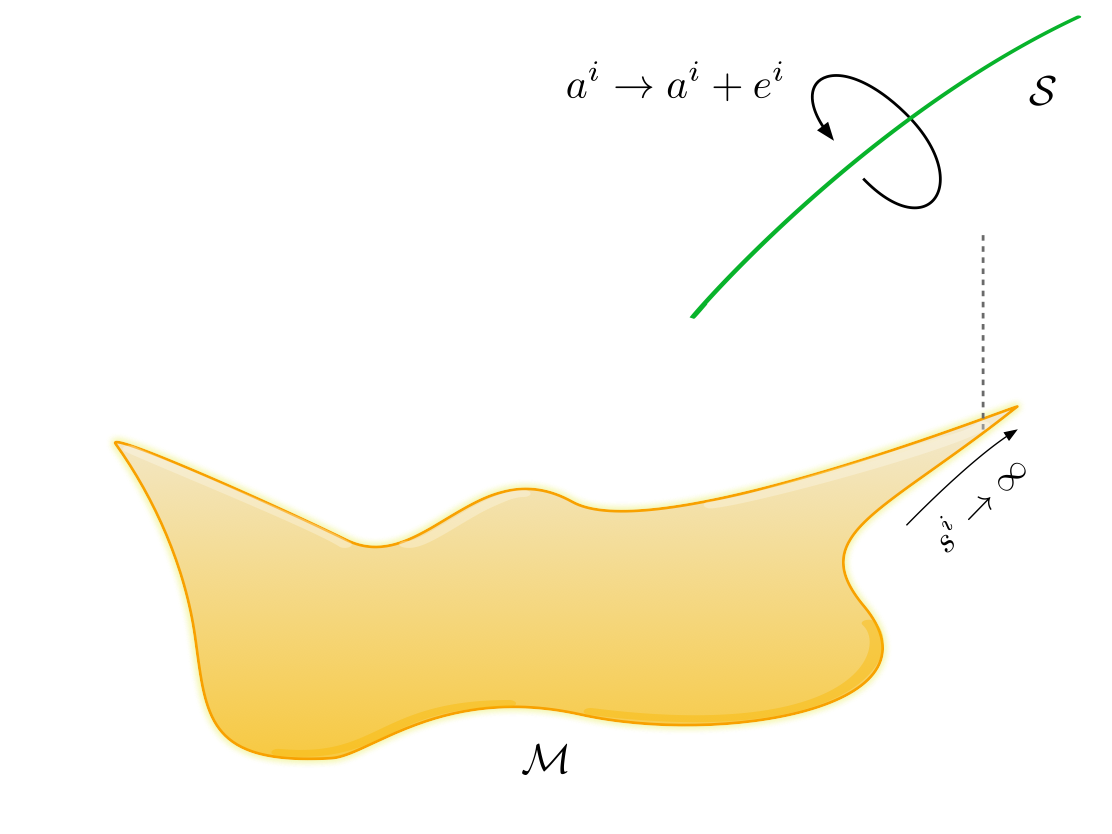}

\end{center}

\end{importantbox}

The evidence for this proposal in string compactifications will be studied in detail in \cite{Lanza:2021qsu}. As usual, most of the evidence comes from supersymmetric string/M-theory compactifications and it is a non-trivial step to assume its generality without supersymmetry. However, at least for spontaneous supersymmetry breaking, one could argue that close enough to the string core the solution is described by an
 EFT at a regime in which supersymmetry is effectively restored, so that the above arguments apply. The same argument applies for any mass deformation of the theory if the induced mass is well below the EFT cut-off $\Lambda$. Examples of this are flux-induced mass terms for the scalars $t^i$, which we will study in section \ref{sec:DWandEFTf}, and also $\cB_{2} F$ couplings, in agreement with the scenario considered in \cite{Reece:2018zvv}

Notice that the non-trivial statement of the conjecture does not come from assuming that continuous shift symmetries are only restored at infinite field distance but from claiming that there is always a continuous axionic shift symmetry being restored at every infinite field distance limit. Arguments in favour for this as well as possible caveats will appear in \cite{Lanza:2021qsu}. In this paper, we will only discuss this conjecture and its  implications in the context of 4d $\cN=1$ EFT's, although one could consider extending it to higher dimensional cases.
 In this respect, notice the key ingredient of our proposal is the codimension and not the dimension of the object, which suggests that higher dimensional  generalisations of the conjecture should involve codimension-two objects instead of strings.

One may compute the field distance in $\MM$ that corresponds to a radial saxionic flow \eqref{imt} by first parametrising it as $s^i(\sigma)=s^i_0+\sigma e^i$, where we have defined $\sigma = (2\pi)^{-1} \log (r_0/r)$.  For a given starting point $s^i_0$,  let us denote by $\sigma_*$ the value of $\sigma$ at which the flow reaches the boundary of the moduli space. Then the field space distance travelled by the radial flow is given by
 \be\label{d*}
 \begin{aligned}
 {\rm d}_*&=\int_{\rm flow}\sqrt{\cG_{ij}\d s^i\d s^j}=\int^{\sigma_*}_0\d\sigma \sqrt{e^i e^j\,\cG_{ij}(\sigma)} =\frac{1}{M_{\rm P}}\int^{\sigma_*}_0 \cQ_{\bf e}(\sigma)\,\d\sigma\, ,
 \end{aligned}
 \ee
 with $\cQ_{\bf e}$ the physical string charge \eqref{strten}, which along the flow can be expressed as
 \be\label{flowQ2}
\cQ_{\bf e}(\sigma)=M_{\rm P}\sqrt{\frac12 \frac{\d^2 K(\sigma)}{\d\sigma^2}}\, .
\ee

For 4d axionic strings one expects that $\cQ \rightarrow 0$ as $\sigma \rightarrow \sigma_*$, because otherwise an exact axionic shift symmetry would not be recovered at $s^i(\sigma_*) \in \partial\MM$, due to non-vanishing instanton corrections.\footnote{One way to see this is by means of the axionic WGC, which reads $\cQ \, \Im t \leq M_{\rm P}$ for the case of one axion, so taking $\Im t \rightarrow \infty$ implies that $\cQ \rightarrow 0$. A similar statement holds in the presence of more axions, see e.g. \cite{Rudelius:2015xta,Montero:2015ofa,Brown:2015iha,Hebecker:2015rya}.} Also, by quantum gravity arguments we expect that ${\rm d}_* = \infty$, so the decreasing monotonic behaviour of $\cQ$ implies that  $\sigma_* =\infty$. Furthermore, by taking the $\sigma$-derivative of the string tension \eqref{strtenb}
and recalling the general definitions \eqref{dualfields} and \eqref{GGmetric} we get 
\be\label{monoT}
\frac{\d\calt_{\bf e}(\sigma)}{\d\sigma}=-\calq^2_{\bf e}<0\, ,
\ee
which shows that, generically, the string tension decreases along its own RG-flow as we increase the cutoff scale $\Lambda$. 
Finally, we can rewrite \eqref{monoT} in a slightly different way
\be\label{QTstring}
\frac{\calq_{\bf e}^2}{\calt^2_{\bf e}}=\frac{\d}{\d\sigma}\left(\frac{1}{\calt_{\bf e}}\right)\, ,
\ee
so that $\calq_{\bf e}/\calt_{\bf e}=\gamma M_{\rm P}^{-1}$ with constant $\gamma$, if and only if $\calt^{-1}_{\bf e}$ scales precisely as $\calt^{-1}_{\bf e}=\sigma(\gamma M_{\rm P}^{-1})^2 +\text{const}$. This relation will be useful when discussing the WGC for strings, in section \ref{sec:WGCstr}. In all string theory examples analysed in \cite{Lanza:2021qsu} the relation $\calt^{-1}_{\bf e}=\sigma(\gamma M_{\rm P}^{-1})^2 +\text{const.}$ is realised asymptotically, due to 
a K\"ahler potential which asymptotically takes the form
\be
K = - \log  P(s) +\ldots\, ,
\label{Klog}
\ee
up to subleading corrections, with $P(s)$ some homogeneous saxionic function of integral positive degree. Indeed, in this case the dual saxions $\ell_i$ entering the string tension \eqref{strten} decrease as $(s^i)^{-1}$ along the trajectory, from where the statement follows.

String flows with $\sigma_* = \infty$ do not occur for any string charge, but under certain assumptions one can characterise this property in terms of discrete cones of $\half$BPS charges. Indeed, let $\Delta$ be the saxionic domain in the asymptotic regime of interest, and let ${\cal P}$ the same region expressed in dual saxionic variables. As argued in \cite{Lanza:2021qsu}, it is natural to assume that $\Delta$ has the structure of a cone, while the form of the K\"ahler potential \eqref{Klog}  guarantees that then ${\cal P}$ is also a cone. Then $\Delta\subset \calp^\vee$ and $\calp\subset \Delta^\vee$, where $\calp^\vee$ and $\Delta^\vee$ are the dual cones of $\calp$ and $\Delta$ respectively. Furthermore, let $N_\bbZ$ be the lattice of possible string charges in that region and $M_\bbZ = N_\bbZ^\vee$ the dual lattice of instanton charges. Then one can define the following discrete cones \cite{Lanza:2021qsu}
\be
\begin{aligned}
&\cC^{\text{\tiny EFT}}_{\rm S} = \ \overline{\Delta} \cap N_\bbZ \quad \subset\quad \cC_{\rm S} \ = \ \cP^\vee \cap N_\bbZ \, ,\\
&\cC^{\text{\tiny EFT}}_{\rm I} = \ \overline\cP \cap M_\bbZ \quad \subset\quad \cC_{\rm I} = \ \Delta^\vee \cap N_\bbZ \, ,
\end{aligned}
\ee
where $\cC_{\rm S}$ and $\cC_{\rm I}$ stand for the cone of mutually BPS string and instanton charges, respectively. For string flows generated by ${\bf e} \in \cC^{\text{\tiny EFT}}_{\rm S}$ as we approach the string core the saxions will be driven to the weakly-coupled region $s_\infty^i$ in \eqref{infs}, where all instantons in $\cC_{\rm I}$ are suppressed. Differently, for flows generated by ${\bf e} \in \cC_{\rm S} - \cC^{\text{\tiny EFT}}_{\rm S}$ a finite distance boundary point of $\Delta$ will be reached at a finite value of $\sigma_*$, and there will be instantons with charges in $\cC_{\rm I} - \cC^{\text{\tiny EFT}}_{\rm I}$ whose corrections will become significant before that happens. As such, $\half$BPS strings with charges on ${\bf e} \in \cC_{\rm S} - \cC^{\text{\tiny EFT}}_{\rm S}$ cannot be described from the viewpoint of a weakly-coupled EFT as in the simple model of section \ref{sec:stringflow}. In particular, since the said instanton effects strongly modify \eqref{tsol} and break the shift symmetry associated to \eqref{tmon}, the metric solution is not radially symmetric near the string core and it does not correspond to a fundamental axionic strings. Interestingly, this characterisation of fundamental axionic string shares some similarities with the notion of \emph{supergravity strings} of \cite{Katz:2020ewz}, used to constrain 5d $\cN=1$ supergravity theories. It would be very interesting to further explore this connection.

As said, for strings flows generated by ${\bf e} \in \cC^{\text{\tiny EFT}}_{\rm S}$ one should have that $\cQ_{\bf e} \rightarrow 0$ as $\sigma \rightarrow \infty$ but still ${\rm d}_* \rightarrow\infty$. The simplest asymptotic behaviour that reproduces this feature is when the physical string charge decreases like $\sigma^{-1}$ along the flow, and therefore ${\rm d}_*$ diverges logarithmically as $\sigma_*\rightarrow\infty$, which is guaranteed by a K\"ahler potential of the form \eqref{Klog}. We will dub {\em non-degenerate} string flows those in which $P(s)$ can be approximated by a monomial very close to the string core, and {\em degenerate} those where this approximation does not hold. In this work we will only consider non-degenerate flows, leaving a more general discussion to \cite{Lanza:2021qsu}, in which examples of degenerate flows will be analysed. To simplify our discussion we will also assume that the saxionic domain $\Delta$ is a simplicial cone defined by $s^i >0$, and that $\cC^{\text{\tiny EFT}}_{\rm S}$ is generated as a positive sum $e^I{\bf v}_I$ of linearly independent vectors ${\bf v}_I$ that match the number of axions of the theory. The strings that correspond to the generators ${\bf v}_I$ will be dubbed {\it elementary} axionic strings, since their charges cannot be decomposed as positive linear combinations of other charges in $\cC^{\text{\tiny EFT}}_{\rm S}$. We refer the reader to \cite{Lanza:2021qsu} for a more general discussion.

\subsection{Membranes ending on strings}
\label{sec:StrMem}

We have defined the perturbative regions of our EFT as those in which the saxionic variables $s^i\equiv \Im t^i$ are large within their domain $\Delta$.  Alternatively, one can describe this regime in terms of the complex  variables
$e^{2\pi\ii\, t^i}$. Assuming a simplicial saxionic domain $\Delta=\{s^i>0\}$,  the loci $\cald_i=\{e^{2\pi\ii t^i}=0\}$ are asymptotic divisors in field space, associated with each perturbative limit.  As discussed above, we expect that each of these limits is associated to the RG flow of a string becoming tensionless, such that the periodicity
\be\label{tper}
t^i\simeq t^i+1
\ee 
is extended to approximate continuous shift symmetries of the K\"ahler potential, while only the discrete shift $t^i \rightarrow t^i + 1$ is exactly preserved as it corresponds to a gauge redundancy. As in all known string theory examples, we will assume that in the perturbative regime the leading contribution to the K\"ahler potential takes the form \eqref{Klog}. It is then easy to see that the divisors $\cald_i$ are located at infinite distance in field space, as expected. 

On top of non-perturbative effects, the continuous shift symmetry can also be broken by a fluxed-induced superpotential, which in perturbative regimes is of the form \eqref{supoflux}
\be\label{compsup}
W= f\cdot \Pi(t,\chi)+\dots \equiv f_A \Pi^A(t,\chi)+\dots
\ee
where the dots denote non-perturbatively suppressed corrections by appropriate powers of $e^{2\pi\ii\, t^i}$, and $\chi$ denotes the complex fields that do not develop an approximate continuous shift symmetry. 

Requiring that \eqref{tper} is a gauge symmetry of the theory (that is, a duality) and in particular of the superpotential, implies that 
the divisors $\cald_i$ can be seen as branch-loci of the periods $\Pi^A(t,\phi)$. By encircling $\cald_i$ by a shift in the axion $a^i\equiv {\rm Re}\, t^i$, we obtain a monodromy transformation 
 \be\label{monper}
\Pi^A(t^i+1,\ldots)=({\cal R}_i)_B{}^A \Pi^B(t,\ldots)\, .
 \ee
 which is then compensated by a transformation of the fluxes $f_A$:
\be
\label{monf}
f_A\quad\rightarrow\quad ({\cal R}_i^{-1} \cdot f)_A\, .
\ee
These generate the monodromy group $\calg:\Gamma_{\rm F}\rightarrow\Gamma_{\rm F}$ on the full  lattice of fluxes $\Gamma_{\rm F} = \{f_A\}$, which is a subgroup of all the transformations that leave invariant the K\"ahler potential. The EFT will then be invariant under the combined transformation \eqref{monper} and \eqref{monf}, that is $(t^i,f)\rightarrow (t^i+e^i,({\cal R}_i)^{-e^i}f)$ with $e^i$ some integers. Guided by the monodromy theorem of Schmid \cite{NOT}, we assume that the monodromy transformations are represented by quasi-unipotent matrices satisfying
    \beq
    ({\cal R}_i^{\hat m_i}-1)^{m_i+1}=0 \qquad \forall i\, ,
    \eeq
with $\hat m_i,m_i$ some integers. The integer $\hat m_i$ may be reabsorbed in a reparametrisation of the coordinates $t_i$, while $m_i$ is upper bounded  in string theory compactifications.\footnote{In typical string theory examples based on Calabi-Yau compactifications, this integer $m_i$ is upper bounded by the dimension of the internal space. For instance, as explained in \cite{Grimm:2018ohb,Grimm:2018cpv,Corvilain:2018lgw,Grimm:2019bey,Cecotti:2020rjq}, in Calabi--Yau manifolds CY$_D$ one can have $m_i=0,1,\dots D$. More generally, it is bounded by the weight of the Hodge structure.}  If $m_i\geq 1 $, the monodromy is of infinite order, and we can always define a nilpotent operator $N_i\equiv\log {\cal R}_i$ whose nilpotency order is precisely $m_i$.

Note that the axion shifts \eqref{tper} are precisely those in \eqref{tmon} describing elementary axionic string charges, and that the latter can be identified with the divisors $\cald_i$. The difference with respect to our previous analysis is that now such axions appear in the superpotential \eqref{compsup} through the periods $\Pi^A(t, \chi)$, and therefore in the F-term scalar potential. Under these circumstances, the axionic string solution of section \ref{sec:infinite} is no longer valid, and a membrane must end on such a string to render the configuration consistent. Indeed, as we know when moving around a string of charges $e_i$, the axions change as $a^i \rightarrow a^i + e^i$. In the presence of fluxes, this does not correspond to the same point in field space, whenever the monodromy ${\cal R}_i^{e_i}$ acts non-trivially in the flux vector $f_A$. In this case the string is considered anomalous,  and the problem is solved by inserting a membrane of charge $q_A=({\cal R}_i^{-e_i}- 1)_A{}^Bf_B$ ending on the string. Hence, when moving around the string, we are also crossing a domain wall, and the EFT remains invariant.\footnote{In string theory compactifications, the lack of gauge invariance is seen microscopically as a Freed-Witten anomaly induced  on the string by the presence of fluxes \cite{BerasaluceGonzalez:2012zn,Herraez:2018vae}, which is cured by the membrane ending on it.}

The fact that membranes can break by generating holes associated to the strings signals that the corresponding  3-form gauge fields are becoming massive because of the interaction with the axions, which spontaneously break the gauge symmetry of the 3-forms. This is particularly easy to see in the case in which the monodromy generators $N_i$ have nilpotency order one, since in that case we can reabsorb the axionic shift by crossing a single membrane, or several copies of it. Restricting ourselves to the lattice $\Gamma_{\rm EFT}$, such a configuration is captured by the gauging \cite{Marchesano:2014mla} 
\beq
\label{Lstuck}
\mathcal{L}\supset \ |{\rm d}\cB_{2\,i} + c_{a\,i} C^a_3|^2\, ,
\eeq
which can be obtained by supersymmetrically dualising fluxes and axions simultaneously \cite{Lanza:2019xxg, Lanza:2019nfa}. Gauge invariance requires the following transformation
\beqa
C_3^a\rightarrow C_3^a+ {\rm d}\Lambda^a_2\ ,\quad \cB_2^i\rightarrow \cB_{2\,i} -c_{a\,i} \Lambda^a_2\, ,
\eeqa
which signals that the 2-form is gauged. The gauge invariant Wilson surface operator is given by
\beq\label{WSO}
e^i \int_{\cS} \cB_{2\,i} + q_a \int_{\cW} C_3^a\, ,
\eeq
with $q_a=c_{a\,i} e^i$ and $\cS=\partial \cW$. The case of higher nilpotency order is trickier since  it is not possible to write an Abelian  coupling of the form \eqref{Lstuck} and there is no simple way of writing a  gauge invariant counterpart of the form \eqref{WSO}, but the membrane charges should satisfy  
\beq
\label{qstuck}
q_a=(e^{-N_i e^i}-1)_a{}^bf_b \, 
\eeq
by consistency with the discrete axionic shifts.

As discussed in section \ref{sec:infinite}, the DASC describes a one-to-one correspondence between strings that belong to $\cC^{\text{\tiny EFT}}_{\rm S}$ and infinite field distance limits in the EFT moduli space. Each string selects a perturbative direction parametrised by an RG flow of scalars generated by approaching the string core. In the presence of fluxes, some of these scalars will become massive. Correspondingly, some
strings will become anomalous, and will need of membranes ending on them to describe gauge-invariant operators. Since the strings corresponding to $\cC^{\text{\tiny EFT}}_{\rm S}$ are fundamental,  it is expected that the membranes ending on it satisfy \eqref{EFTregime} as well, or else it would not be possible to describe a gauge-invariant localised operator. Moreover, this should be true as we change the cut-off scale, and both the anomalous string and the membranes ending on it vary their saxion-dependent tensions. 

It was observed in \cite{Lanza:2019xxg} that the conditions \eqref{EFTregime} are indeed correlated for membranes ending on strings, by looking at a few examples of string compactifications. In the following subsection we will argue that this is a  general feature, and that it can be seen as a direct consequence of the DASC and general properties of the periods $\Pi^A (t,\chi)$.  

As we will see, for each choice of string/perturbative limit, a different lattice of fundamental membranes $\Gamma_{\rm EFT}$ is selected by the flow of scalars. On flows generated by anomalous strings, $\Gamma_{\rm EFT}$ is such that the membranes in it can cure the anomaly, and so a gauge invariant localised operator can be defined at the level of the EFT.  In particular, $\Gamma_{\rm EFT}$ will be non-empty whenever the flowing scalars have a mass below $\Lambda$, so that this flow can be considered as a genuine field space direction of the EFT. Note that if the mass of the flowing saxionic scalars is above $\Lambda$, the approximations leading to the solution \eqref{solsplit} are no longer valid, and the corresponding string should no longer be an EFT operator.  

\subsection{Domain walls and EFT fluxes}
\label{sec:DWandEFTf}

Typically, in $\caln=1$ string compactifications to four dimensions the EFT is under control in some perturbative regime associated with some asymptotic region of the field space $\calm$. Generically, the presence of a flux-induced superpotential comes together with the presence of membranes describing dynamical transitions between different flux vacua. However, as emphasised in \cite{Lanza:2019xxg} and further elaborated in section \ref{sec:EFTmem}, the membranes entering the EFT are selected by the perturbative regime, and not all fluxes can be promoted to \emph{dynamical} variables for a given EFT. 

In this section, we will explain how to determine the total flux lattice $\Gamma_{\rm F}$  associated to each perturbative regime, and how to identify the sublattice $\Gamma_{\rm EFT}\subset \Gamma_{\rm F}$ of dynamical fluxes. Only fluxes on $\Gamma_{\rm EFT}$ can be promoted to 3-form gauge fields with electrically charged membranes which satisfy the EFT conditions discussed in section \ref{sec:EFTmemlat}. In particular, this guarantees that the energy scales of the induced potential are compatible with the EFT regime. As explained in section \ref{sec:EFTmem}, this sublattice should contain membranes ending on the anomalous strings associated to the axions becoming massive, to render the EFT consistent.

Throughout our analysis we will assume certain properties of the periods $\Pi^A$ describing the superpotential \eqref{compsup}, like the Nilpotent Orbit Theorem \cite{NOT}. These properties are based on the theory of asymptotic Mixed Hodge Structures \cite{Cattani:1982a,Cattani:1989a} whose key ingredient is precisely the monodromy transformation \eqref{monper}. They are well-established in the context of compact K\"ahler manifolds and have been recently applied to test some swampland conjectures in supersymmetric EFTs arising from Calabi--Yau compactifications in \cite{Grimm:2018ohb,Grimm:2018cpv,Corvilain:2018lgw,Grimm:2019bey,Grimm:2019ixq,Cecotti:2020rjq,Gendler:2020dfp}. In \cite{Grimm:2019ixq} they were also used to classify and analyse the possible flux-induced asymptotic scalar potentials arising from $\cN=1$ F-theory CY compactifications to four dimensions, although this machinery is certainly not restricted to CY's. It has moreover been argued that these algebraic  properties are not particular of certain microscopic EFT descriptions, but in fact 
inherent to the vector multiplet sector of any $\cN=2$ EFTs  consistent with quantum gravity \cite{Cecotti:2020rjq}.   
Our working assumption will be that  some relevant aspects of this statement also apply to the asymptotic regimes of the $\cN =1$ EFTs under consideration.

\subsubsection{EFT Flux lattice}
\label{sec:EFTlattice}

Let us begin with characterising the total flux lattice $\Gamma_{\rm F}$ associated to each perturbative regime, combining the results of \cite{Lanza:2019xxg} and \cite{Grimm:2019ixq}. First, using the Nilpotent Orbit Theorem, near the divisor $\cald_i$ the periods can  be written as 
\beq
\label{nil}
\Pi(t,\chi)=e^{t^i N_i^T}\Pi_0(\chi)+\mathcal{O}(e^{2\pi \ii t})\, ,
\eeq
with $\Pi_0$ some period independent of the fields $t^i$. The perturbative part of the superpotential can then be written as  $W_{\rm pert}= e^{t^i N_i}f\cdot \Pi_0(\chi)$.
For later convenience, it is useful to define an ``effective" nilpotency order  $d_i \in \mathbb{N}$ for each generator $N_i$, as
\beq
\label{Nd}
(N_i^T)^{d_i} \Pi_0(\phi)\neq 0\ , \qquad (N_i^T)^{d_i+1} \Pi_0(\phi)= 0\, .
\eeq
When expanding the exponential in \eqref{nil} these integers $d_i \leq m_i$ provide the highest power of the fields $t^i$ appearing in the period. They also characterise the type of asymptotic/perturbative limit and its singular geometry around the corresponding divisor $\cald_i$, so they are usually referred as the singularity type of that limit \cite{Grimm:2018ohb}.

For simplicity, let us focus on the case in which there is a single complex variable $e^{2\pi \ii t}$, which implies a single axion and therefore, a single nilpotent monodromy operator $N$. The generalisation to multi-moduli limits will be discussed in Appendix \ref{app:multimoduli}. By assumption, the leading contributions to the asymptotic expansion of the K\"ahler potential can be written as
\be
K=-n\log \Im t+\hat K(\chi,\bar\chi)+\ldots
\label{K1}
\ee
up to subleading corrections in $1/\Im t=s^{-1}$. Note that if the K\"ahler potential can be written as  $K=-\log (\Pi^A\eta_{AB}\bar\Pi^B)$ for some non-degenerate (anti-)symmetric bilinear form $\eta_{AB}$,\footnote{This occurs for instance in the complex structure moduli space of Calabi--Yau compactifications, and also in the mirror dual K\"ahler moduli space.} then one has $n=d$, so that the homogeneous degree of the K\"ahler potential  is fixed by the effective nilpotency order $d$ defined in \eqref{Nd}. Nevertheless, in the following we will leave $n$ as a free parameter, to keep the discussion as general as possible.
  
\begin{center}
	\begin{figure}[h!]
		\centering
		\includegraphics[width=6.5cm]{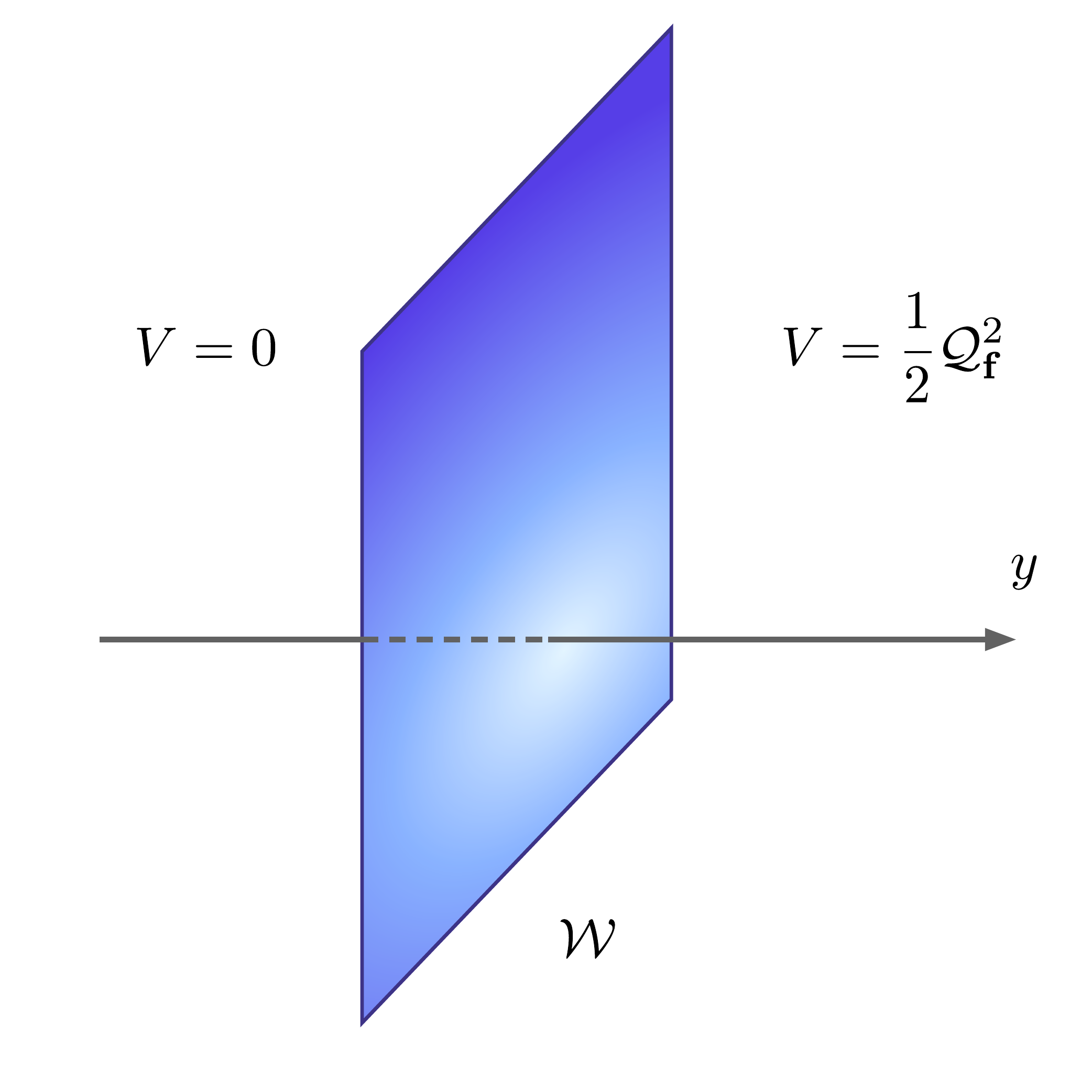}
		\caption{A generating membrane interpolates between a fluxless configuration and a background with flux {\bf f}, with 
		a potential $V = \frac12 \mathcal{Q}^2_{\bf f}$ for the latter.\label{Fig:MemGen}}
	\end{figure}
\end{center}

Membrane tensions are given by the expression \eqref{memT} and thus employing \eqref{nil}  yields
  \be
  \label{Tgrowth}
\calt_{{\bf q}}= 2M^3_{\rm P}\, e^{\frac12 K}\left| \sum_{k=0}^{d/2}\frac{1}{k!}(\ii s)^k\, N^k\boldsymbol{\rho}(\textbf{q},a)\cdot \Pi_0\right|\, ,
\ee
where $\boldsymbol{\rho}(\textbf{q}, a) = e^{a N} {\bf q}$  and the sum always involves only a finite number of terms, as $N$ is nilpotent. This nilpotent operator induces a monodromy filtration on the flux lattice constructed from the images and kernels of $N$, which allows us to define a Deligne splitting of the total flux vector space at the singular limit. 
This implies that the total flux lattice $\Gamma_{\rm F}$ splits into orthogonal subspaces\footnote{At the level of the corresponding vector spaces, the asymptotic splitting \eqref{split-Vell}  can be derived in the context of limiting Mixed Hodge Structures. In the perturbative limit $s\rightarrow \infty$ one can define the inner product of charges $\langle q_a,*q_b \rangle_{\infty}\equiv q_a T^{ab}_\infty q_b$, where $T^{ab}_\infty$ is the 3-form kinetic matrix $T^{ab}$ with the divergent saxionic dependence extracted. Then a charge $q_r\in \Gamma_r$ satisfies $( q_r, *q_{r'})_\infty\neq 0$ only if $r=r'$.   See Appendix \ref{app:multimoduli} for more details and \cite{Grimm:2018ohb,Grimm:2018cpv,Grimm:2019ixq} for explicit examples (to compare with these refs. replace $r\rightarrow \ell - D$). A working assumption in all these works is that the vector space splits also over the integers. For simplicity, we will mostly adopt the same working assumption, which will be particularly relevant in section \ref{sec:SC}. However, several of the following results hold more generally.} 
\beq \label{split-Vell}
\Gamma_{\rm F} = \bigoplus_{r} \Gamma_{r}\, ,\qquad  -d\leq r \leq d\,,
\eeq
satisfying  $\text{dim}\,  \Gamma_{r} =   \text{dim}\,  \Gamma_{ - r}$ and $N\, \Gamma_r \subset \Gamma_{r-2}$. As such, we can always find a basis of charges {\bf q} adapted to the asymptotic splitting such that  the leading contribution to the membrane  tension behaves as 
  \be\label{Tsinglesax}
\calt_{{\bf q}_r}\simeq\, M^3_{\rm P} \cT_0(\chi) \rho_{r}(q,a) \ s^{r/2- (n-d)/2}  \, , 
\ee
where $\cT_0 \rho_r(a)  = N^{(r+d)/2} \boldsymbol{\rho}(\textbf{q}_r,a) \cdot \Pi_0$, the vector ${\bf q}_r$ takes values in $\Gamma_r$, and $\cT_0(\chi)$ is constant along the limit. 
Remarkably, this result works even for multi-moduli limits up to polynomially suppressed corrections, as discussed in Appendix \ref{app:multimoduli}. For the case at hand in which we have a single chiral field, one can simply identify
\beq
\label{kr}
r=2k-d\, .
\eeq
Recall that  the charge $\cQ_{\bf q}$  (cf. \eqref{memT}) of a membrane with a quantised charge {\bf q} is given by
  \begin{equation}
  \label{rho}
\cQ_{\bf q}^2\equiv   q_a T^{ab} q_b=\rho_a Z^{ab}\rho_b\, ,
 \end{equation}
 with $T_{ab}$ the gauge kinetic function of the 3-form gauge fields, cf. \eqref{dualF4lagr}, which can always be split into a saxionic and axionic dependence as $T(s,a)=e^{aN}Z(s)e^{-aN}$ \cite{Herraez:2018vae,Grimm:2019ixq}.
 The leading order behaviour of $\cQ_{\textbf{q}_r}$ when ${\bf q}_r$ takes values in a single subspace $\Gamma_r$ reads
 \beq
  \label{growth}
 \cQ^2_{\textbf{q}_r}
 \ \simeq\ M^4_{\rm P} \cQ^2_0(\chi)\,  \rho_{r}(q,a)^2\, s^{r- (n-d)} \ .
 \eeq
The orthogonality properties of $\Gamma_r$ in \eqref{split-Vell} imply $\cQ^2\simeq \sum_r \cQ_{\textbf{q}_r}^2$ to leading order in the perturbative expansion.
As discussed in section \ref{sec:N=1EFT}, strictly speaking the 3-form kinetic function is only defined for the dynamical fluxes of $\Gamma_{\rm EFT}$. Nevertheless, it proves useful to extend the definition of physical charge to the whole flux/membrane lattice. Alternatively, one may understand \eqref{growth} as an asymptotic expression for the flux-induced scalar potential, which can always be understood as the charge $V=\frac12 \cQ^2$ of a membrane interpolating between Minkowski and a vacuum with fluxes $f_a=q_a$ (see figure \ref{Fig:MemGen}).  In this sense, notice that all the axionic dependence in \eqref{growth} appears in $\boldsymbol{\rho}(\textbf{q}, a) = e^{-a N} {\bf q}$. This is reminiscent of the factorised bilinear structure  found in \cite{Bielleman:2015ina,Carta:2016ynn,Herraez:2018vae} for type IIA flux potentials at large volume, as well as in other setups \cite{Grimm:2019ixq} where the above expression and its generalisation to multiple axions hold. The mass of the axion and of the chiral field $t$ containing it can be estimated by the prefactors multiplying $\rho_r^2$, and therefore varies asymptotically as $M_{\rm P} s^{r/2 - (n-d)/2}$, just like the quotient $\cT_{{\bf q}_r}/M_{\rm P}^2$. In general, a flux whose dual membrane's tension becomes transplanckian in a perturbative limit will generate transplanckian mass terms for several chiral fields in that regime. Generically, the mass term for $a$ arises from the linear terms on the axion in ${\rho}_r(q,a)$, so it is induced by a flux in ${\bf q}_r$ whenever  $N {\bf q}_r \neq 0$. Relating \eqref{nil} to the discussion of anomalous strings in section \ref{sec:StrMem}, one can show that this happens whenever the chiral field $t$ appears in the string flow of an anomalous string. The anomaly is created by a flux ${\bf q}_r$, and a membrane with charge  ${\bf q}_{r-2} \in \Gamma_{r-2}$ ends on the string to cure the anomaly, since $N{\bf q}_r\in \Gamma_{r-2}$. The same story applies if we look at those membranes that satisfy $\cT_{{\bf q}_r}/M_{\rm P}^2 < \Lambda$, or in other words to fluxes that belong to $\Gamma_{\rm EFT}$. Such dynamical fluxes generate anomalies for certain strings, and mass terms below $\Lambda$ for the scalars that couple to them. Then, the anomaly is cured by a lighter membrane which therefore also corresponds to $\Gamma_{\rm EFT}$, and thus one can describe the gauge invariant string-membrane configuration at the EFT level. It also follows that a flux-induced scalar mass below $\Lambda$ can only be generated if $\Gamma_{\rm EFT}$ is non empty, as anticipated.

In order to determine the leading behaviour of the tension and charge of a membrane with charge $\bf q$, one first needs to span the charge into a basis adapted to the asymptotic splitting and then sum over the different contributions using the asymptotic behaviour in \eqref{Tsinglesax} and \eqref{growth}. From now on, we will denote as \emph{elementary saxionic} membranes those with a charge $\textbf{q}_r$ which belongs to a single subspace $\Gamma_r$.

For each perturbative limit, the splitting \eqref{split-Vell} allows us to separate the lattice of fluxes on two sublattices, regarding the asymptotic growth of the corresponding membrane tension and charge (or flux potential) in such a limit. We have that
\beq
\Gamma_{\rm F}=  \Gamma_{\rm light} \oplus  \Gamma_{\rm heavy}\, ,
\label{GLH}
\eeq
where we have defined 
\bea
     \Gamma_{\rm light} =  \bigoplus_{r}  \Gamma_{ r}\  \quad \text{with }r <n-d\, , \quad \quad 
       \Gamma_{\rm heavy} =  \bigoplus_{r}  \Gamma_{ r}\  \quad \text{with }r >n-d \, . 
\eea
By construction, membranes with charge ${\bf q} \in  \Gamma_{\rm light}$ are lighter than $M_{\rm P}^3$ in the perturbative regime, while membranes with ${\bf q} \in  \Gamma_{\rm heavy}$ develop a transplanckian tension, cf. \eqref{Tsinglesax}. Equivalently, this distinguishes \emph{electric} three-form gauge fields with a small gauge coupling in the perturbative regime (and therefore a subplanckian flux-induced potential) from \emph{magnetic} fields with a large gauge coupling (so a transplanckian flux-induced potential). It is then clear that charges in $\Gamma_{\rm heavy}$ are excluded from the EFT regime. By analogy to the case of particles, where it is not possible to have a Lagrangian description with both electric and magnetic variables, one is not able to promote the fluxes to three-form gauge fields both for electric and magnetic ones. Now, while $\Gamma_{\rm light}$ naturally contains the EFT lattice defined in \eqref{GammaEFT}, one should not identify these two lattices, since in general it is not sufficient to belong to $\Gamma_{\rm light}$ to guarantee that the energy scales associated to the membrane are compatible with the EFT regime. In fact, it turns out that typically $\Gamma_{\rm EFT}\subset \Gamma_{\rm light}$ strictly, as we explain in the following.

In order to describe a membrane within the EFT regime, its  tension $\cT_{\rm mem}$ has to satisfy:
\beq
\label{TEFT}
\Lambda^3 \leq \cT_{\rm mem} < M_{\rm P}^2 \Lambda\, .
\eeq
As discussed in section \ref{sec:N=1EFT}, the upper bound comes from requiring that $\cT_{\rm mem}/(M_{\rm P}^2\Lambda)$ is small to keep the gravitational backreaction under control, and so the corresponding membrane belongs to $\Gamma_{\rm EFT}$. Indeed, notice that even if $\cT_{\rm mem} < M_{\rm P}^3$ (so the potential is subplanckian), this does not necessarily  imply that the Hubble tension $\cT_{\rm mem}/M_{\rm P}^2$ is small compared  to the cut-off scale. 

We can use the tension of the elementary axionic string associated with  $t\rightarrow t+1$ to estimate whether the upper bound in \eqref{TEFT} can ever be reached. As our limit is described by a fundamental string we have that $\Lambda\leq \cT_{\rm str}^{1/2}$, and in the one-modulus case $\calt_{\rm str}= n M^2_{\rm P}/2s$. Then the condition \eqref{TEFT} implies that $\cT_{\rm mem} < M_{\rm P}^2 \Lambda \leq M_{\rm P}^2 \cT_{\rm str}^{1/2}$, which in turn  becomes
\beq
\frac{\cT_{{\bf q}_r} }{M_{\rm P}^2}\sim M_{\rm P} \cT_0(\chi)\left(\frac{\cT_{\rm str}}{M_{\rm P}^2}\right)^{(n-d-r)/2}\leq \cT_{\rm str}^{1/2}\, .
\eeq
The second inequality can only be satisfied asymptotically if $n-d-r> 1$. The lightest membrane comes from choosing $k=0$ in \eqref{kr}, so it has $r=-d$. Hence, the backreaction of the lightest membrane is mild only if $n>1$. If the backreaction of the lightest membrane cannot be controlled, neither can the backreaction of heavier ones, and so in this case $\Gamma_{\rm EFT}$ is empty.
\begin{center}
	\begin{figure}[htb]
		\centering
		\includegraphics[height=6.5cm]{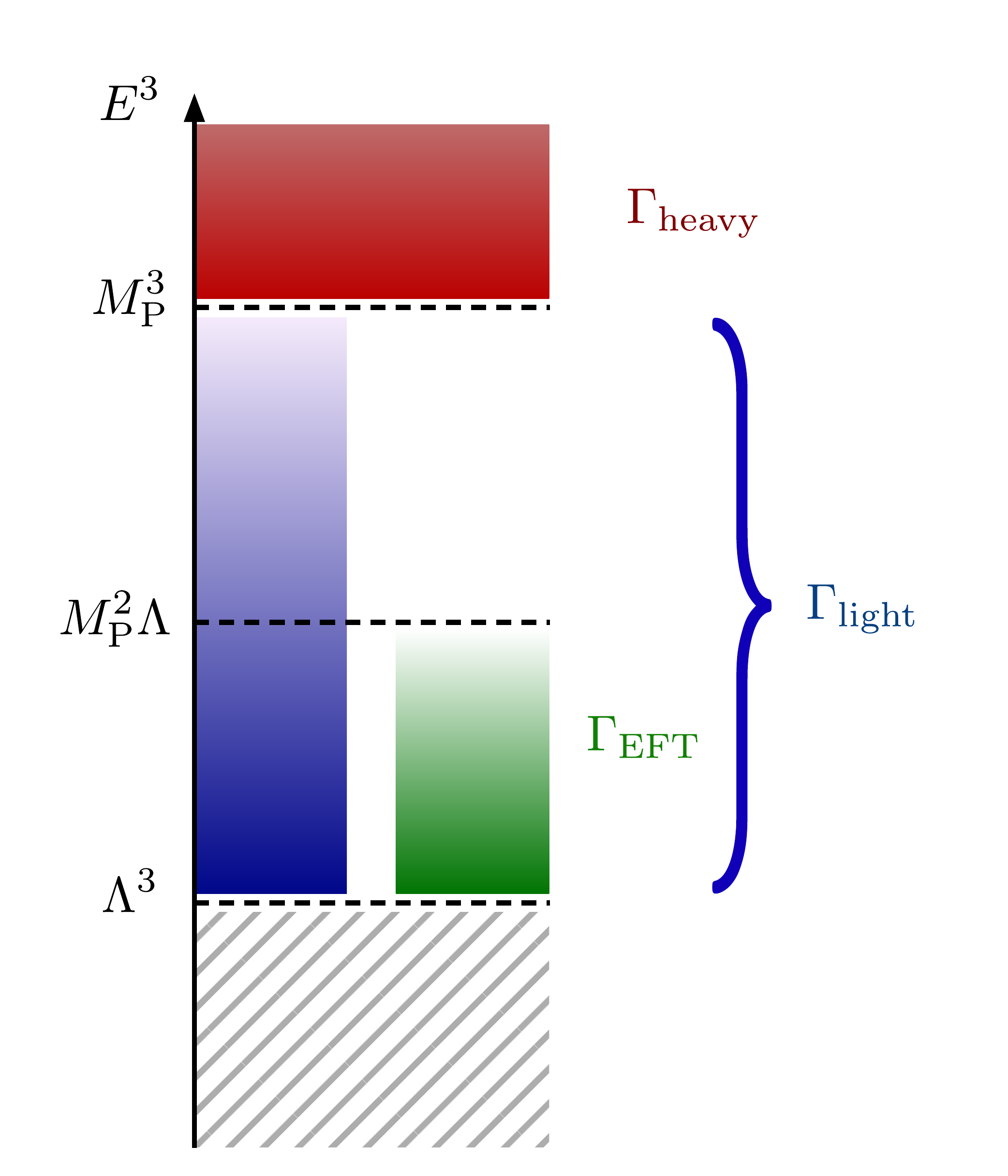}
		\caption{The flux lattices as determined by the energy scales. \label{Fig:Lattices}}
	\end{figure}
\end{center}
\vspace{-1cm}
We have therefore two regimes for the light membranes depicted in Fig.~\ref{Fig:Lattices}, depending on whether the upper bound in \eqref{TEFT} is satisfied (EFT region) or not. The case with $n=1$ would be an example in which $\Gamma_{\rm EFT}$ would be empty as all membranes violate \eqref{TEFT}. In  cases with $n>1$, either all light membranes satisfy \eqref{TEFT} so $\Gamma_{\rm EFT}=\Gamma_{\rm light}$ or only a sublattice of them does, so $\Gamma_{\rm EFT}\subset \Gamma_{\rm light}$.
In order to determine which membranes belong to the EFT lattice, we need more accurate information about the cut-off, which can in fact be smaller than the string tension, $m_{*}=\Lambda< \cT_{\rm str}^{1/2}$. We leave to \cite{Lanza:2021qsu} a more detailed discussion of this point.  For the purposes of this paper, it is enough to notice that the EFT lattice is always a sublattice of $\Gamma_{\rm light}$.

\subsubsection{Scalar flow for EFT Membranes}
\label{sec:memscflow}

In four dimensions, membranes can be related to domain walls interpolating between different EFT vacua. The charge of a membrane interpolating between two vacua with flux  $(t^i,f)$ and $ (t^i,\calr_i^{-n^i}f)$ is fixed to be
\beq
q_a=(1-\calr_i^{-n^i})_a{}^bf_b
\eeq
by charge conservation. As already discussed, the low codimension of these objects make backreaction effects away from the membrane significant, which means that they cannot be freely inserted in an EFT without changing the asymptotic structure of the vacuum. This backreaction can be understood as a classical RG flow as described in section \ref{sec:EFTRG}, such that the EFT eventually breaks down away from membrane. As we will see, this flow may lead the scalars to a weak or strong coupling regime depending on the membrane charges, restricting the sublattice of charges that can be associated to EFT membranes.

The starting point is the off-shell `no-force' condition \eqref{No-fo_Idm} satisfied by any physical charge and tension of a membrane in a 4d $\cN=1$ EFT. Notice that this is not the tension of the physical domain wall but the bare tension $\cT_{\rm mem}$ of the localised membrane, which satisfies the EFT conditions only if $\Lambda$ is high enough. Taking into account the backreaction implies that the tension of the physical domain wall changes away from the membrane location.

Let us investigate the presence of flat domain walls generated by a single membrane with charge ${\bf q}$. 
Assume that the membrane is located at a point $y=\hat y$. The presence of the membrane breaks the spacetime translational symmetry along its transverse direction $y$. A domain wall is an extremal solution of the solitonic equation described by the metric
\be
\label{DWmetr}
\d s^2=e^{2D(y)}\d x^\mu\d x_\mu+\d y^2\,,
\ee
where the warp factor depends only on the transverse direction and respects the $SO(1,2)$ invariance along the directions parallel to the membrane. A \emph{BPS-domain wall} is further characterised by a solitonic solution that preserves half of the bulk supersymmetry. 

In the context of $\cN=1$ supergravity theories with a set of chiral multiplets $\{\phi^\alpha\}$, a $\half$BPS-domain wall is a solution of the following flow equations \cite{Cvetic:1992st,Cvetic:1992sf,Cvetic:1992bf,Ceresole:2006iq,Bandos:2018gjp} 
\begin{subequations}\label{genflow}
\begin{align}
\frac{\d D}{\d y}&=-M^{-2}_P\zeta|\calz|\, ,\label{genflowa}\\
\frac{\d \phi^\alpha}{\d y}&=2M^{-2}_P\,\zeta K^{\alpha\bar \beta}\del_{\bar \beta}|\calz|\, ,\label{genflowb}
\end{align}
\end{subequations}
where $\calz\equiv e^{\frac{K}2}W$ and $\zeta = \pm 1$ specifies the half of  supersymmetry preserved, which is defined by the projector condition (in two-index Weyl notation) \cite{Bandos:2018gjp,Lanza:2019nfa}
\be\label{DWproj}
\epsilon=\ii\zeta e^{\ii\theta}\sigma_3\bar\epsilon\, ,
\ee
with $\theta\equiv {\rm arg}\, \calz$.
Notice that across the zeros of $\calz$ the argument $\theta$ discontinuously shifts by  $\pi$. Hence, to have a continuous Killing spinor satisfying \eqref{DWproj}, also $\zeta$ must discontinuously change sign. For monotonically increasing $|\mathcal{Z}|$, the appropriate choice to preserve half of the bulk supersymmetry is $\zeta=1$ \cite{Bandos:2018gjp,Ceresole:2006iq}.

Let us now investigate domain-wall solutions to the flow equations \eqref{genflow}. We are going to restrict ourselves to the perturbative regimes of  section \ref{sec:infinite}, selected by the string RG flows. As in there, we define the perturbative asymptotic limit  by a subset of these chiral fields $t^i=a^i+\ii s^i$, $ i=1,\dots, I$, whose saxions $s^i$ take large values \eqref{infs}, while other chiral fields $\chi$ are kept finite. In the perturbative regime, the K\"ahler potential is approximately invariant under continuous shifts of the axions $a^i$. As before, we may  assume that it takes the asymptotic form\footnote{The factorisation of the K\"ahler potential on the fields $t^i$ and $\chi^\kappa$  is actually not necessary, see Appendix \ref{app:multimoduli}.}
\be
K= -\log P(s) +\hat K(\chi,\bar\chi) \, ,
\ee
where $P(s)$ is a  function of the saxions whose leading order term can be parametrised as $P(s)=s_1^{n_1}s_2^{n_2-n_1} \cdots s_n^{n_I-n_{I-1}}+\dots$, with  $n_1\leq n_2\leq \dots\leq n_I$, so that $n_I$ is the homogeneity degree of $P(s)$. Assuming that $n_i-n_{i-1} > 0$ $\forall i$ implies that to leading order 
\beq
\label{Ki}
 K_i \equiv \del_i K =-\frac{n_i-n_{i-1}}{2\ii s_i}\, ,\quad\quad~~~ \ell_i=\frac{n_i-n_{i-1}}{2 s^i}\quad~~~~~(\text{with $n_0=0$})\, .
\eeq
More complicated cases in which some $n_i=n_{i-1}$ lead to the degenerate string flows defined below \eqref{Klog}, because then $P(s)$ cannot be approximated as the same monomial on every path. In that case, there can be several strings becoming tensionless in the asymptotic limit, see \cite{Lanza:2021qsu}.
The general formula for the tension of the membranes is given by (see appendix~\ref{app:multimoduli}) 
\beq
\label{Tgeneral}
\calt_{{\bf q}}\simeq M^3_{\rm P}T_0(\chi,\bar\chi)\sum_{\bf r}\rho_{\bf  r}(q,a^i) s_1^{\frac{\hat r_1}2}s_2^{\frac{\hat r_2-\hat r_1}2}\dots s_n^{\frac{\hat r_n-\hat r_{n-1}}2} \, ,
\eeq
which for a single charge and a single saxion $s$ reduces to \eqref{Tsinglesax}. To simplify the notation, we have defined $\hat r_i\equiv  r_i-(n_i-d_i)$ in the exponents of the saxions.

Applying \eqref{genflow} to the fields $\{t^i\}$, we can study the flow equations for the saxions. These can be simplified by employing the dual saxions $\ell_i$, in terms of which they read 
\beq
\label{genflow3b}
\frac{{\rm d} \ell_i}{{\rm d}D} = \frac{2}{M^{2}_P} \frac{\partial}{\partial s^i}\log |\mathcal{Z}|\,.
\eeq
They are accompanied by the equations $\frac{\d }{\d y}a^i=0$, which imply that the axions remain constant along the flow. For simplicity, in the following we set $a^i = 0$. We also assume that the scalars $\chi^\kappa$ are fixed to the value $\chi^\kappa_*$ that extremises $|\mathcal{Z}|$, so that they do not flow away from the membrane.

The flow equations need to be supplemented by appropriate boundary conditions.  Recall that a domain wall has to interpolate between a vacuum on the far left of the membrane to one on its far right. Therefore, asymptotically both vacua have to be reached. Furthermore,  the membrane itself imposes some gluing conditions for the fields and the warping. In fact, at the location of the membrane $y = \hat y$ one could set
\beq
D(\hat y)=0\, ,
\eeq
and impose continuity of the scalars  $(t^i,\chi^\kappa_*)$, albeit their derivatives may develop discontinuities.

In order to get explicit solutions for the warp factor $D(y)$ and the saxions $s^i$, one needs to specify the covariant superpotential ${\mathcal{Z}}$. For simplicity, let us consider the case in which we have a membrane separating a Minkowski vacuum to the left, where the flux quanta are trivially vanishing, i.e. $f_A=0$ which implies ${\mathcal{Z}}=0$, and a vacuum characterised by $f_A\neq 0$ to the right side. Then, the membrane charge becomes equal to the potential on the right, so
\beq
\label{genmemV}
M_{\rm P}^{-2} V = \half \cQ_{\bf q}^2= \half \cQ_{\bf f}^2=   M_{\rm P}^{-4} (4\|\partial |\mathcal{Z}|\|^2-3 |\mathcal{Z}|^2)\, ,
\eeq
where we have used that $\cT_{\bf f}=2|\Delta \mathcal{Z}|=2M_{\rm P}^{3} |\mathcal{Z}|= 2e^{K/2}|f_A\Pi^A(t)|$. We will refer to such membranes that generate the full flux superpotential as \emph{generating membranes}.  We assume that the potential \eqref{genmemV} admits a supersymmetric extremum $\phi^\alpha_0=(t^i_0, \chi^\kappa_0)$ at which $\partial_\alpha |\mathcal{Z}|_{\phi^\alpha_0}$ = 0.

Let us focus on {\em elementary saxionic} flows, to illustrate the different behaviours of the RG flow for membranes with only electric or magnetic charges. Recall that elementary saxionic membranes are those with a charge $\textbf{q}_{\textbf r}\in \Gamma_{\bf r}$ as defined in section \ref{sec:EFTlattice}, so that we keep only one monomial term in the sum \eqref{Tgeneral}. In this case, the saxionic dependence of the tension satisfies
\beq
\partial_{i} \cT_{q_r}= K_i \ \sigma^{(r)}_i\ \cT_{q_r}\, ,
\label{Tsigma}
\eeq
where $\sigma_i^{(r)}$ is determined in terms of the perturbative limit discrete data, as
\beq
\label{sigma}
\sigma^{(r)}_i = -\frac12\frac{\hat r_i-\hat r_{i-1}}{n_i-n_{i-1}}\, ,
\eeq
where we have used \eqref{Tgeneral} and \eqref{Ki}, and $\hat r_0\equiv 0$. 
Notice that $\textrm{sign}(\sigma_i^{(r)})=-\textrm{sign}(\hat r_i-\hat r_{i-1})$ because $n_i-n_{i-1}$ is definite positive. We can then see that $\sigma_i^{(r)}>0$ implies that the saxion $s_i$ appears with a negative exponent in the asymptotic behaviour of the tension \eqref{Tgeneral}. Hence, membranes with $\sigma_i^{(r)}>0$ $\forall i$ will be light ($\cT_{q_r}<M_{\rm P}^3$) along any path in the perturbative regime.

The flow equation \eqref{genflow3b} now gives
\beq
\label{flowdill}
\frac{{\rm d}\ell_i}{{\rm d}D}=-4 \ \sigma^{(r)}_i \ell_i\ .
\eeq
which can be easily solved as
\beq
\ell_i=\ell^*_i\ e^{-4\sigma^{(r)}_iD} \, , 
\label{lflow}
\eeq
where $\ell^*_i$ is the value of the dual saxions at the location of the membrane. This shows that the flow of the dual saxions $\ell_i$ is completely determined by the flow of the warp factor $D$,\footnote{Note that this is exactly the right dependence such that a probe string  parallel to the domain wall and of arbitrary admissible charges $e^i$ does not feel any force.}
which in turn can be obtained by solving \eqref{genflowa}
\beq
\label{warp}
e^D=\left(1-{\alpha^2}\frac{|\mathcal{Z}_*|}{2 M_{\rm P}^{2}}(y-\hat y)\right)^{\frac{1}{{\alpha^2}}}\, ,
\eeq
where $\calz_*\equiv \calz(s_*,\chi_*)$ and
\beq
\label{kflow}
\frac{\alpha^2}{2} \equiv - 2  \sum\limits_{i=1}^{I} (\hat r_i-\hat r_{i-1}) \sigma_i^{(r)}=\sum_i^I \frac{(\hat r_i-\hat r_{i-1})^2}{n_i-n_{i-1}}\, .
\eeq
In the last step we have used \eqref{sigma} which obviously implies $\alpha^2>0$.\footnote{We have conveniently chosen the symbol $\alpha$ to define this quantity as it will precisely correspond to the dilatonic factor appearing in the extremality bound when studying the WGC in section \ref{sec:WGCmem}.} Recall that we have assumed that the rest of the scalars $\chi^\kappa$ do not flow as they are fixed to the values $\chi^\kappa_*$ extremising the potential. We could relax a bit this assumption in the case in which they only appear in $K$ but not in $W$. In that case, the modification of the flow of the warp factor is minimal and one can simply replace $\alpha^2 \rightarrow \alpha^2 + 2 \hat K^{\kappa\bar\lambda}\hat K_\kappa \hat K_{\bar\lambda}$ in \eqref{warp}.
 
Finally, by plugging the warp factor solution into the flow of the dual saxions \eqref{flowdill}, we  get 
\beq
\ell_i=\ell^*_i\, \left[1-{\alpha^2}\frac{|\mathcal{Z}_*|}{2M_{\rm P}^{2}}(y-\hat y)\right]^{-\frac{4\sigma^{(r)}_i}{{\alpha^2}}}  \ .
\label{lflow2}
\eeq
Clearly, the direction of the flow depends on the sign of $\sigma_i^{(r)}$. If $\sigma^{(r)}_i>0$, the dual saxion $\ell_i$ will grow when going away from the membrane. The flow then breaks down at a distance $y-\hat y = 2M_{\rm P}^{2} ({\alpha^2} |\mathcal{Z}_*|)^{-1} = \frac{4M_{\rm P}^{2}}{{\alpha^2} }  \cT_r^{-1}$ from the membrane, where $e^D=0$. However, by assumption our dynamical membranes satisfy the small $\cT_r/(M_{\rm P}^{2}\Lambda)$ condition, where $\Lambda$ is the EFT UV cut-off. Therefore the break-down of the solution is indeed at a distance large compared to  the minimal EFT length. This is consistent with the fact that such a membrane has an EFT description. Indeed, $\ell_i\rightarrow \infty$ at the strong coupling distance $2M_{\rm P}^2 ({\alpha^2} |\mathcal{Z}_*|)^{-1}$ and then our weakly-coupled EFT stops to be reliable before the singular point. 
The particular case in which the saxion does not appear in the superpotential would correspond to set $\hat r_i=-n_i$ so that $\sigma_i=1/2>0$.

If instead the membrane belongs to $\Gamma_{\rm light}$ but not to $\Gamma_{\rm EFT}$, we still have $\sigma^{(r)}_i>0$ and  $ \calt_{\rm mem}\leq M^3_{\rm P}$, but $\calt_{\rm mem}$ is larger or just slightly smaller than $\Lambda M^2_{\rm P}$. Hence the flow exits the perturbative regime already at a spatial distance which is smaller (or just slightly bigger) than $\Lambda^{-1}$. Hence such flows cannot be described within our EFT.

Conversely, if $\sigma^{(r)}_i<0$, the dual saxions flow in the opposite direction, i.e. are sent to smaller values $\ell_i \rightarrow 0$, getting deeper into the perturbative regime.  However, we do not expect to find any extremum for $\mathcal{Z}$ when approaching the asymptotic limit $t^i\rightarrow \ii\infty$ and thus the saxions will continue flowing to infinity if all $\sigma^{(r)}_i<0$, reaching the asymptotic infinity at a finite spatial distance $y-\hat y=\frac{4M^2_{\rm P}}{\alpha^2}\calt^{-1}_{r}$. Membranes with $\sigma^{(r)}_i<0$ $\forall i$ are heavy membranes with $\cT_{q_r}>M_{\rm P}^3$ along any path in the perturbative regime. Hence, these flows  completely degenerate at a subplanckian distance and, since we are in a perturbative regime, no non-perturbative corrections can come to the rescue. The absence of such an extremum indicates that there is no solution for the heavy membrane. This further motivates that only membranes with charges in $\Gamma_{\rm light}$ might eventually be described as dynamical objects in the EFT.

We can easily provide the effective tension of the domain wall at some distance $y$ away from the membrane. By plugging \eqref{lflow2} into \eqref{Tgeneral} we get 
\beq
\calt^{\rm eff}_{q_{r}}(y)\simeq \frac{\cT^*_{q_r}}{ 1-{\alpha^2}\frac{|\mathcal{Z}_*|}{2M_{\rm P}^{2} }(y-\hat y) }\,,
\label{incTmem}
\eeq
where we have identified the membrane bare tension with $\calt_{\bf q}=2|\calz_*|$ by matching this solution with the left-hand side. 
Therefore, $\calt_{q_{r}}(y)$ always increases moving away from the membrane, as anticipated in \eqref{memTeff}. The effective membrane tension in \eqref{Tefflambda} is nothing but \eqref{incTmem}, from where the identification of $\Lambda_{\rm strong}$ in \eqref{memLstrong} follows directly.
It explicitly shows that, even if the initial values $|s^i_*|$ on the membrane are large enough, in compatibility with the asymptotic perturbative regime, $s^i(y)$ vanishes at $y_{\rm strong}=(\alpha^2\Lambda_{\rm strong})^{-1}$ where the effective coupling diverges.

\section{Swampland conjectures for extended objects}

\label{sec:SC}

Some swampland conjectures, like the WGC, constrain properties of the states of the theory. Others, like the asymptotic de Sitter conjecture, constrain the form of the EFT action. When referring to low codimension objects, all these conjectures get interlinked, as the 2-form and 3-form gauge couplings determine both the behaviour of the effective action of the scalars as well the behaviour of the charge of the low codimension objects.
One of the advantages of the dual formulation of the EFT Lagrangian in terms of 2-form and 3-form gauge fields is that we can translate swampland conditions on the axionic kinetic terms and scalar potentials to properties of the strings and membranes. And vice versa, constraints on the charge and tension of strings and membranes translate to particular behaviours of the scalar kinetic terms and scalar potential. In this section, we will  discuss these connections and show that the WGC for strings implies the SDC with an exponential mass rate fixed by the extremality bound, while WGC-saturating membranes generate a scalar potential that satisfies the de Sitter conjecture.

\subsection{Repulsive Force Conjecture}
\label{sec:RFC}

Given our EFT viewpoint on strings and membranes, we can consider the case in which the supersymmetry is spontaneously broken and, at a lower cutoff scale, the EFT becomes non-supersymmetric. In that case we expect  no-force identities not to be valid anymore and, if at such cut-off scale they still belong to the spectrum of possible EFT operators, one may formulate a Repulsive Force Conjecture (RFC) for strings and membranes. By analogy with the particle case \cite{Palti:2017elp,Heidenreich:2019zkl}, it is natural to demand that whenever the no-force identities fail identical strings and membranes repel each other, resulting in the following inequalities:
\be\label{RFC}
\boxed{
\begin{aligned}
\|\del\calt_{\rm str}\|^2&\leq M^2_{\rm P}\calq^2_{\bf e}\quad~~~~~~~~~~~~~~~~~~~\text{(strings)}\\
\|\del\calt_{\rm mem}\|^2&\leq M^2_{\rm P}\calq^2_{\bf q}+\frac32 \calt_{\rm mem}^2\quad~~~~\text{(membranes)}
\end{aligned}
}
\ee
Imposing these RFC at all scales will translate into non-trivial conditions on the underlying ${\cal N}=1$ EFT. Let us briefly discuss several instances of such constraints.

Given a UV cut-off $\Lambda$, the scalar-mediated forces
captured by the lhs of \eqref{RFC} should involve scalars with masses $m \ll \Lambda$. Therefore, as we lower the cut-off and hit some scalar mass threshold, the terms of the form $\|\del\calt\|^2$ will lower their value, as some of the scalars will be integrated out and will not contribute to the derivatives. A priori this automatically satisfies the strict inequalities in \eqref{RFC}, if their rhs does not change. However, one can easily see that in some cases the RFC inequalities should still be saturated as we cross the threshold. Indeed, if after integrating out some scalars the theory is still supersymmetric and its shift symmetries have not been modified, by the arguments of section \ref{sec:N=1EFT} and appendix  \ref{app:RFC_Id} the no-force identities \eqref{No-fo_Ids} and \eqref{No-fo_Idm} should still hold, irrespective if we evaluate them on a non-supersymmetric vacuum or even off-shell. Therefore, by consistency the quantities involved in the rhs of \eqref{RFC} should vary as well. The most natural possibility would be that integrating out scalars also removes string and membrane charges from the spectrum, or in other words that it reduces the lattices $\cC^{\text{\tiny EFT}}_{\rm S}$ and $\Gamma_{\rm EFT}$.  In the case of membranes, this picture matches well with the definition \eqref{GammaEFT} and with the relation of $\Gamma_{\rm EFT}$ to 3-form multiplets. Indeed, as seen in \cite{Lanza:2019xxg} such 3-form multiplets account for the dynamical fluxes of the EFT and always contain scalars, so if we remove some of the latter in a supersymmetric fashion we need to remove the whole 3-form multiplet, necessarily reducing $\Gamma_{\rm EFT}$. Alternatively, if we take the cut-off below the supersymmetry-breaking scale $\Lambda_{\rm SUSY}$, then we do not need to integrate out entire 3-form multiplets. In this sense, the RFC above suggests that scalars could be integrated out without removing dynamical fluxes and their corresponding membranes from the spectrum. For instance, one may consider the case where all the scalars that couple to membranes get a mass around $\Lambda_{\rm SUSY}$ or above, while part of the membranes satisfy $\cT_{\rm mem}/M_{\rm P}^2 \ll \Lambda_{\rm SUSY}$. Then, by integrating out such scalars one should get an EFT with membranes which interact only gravitationally and electrically, and then repel each other, independently of the value of their tension.\footnote{In particular, if the mutual electric forces are repulsive, namely $\calq^2_{\rm mem} > 0$, then integrating out the scalars trivially leads to a self-repulsion condition. On the other hand, if $\calq^2_{\rm mem} < 0$, \eqref{RFC} implies that the gravitational repulsion ought to overcome the electric attraction.} Notice that this last scenario will not occur if all the scalars that couple to a given membrane have a mass comparable to $\cT_{\rm mem}/M_{\rm P}^2$.

An alternative effect that may lead to a violation of the no-force identities \eqref{No-fo_Ids} and \eqref{No-fo_Idm} relies on the breaking of the continuous shift symmetries assumed in section \ref{sec:N=1EFT}. This is particularly natural for strings for which, as we have seen in section \ref{sec:EFTRG}, lowering the cut-off $\Lambda$ is equivalent to flow to a region in moduli space in which non-perturbative effects start to become relevant. As the derivation of the string no-force identity \eqref{No-fo_Ids} relies on the description in terms of dual linear multiplets, the string RFC inequality becomes non-trivial once that non-perturbative effects are taken into account.  When that is the case, corrections to the classical string tension due to non-perturbative effects are expected, see e.g. \cite{Mayr:1996sh,Baume:2019sry}, and therefore to the lhs of the string RFC. Therefore, to test whether the string RFC is satisfied one should compute the appropriate form for the string physical charge in this regime. 

A different manner in which continuous shift symmetries are broken is by considering corrections to kinetic terms of the scalars that make them dependent on the background fluxes. In string theory setups, such kind of corrections may for instance appear from warping effects \cite{Martucci:2016pzt}, or when higher-derivative corrections are taken into account in flux compactifications \cite{Ibanez:2014swa,Ciupke:2015msa,Bielleman:2016grv,Landete:2017amp}.  Because the flux dependence must respect the discrete shift symmetries of the compactification, the dependence of the K\"ahler potential on the fluxes must appear through the gauge invariant combination of quantised fluxes and axions, that correspond to the quantities $\rho_r$ in \eqref{growth} (see \cite{Bielleman:2015ina,Carta:2016ynn,Herraez:2018vae,Grimm:2019ixq} for more details). Hence, the kinetic terms will be axion-dependent and the corresponding continuous shift symmetry will be broken. While the derivation of the no-force identity \eqref{No-fo_Idm} on Appendix \ref{app:RFC_Idm} does not rely on the existence of axionic shift symmetries in the kinetic terms, the dualisation to a 3-form Lagrangian of the form \eqref{dualF4lagr} assumes that they do not depend on the fluxes. A flux dependence on the K\"ahler potential will necessarily modify the expressions for the 3-form kinetic terms $T_{ab}$, becoming non-canonical, and therefore the physical membrane charge $\cQ_{\bf q}$. As a result, one would expect that the no-force identity \eqref{No-fo_Idm} no longer holds in non-supersymmetric setups, and the RFC for membranes becomes a non-trivial constraint on the resulting EFT.

\subsection{Weak Gravity Conjecture}
\label{sec:WGC}

The Weak Gravity Conjecture states that, given a gauge theory weakly coupled to Einstein gravity, there must exist some electrically charged state with tension $\cT$ and charge $\cQ$ satisfying
\beq
\label{WGC0}
\cQ\, M_{\rm P}\geq \gamma \, \cT\, ,
\eeq
where $\gamma$ is the charge-to-mass ratio in Planck units of an extremal black hole in that theory, $\gamma \equiv \cQ/\cT\big|_{\rm extremal}$. It is usually an order one numerical factor, but if the theory contains massless scalars, they can also contribute to the value of $\gamma$. In particular, whenever the gauge kinetic function for the $p$-form gauge field has the form $F(\hat s) \sim e^{\alpha_i \hat s^i}$ in terms of the canonically normalised scalars $\hat s^i$, the black brane solution is dilatonic and the extremality factor can be computed \cite{Horowitz:1991cd} (see also \cite{Lee:2019wij,Heidenreich:2015nta,Lee:2018spm,Gendler:2020dfp}) to be
\beq
\label{dilatonic}
\gamma^2= \frac{p(2-p)}{2}+\frac{|\vec{\alpha}|^2}{4}\, ,
\eeq
in four space-time dimensions. The contribution of the scalars is then purely encoded in $\alpha$, known as the dilatonic factor, which is a constant given by 
\beq
\label{alpha}
|\vec{\alpha}|^2= \frac{\cG^{ij}\partial_i F\partial_j F}{F^2}\equiv \frac{\|\partial F\|^2}{F^2}\, ,
\eeq
with $\mathcal{L}\supset \frac1{2} F\, \mathcal{H}_{p+1}^2+ \frac{M_{\rm P}^2}2 \cG_{ij}\partial s^i\partial s^j$.

\begin{center}
	\begin{figure}[htb]
		\centering
		\includegraphics[width=5cm]{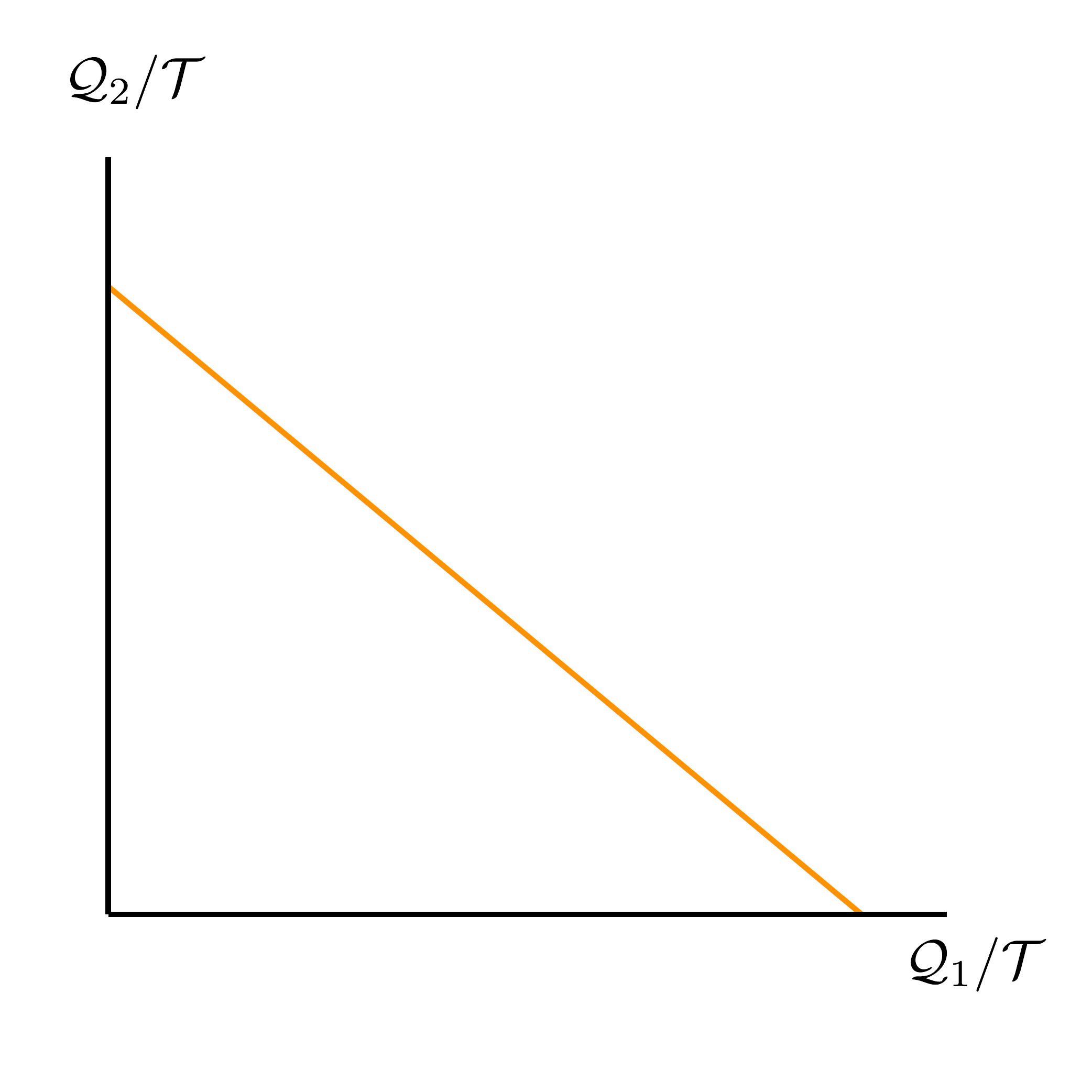}\hspace{1cm}
		\includegraphics[width=5cm]{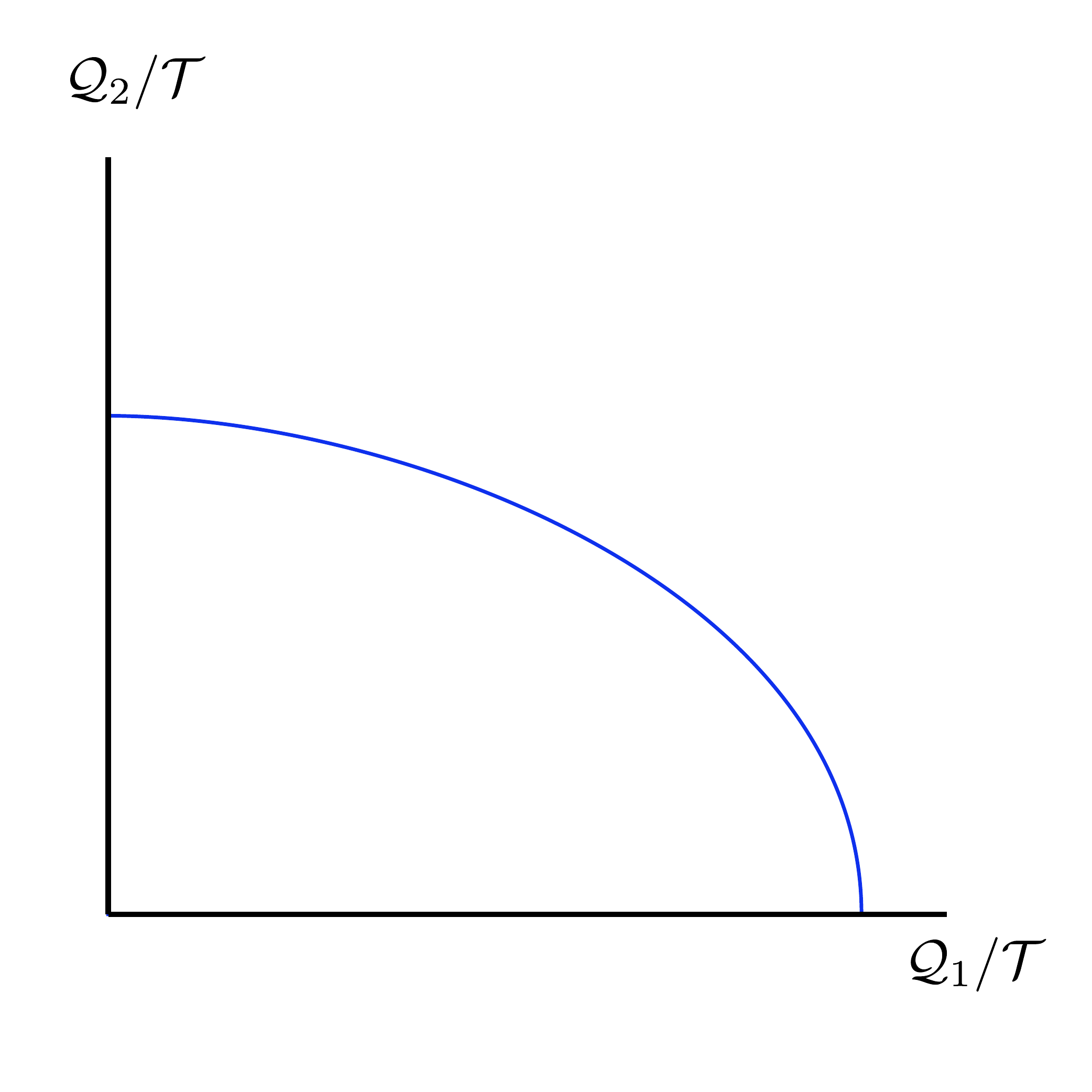}
		\caption{Extremal region for mutually BPS (left) and non-mutually BPS states (right).  \label{Fig:BPSvsExtr}}
	\end{figure}
\end{center}

The generalisation to multiple gauge fields can be phrased as the condition that the convex hull of the charge-to-mass ratio of the states includes the extremal region \cite{Cheung:2014vva}. For this, one needs to define the charge-to-mass ratio vectors:
\beq
\label{CH}
\vec{z}\equiv \frac{\vec Q}{\cT}\, ,
\eeq
with $\vec{Q}=(Q_1,Q_2,\dots)$ such that $|\vec{Q}|^2=Q_1^2+Q^2_2+\dots=q_IF^{IJ}q_J=\|\cQ\|^2$, where $q_I$ are the quantised charges and $F^{IJ}$ the inverse gauge kinetic function. When plotted over the hyperplane with axis defined by $z_i$, the extremal region is a codimension one hyperplane.
In the absence of scalars, this extremal region is just a ball of radius $\gamma$. However, the presence of scalars modifies this region and the shape needs to be determined by studying the extremal solutions of the corresponding theory. Since the states will generically be charged under several gauge fields with a different scalar behaviour of the gauge kinetic function, the simple dilatonic formula \eqref{dilatonic} is not valid anymore. If the states have a tension linear in the quantised charges (as occurs with BPS states), then this extremal region takes the form of an ellipsoid or a hyperplane, depending respectively on whether the states feel any force among each other or not \cite{Heidenreich:2015nta,Gendler:2020dfp}. These possibilities have been represented in Fig.~\ref{Fig:BPSvsExtr}. If the state has a charge to tension ratio inside (outside) the extremal region, it is called subextremal (superextremal) and violates (satisfies) the WGC. Hence, the particular value of $\gamma$ in \eqref{WGC0} depends on the charge direction $\vec{z}$ considered.

The WGC has been tested in multiple string theory setups for codimension $> 2$ objects, as e.g. particles in four dimensions \cite{Heidenreich:2015nta,Grimm:2018cpv,Lee:2019tst,Enriquez-Rojo:2020pqm,Gendler:2020dfp}. The question that we are interested in here is how to properly apply the WGC to low codimension objects, i.e. strings and membranes in four dimensions, since the tension becomes scale dependent. Should then the WGC hold at any scale or is it a constraint only on the UV or IR values of $\cT,\cQ$? We have seen that, from the EFT perspective, $\cT$ must be understood as the tension of these objects evaluated at a given cut-off scale $\Lambda$, which suggests that it is more natural to impose the WGC at the UV. 
This is in contrast to the case of higher codimension objects where the WGC is typically imposed on the asymptotic IR values of the charge and tension.\footnote{More precisely, in the presence of several charged objects, it is natural to evaluate the WGC at the IR tension of the each object, see e.g. \cite{Cheung:2014vva}.} This is not possible for low codimension  objects as their backreaction destroys the asymptotic structure of the vacuum.\footnote{An analogy in terms of particles would be to try to define a WGC for a gauge group that confines in the IR.} The perspective we will pursue in this paper is therefore that the proper interpretation of the WGC corresponds to impose that
\beq
\label{WGC}
  \cQ(\Lambda) \, M_{\rm P} \geq  \gamma\ \cT(\Lambda)\, ,
\eeq
for any  cut-off scale $\Lambda$, as long as EFT description does not break down. By varying $\Lambda$, we will be probing the whole RG flow generated by these objects, or in other words the backreaction of the physical objects. In this sense, our perspective is analogous to the Local Weak Gravity Conjecture proposed in \cite{Klaewer:2016kiy}, where it was applied to point-like objects. Notice that imposing \eqref{WGC} is conceptually stronger than requiring the WGC to hold for strings and membranes in the probe approximation. This goes along with the intuition that the WGC should apply to physical objects, i.e.  upon taking into account the classical backreaction inducing a non-trivial profile EFT for the fields. Since the scalar flow forces the EFT to enter into a strongly-coupled regime at some distance $\Lambda_{\rm strong}^{-1}$ from the object, we will only be able to check if the objects satisfy the WGC in the weakly coupled regime near the string/membrane. 

The generalisation to multiple gauge fields will again involve to replace $\gamma$ by the corresponding extremal region, which we require to be independent of $\Lambda$.  Notice that not any BPS string or membrane saturating the no-force conditions \eqref{RFC} will also necessarily satisfy the WGC \eqref{WGC}. This is going to crucially depend on the behaviour of the K\"ahler potential.  Hence, the fact that the objects satisfy the WGC in the UV perturbative regime already yields interesting implications for the EFT Lagrangian,  as we will show later on.

Before getting into the details, a further comment is in order. The definition of the WGC relies on the extremality bound for black holes, which is essential to compute the value of $\gamma$. However, the notion of extremality becomes confusing when discussing low codimension objects, as the strong classical backreaction makes impossible to have extremal asymptotically flat solutions. So there are not usual black brane solutions of low codimension and we cannot define extremality in terms of the asymptotic values of the tension and charges.  However, there is still a notion of extremality (and hence, a WGC bound) that survives:\footnote{Analogously, the WGC applied to axions is missing a black hole interpretation, but still seems to be realised in string theory. In string theory, we expect that if the WGC applies for particles, it should also apply to instantons and low codimension objects, as they can be related by string dualities.} an extremal $p$-brane solution is a solitonic solution to the flow equations which preserves the $SO(1,p)\times SO(d-p-1)$ isometries, i.e strings that satisfy the metric Ansatz \eqref{strmetr} and membranes with a metric \eqref{DWmetr}. For instance, the metric Ansatz for membranes corresponds to that of a flat domain wall, so that this type of membranes would be candidates to describe domain walls solutions if the scalars reach a minimum of the potential when flowing to the IR. Extremality would  then be associated to flat domain walls with infinite area, while superextremal solutions would correspond to bubbles of finite radius that expand allowing for non-perturbative transitions. Even if we cannot guarantee the existence of the solution in the IR, we can show that they satisfy some good features in the UV,  in the sense that the extremal region is independent of $\Lambda$ and reduces to the value set by the dilatonic formula \eqref{dilatonic} for single-charged objects. To call it extremal we also require that the flow equations do not break down in the UV regime where we can explore them, by entering in a non-physical regime for instance, which will already allow us to discard some cases for degenerate string flows. 
We believe that these conditions are necessary so we will call these solutions extremal from now on, but notice that these conditions might not be sufficient to guarantee actual extremality.

Since the  flow of the scalars works differently for strings and membranes, we will deal with both cases separately.

\subsubsection{WGC for strings}
\label{sec:WGCstr}

Let us recall that the tension and charge of the EFT strings read
\beqa
\label{TQagain}
\cT_{\bf e}(\Lambda)=M_{\rm P}^2 e^i\ell_i(r_\Lambda)\, ,\quad  \quad \cQ^2_{\bf e}(\Lambda)=M_{\rm P}^2 \cG_{ij}(\ell(r_\Lambda))e^i e^j\, ,
\eeqa
where $r_\Lambda=\Lambda^{-1}$ is the cut-off distance  from the localised string, and we recall that    $\ell_i\equiv -\frac12 \frac{\partial K}{\partial s^i}$  and $\cG_{ij}$ is the inverse of the of the 2-form gauge kinetic function,  given by \eqref{GGmetric}.

Although these strings always satisfy automatically the no-force (BPS) condition
$\cQ^2_{\bf e}M_{\rm P}^2= ||\partial \cT_{\bf e}||^2=\cG_{ij}\partial_{\ell_i} \cT_{\bf e} \partial_{\ell_j} \cT_{\bf e}$,  as anticipated they do not necessarily satisfy the WGC.
This depends on the specific form of the 2-form gauge coupling, i.e. the K\"ahler field metric. We will now show that the extra condition that one needs to impose to ensure these strings to be extremal is to be weakly coupled, i.e. to have charges ${\bf e}\in \cC^{\text{\tiny EFT}}_{\rm S}$.

As we have seen, the RG flow of weakly coupled EFT strings selects a perturbative/asymptotic regime in moduli space. If the string has charges $e^i$, the following saxionic fields are sent to infinity at the core of a string located at $r=0$,
 \beq
 \label{flowWGC}
 s^i=s^i_0+e^i \sigma\, , \quad \quad \sigma= \frac{1}{2\pi}\log\frac{r_0}{r}\rightarrow \infty\,.
 \eeq
 As mentioned around \eqref{Klog}, in typical string theory examples the K\"ahler potential takes the asymptotic form $K=-\log P(s)$, with $P(s)$ a homogeneous function of  integral positive degree. Possible perturbative corrections scaling with a lower degree under an overall rescaling of the saxions are subleading in the saxionic region that corresponds to the vicinity of the string core, since there we probe the regime of large $s^i$.
 In the following we will assume a K\"ahler potential of this form, even if the key results of the following discussion do not depend on that assumption.

Let us first consider the case of an elementary string charged under a single 2-form gauge field (i.e.\ a single-charged string). If there is only one saxion $s$, we can set $P(s)=s^n$ for some $n>0$. Then the dual saxion $\ell$ and the field metric take the form
\beq
\ell =\frac{n}{2s}\, ,\quad  \quad \cG_{\ell\ell}=\frac{n}{2s^2}\, .
\label{GAA}
\eeq
In this case, the corresponding axionic string can be regarded as  a dilatonic string  \cite{Dabholkar:1990yf,Horowitz:1991cd} whose extremality factor $\gamma$ can be derived from the gauge kinetic function of the 2-form gauge fields using \eqref{dilatonic} for  $p=2$:
 \beq
 \label{gastring}
 \gamma^2= \frac{\alpha^2}{4}
\quad
\text{ with }
\quad 
 \alpha=\frac{\|\partial \cG_{\ell\ell}\|}{\cG_{\ell\ell}}= 2 \frac{\|\partial \cQ_{\bf e}\|}{\cQ_{\bf e}}\, ,
 \eeq
where $\cG_{\ell\ell}$ is the gauge kinetic function of the two-form $\cB_2$ under which a string is charged, with charge $e$. Using \eqref{GAA} we get that
 \beq
 \label{gstring}
 \gamma^2= \frac{2}{n}\, .
 \eeq

On the other hand, by plugging \eqref{GAA} into \eqref{TQagain}, the charge-to-mass ratio of the string satisfies
\beq
\label{QTelem}
M_{\rm P}^2\cQ_{\bf e}^2=\|\partial \cT_{\bf e}\|^2= \frac{2}{n} \cT^2_{\bf e}\, ,
\eeq
 which indeed coincides with the extremality factor $\gamma$ in \eqref{gstring}. Hence elementary weakly coupled BPS strings of this type saturate the WGC.

For multiple 2-form gauge fields $\cB_{2\, i}$ there are several charges involved and one needs to check the convex hull condition, as discussed around \eqref{CH}. The dilatonic formula is not valid anymore, but we expect the extremal region to form a hyperplane in the space spanned by the charge-to-mass ratio vectors. This is simply due to the fact that the strings are mutually BPS, so that  $\cT_{e_1+e_2}=\cT_{e_1}+\cT_{e_2}$. Since the tension is linear in the charges this will always form a straight line in the charge-to-mass ratio plane.\footnote{Consider two strings with tensions and charges $(\cT_1,Q_1)$ and $(\cT_2,Q_2)$. A string given by taking $n_1$ copies of the first and $n_2$ of the second one will have a tension and charge $(n_1\cT_1+n_2\cT_2,n_1Q_1+n_2Q_2)$. The charge-to-mass ratio vector is $\vec{z}=(n_1Q_1,n_2Q_2)/(n_1\cT_1+n_2\cT_2)$. When plotting all possible values of $n_1,n_2$ one always gets a straight line in the $\vec{z}$-plane of the form $z_2=az_1+b$ with $a=-\frac{Q_2\cT_1}{\cT_2Q_1}$ and $b=Q_2/\cT_2$.} If the single-charged strings $\cT_{e_1}$ and $\cT_{e_2}$ are extremal, with factors $\gamma_1$ and $\gamma_2$, then also $\cT_{e_1+e_2}$ will be so, yielding 
\beq
\label{strex_vecz}
\vec{z}=\frac{\vec{Q}}{\cT}=\frac{(\gamma_1 \cT_{e_1}, \gamma_2 \cT_{e_2})}{\cT_{e_1}+\cT_{e_2}}\, ,
\eeq
which is indeed the equation of a  straight line with slope $-\gamma_2/\gamma_1$ in the $(z_1,z_2)$-plane. Therefore, once two extremal single-charged strings are identified, it is easy to draw the extremal region associated to the charge directions in between. The elementary strings above can be used as these reference points (with $\gamma^2_i=2/n_i$ if the metric is diagonal to leading order along the flow). Clearly, this can be trivially generalised to any number of gauge fields, obtaining a hyperplane.

\bigskip
\noindent
\textbf{An example}

\begin{center}
	\begin{figure}[ht]
		\centering
		\includegraphics[width=7.5cm]{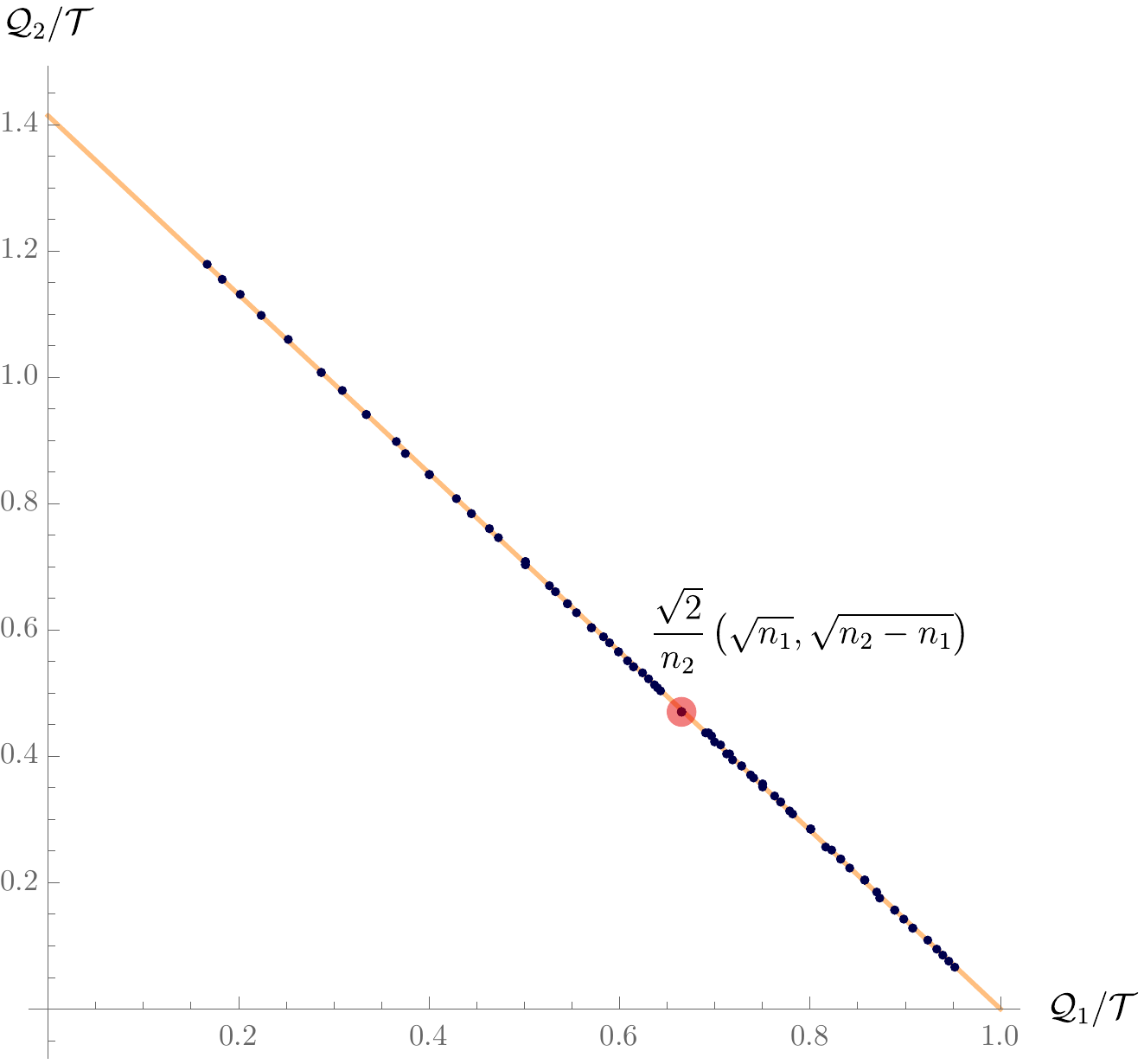}\hspace{0.5cm}
		\includegraphics[width=7.5cm]{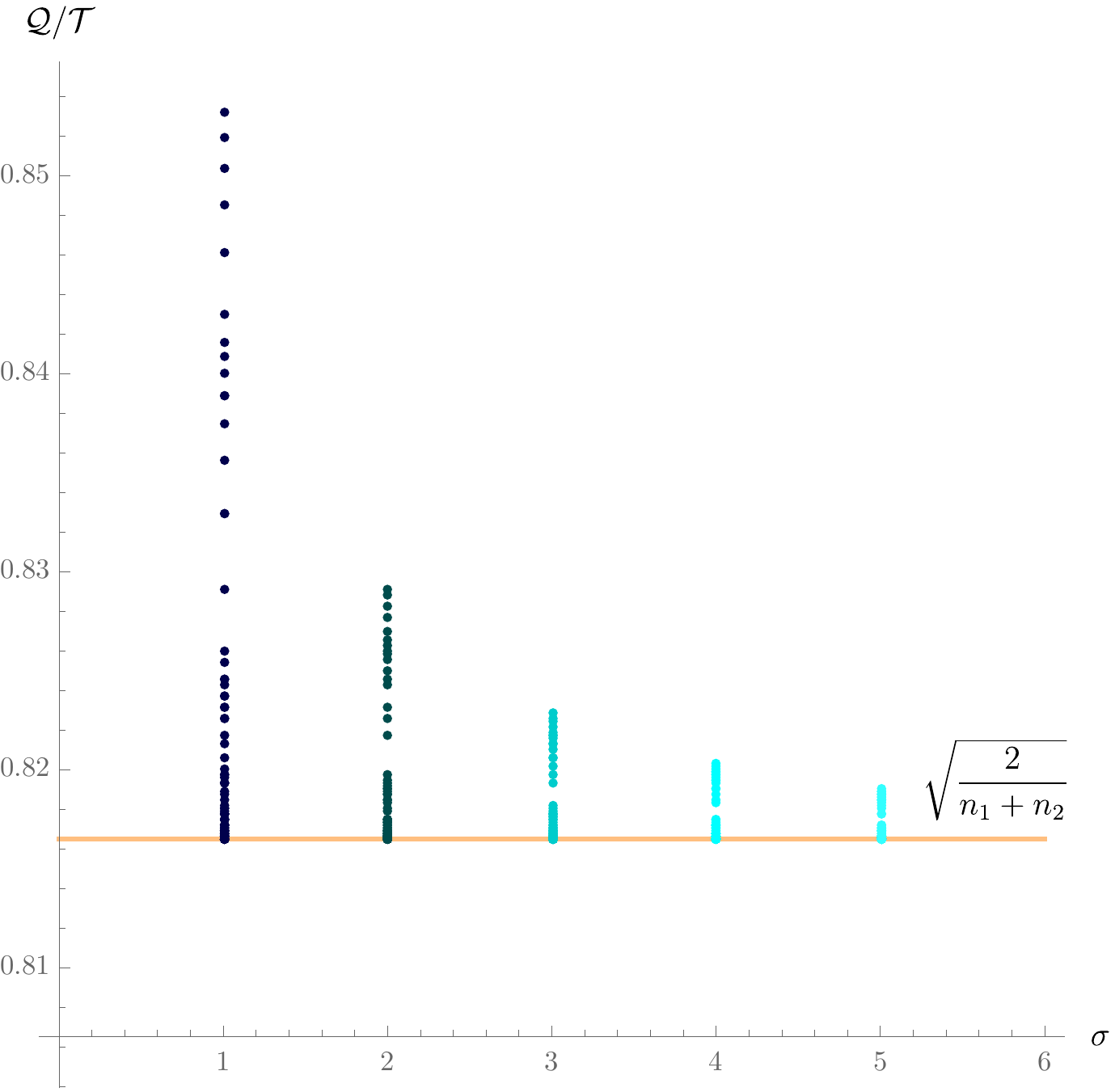}
		\caption{The charge-to-tension ratios of BPS strings in a two modulus limit with $n_1=2$ and $n_2=3$, i.e. $K=-\log (s_1^2 s_2)$. To the left it is plotted  the charge-to-tension vectors \eqref{strex_vecz}, which draw a straight line independent of the cut-off $\Lambda$. To the right it is plotted the charge-to-tension ratio as we move along the flow, which is lower bounded by the extremality factor \eqref{gaminex}. \label{Fig:QT_BPS}}
	\end{figure}
\end{center}

Let us illustrate this construction with an example. Consider an EFT with two saxions and a K\"ahler potential of the form:
\beq
K=-\log(s_1^{n_1}s_2^{n_2-n_1})\, ,
\label{Kex}
\eeq
so that $P(s)$ in \eqref{Klog} is an homogeneous monomial of degree $n_2$. As already mentioned, this monomial can appear as the leading order contribution when approaching an asymptotic limit in moduli space along a non-degenerate string flow. To leading order the metric is diagonal, which implies that  $z_i=\calg_{ii}^{1/2}w_i$. Using \eqref{QTelem}, we get that the elementary strings have the following charge-to-tension ratio:
\beq
z_i|_{e_i}=\sqrt{\frac{2}{n_i-n_{i-1}}}\, ,
\eeq
with $n_0=0$. The extremality bound is a straight line joining the two points $(\sqrt{\frac{2}{n_1}},0)$ and $(0,\sqrt{\frac{2}{n_2-n_1}})$, plotted as an orange line in figure \ref{Fig:BPSvsExtr}. The charge-to-tension ratio $\vec{z}(\Lambda)$ of the strings charged under both gauge fields is given by
\beq
\vec{z}(\Lambda)=\frac{\left(\sqrt{\frac{2}{n_1}}e_1\ell_1,\sqrt{\frac{2}{n_2-n_1}}e_2\ell_2\right)}{e_1\ell_1+e_2\ell_2}\Big|_{r=\Lambda^{-1}}\, ,
\label{zvecLam}
\eeq
with $\ell_i(r)=\frac12 (n_i-n_{i-1})(s_0^i-\frac{e_i}{2\pi}\log\frac{r}{r_0})^{-1}$.  These charge-to-tension ratios evaluated for a fixed $\Lambda$  have been plotted as dots in figure \ref{Fig:QT_BPS}. They lie on the straight line and approach the UV attractor point $\gamma_{\rm min}$ when $\Lambda\rightarrow \infty$. Following a similar reasoning to the one used in \cite{Gendler:2020dfp}, one can see that whenever the field metric can be diagonalised, $\gamma_{\rm min}$ is given by
\beq
\gamma^{-2}_{\rm min}=\sum_i |z_i|_{e_i}^{-2}= \sum_i \frac{n_i-n_{i-1}}{2}=\frac{n_2}{2}\, .
\label{gaminex}
\eeq
This result will be reproduced below from a different perspective.

\vspace{-2em}

Notice that the charge-to-tension ratio of an individual string depends on $\Lambda$. However, when plotting it for all weakly coupled strings of charges $(e^1,e^2)$, they altogether draw a straight line that is independent of the scale $\Lambda$.
In other words, the charge to mass ratio $\cQ(\Lambda)/\cT(\Lambda)$, although field dependent, always takes some value lying on the line parametrised by the extremality bound, and this value shifts along the line as moving along the flow. In the limit $\Lambda\rightarrow \infty$, all points accumulate at
\beq
\vec{z}(\Lambda)\rightarrow \sqrt{2}\frac{(\sqrt{n_1},\sqrt{n_2-n_1})}{n_2}\, ,
\eeq
with $|z(\Lambda\rightarrow \infty)|\rightarrow \gamma_{\rm min}$. As we will further discuss below, this can be easily generalised to any number of charges. Therefore, all EFT strings saturate the WGC bound along their entire RG flow.

\noindent
\textbf{The minimal charge-to-mass ratio}

The above example features a K\"ahler potential of the form $K = - \log P(s)$, with $P(s)$ a monomial on the saxions, which is ubiquitous for non-degenerate string flows in string theory compactifications \cite{Lanza:2021qsu}. Nevertheless, one may highlight the key ingredients of this construction without specifying what the K\"ahler potential is. This turns out to be useful in order to describe geometrically the UV attractor point $\gamma_{\rm min}$ appearing in the example above. As we will see, this minimal charge-to-tension ratio will play a key role in the next subsection, when relating the Swampland Distance Conjecture to the WGC for strings. 

To develop this geometric description, let us define the rescaled charges
\be
{\bf w}\equiv \frac{M^2_{\rm P}\,{\bf e}}{\calt_{\bf e}}\, .
\ee
From the discussion of section \ref{sec:infinite} it follows that this vector {\bf w} belongs to the saxionic cone $\Delta$ if  ${\bf e}\in\cC^{\text{\tiny EFT}}_{\rm S}$. For more general mutually BPS strings, such that ${\bf e}\in \calc_{\rm S}$, we just  have  that ${\bf w}\in\calp^\vee$, where $\calp^\vee$ is dual to the cone $\calp$ spanned by the dual saxions $\ell_i$.  

We can immediately see that the ${\bf w}$-vectors of all mutually BPS strings lie in the hyperplane `orthogonal' to the vector ${\bm\ell}\equiv (\ell_1,\ell_2, \ldots)$ which is at unit distance from the origin, that is, the hyperplane defined by  
\be
\calh_{\bm\ell}\equiv \{{\bf w}\ |\ \langle {\bf w}, {\bm\ell}\rangle = 1\}\, ,
\ee
simply because $\calt_{\bf e}=M^2_{\rm P} \langle {\bf e},{\bm\ell}\rangle$. Hence,  the EFT strings correspond to points  in polytope $ \calh_{\bm\ell}\cap \Delta$ while the BPS (but not necessarily EFT) strings correspond to points in the larger  polytope $\calh_{\bm\ell}\cap \calp^\vee$. Notice that, as we change the saxionic vector ${\bm\ell}$, the hyperplane $\calh_{\bm\ell}$ rotates.

We can also  split ${\bf w}$ into two components, parallel and perpendicular to the hyperplane $\calh_{\bm\ell}$, respectively, with respect to the metric  $\calg_{ij}$: ${\bf w}={\bf w}_\|+{\bf w}_\perp$,  with $\calg_{ij}{\rm w}^i_\|{\rm w}^j_\perp=0$ and 
\be
\begin{aligned}
{\rm w}^i_\perp&\equiv\frac{\calg^{ij}\ell_j}{\|\ell\|^2}\quad~~~ \Rightarrow~~~~\quad \langle {\bf w}_\perp,{\bm\ell}\rangle=1\, ,\quad\quad \langle {\bf w}_\|,{\bm\ell}\rangle=0\, .
\\
\end{aligned}
\ee 
Then, we immediately find that 
\be
\frac{M^2_{\rm P}\calq^2_{\bf e}}{\calt_{\bf e}^2}=\|{\bf w}\|^2\equiv \calg_{ij}{\rm w}^i{\rm w}^j \geq \|{\bf w}_\perp\|^2= \frac{1}{\|{\bm\ell}\|^2}\quad~~~~\text{with}\quad \|{\bm\ell}\|^2\equiv \calg^{ij}\ell_i\ell_j\, .
\label{w}
\ee
Hence, the minimal possible value of the charge-to-tension ratio of the strings is given by
\be
\label{gmin}
\gamma_{\rm min}\equiv \frac{1}{\ \ \|{\bf\ell}\|_{\rm max}}\, , 
\ee
where $\|\ell\|_{\rm max}$ is the maximal dual saxionic norm for any string flow corresponding to $\cC^{\text{\tiny EFT}}_{\rm S}$. The same result  can be obtained by adapting the results of \cite{Gendler:2020dfp} to BPS strings.\footnote{
In \cite{Gendler:2020dfp} it was proven that the charge to mass ratio of BPS states always form a degenerate ellipsoid with up to two non-degenerate directions. It was also given a recipe to compute the principal radii of the ellipse in general, one of them corresponding to the minimal possible value of the charge to mass ratio of light BPS states. To borrow the results of \cite{Gendler:2020dfp} one just needs to replace the central charge by $M_{\rm P}^2 e_i \ell^i$. Since this quantity is real, there is only one non-degenerate direction (that is why the hyperplane) and the ``principal radius'' (the minimal value of $\gamma$) becomes
$\gamma_{\rm min}^{-2}=\calg^{ij}\ell_i \ell_j\, $
which indeed coincides with the result obtained in \eqref{w} and reduces to \eqref{gaminex} if the field metric can be diagonalised.}

One may connect this construction with the description of the charge/tension ratio in terms of the vector $\vec z$ defined above, by identifying $\vec z=(z^1,z^2,...)$ with the components of ${\bf w}=z^A{\bf v}_A$ in an orthonormal basis ${\bf v}_A$. Analogously we may expand ${\bm\ell}$ in the dual orthonormal basis, ${\bm\ell}=l_A{\bm\eta}^A$, such that $\calg_{ij}=\delta_{AB}\eta^A_i\eta^B_i$. For each purpose some choice of orthonormal basis may be more useful then others. Suppose for instance that ${\bm\ell}$ has constant norm $\|{\bm\ell}\|$, as in the example  with $K=-\log [(s^1)^{n_1}(s^2)^{n_2-n_1}]$ discussed above. Then, one can choose a basis ${\bf v}_A$ in which the components $l_{A}$ of ${\bm\ell}$ are constant. In such a basis,  the corresponding components of $\vec{z}$ move along the fixed and moduli independent hyperplane $l_A z^A =1$, representing $\calh_{\bm\ell}$. This is indeed  what happens in the above example, in which we have chosen
\be
{\bf v}_1=\left(\sqrt{\frac{2}{n_1}} s^1, 0\right) \quad \text{and} \quad   {\bf v}_2=\left(0,\sqrt{\frac{2}{n_2-n_1}}s^2\right)\, .
\ee
Since $\ell_1=\frac{n_1}{2 s^1}$ and $\ell_2=\frac{n_2-n_1}{2 s^2}$, then we clearly see that $l_1=\sqrt{\frac{n_1}{2}}$ and $l_2=\sqrt{\frac{n_2-n_1}{2}}$ are constant. This is why $\vec z$ moves along a fixed and moduli-independent hyperplane.

Notice that the lower bound $\|\ell\|^{-2}$ is a constant if and only the K\"ahler potential is no-scale in the saxionic directions. Indeed, we have that $K^{i\bar \jmath}K_iK_{\bar\jmath}= 2\calg^{ij}\ell_i\ell_j$, so the no-scale relation $K^{i\bar\jmath}K_iK_{\bar\jmath}=n$ is equivalent to $\|{\bm\ell}\|^2=\frac{n}{2}$. In string models,  the asymptotic K\"ahler potentials  typically satisfy this no-scale property for some $n$, and so we expect $\gamma_{\rm min} \simeq \gamma_{\text{no-scale}}=\sqrt{2/n}$.  

The  bound \eqref{w} is saturated in the limit in which the normalised charge vector ${\bf w}$ coincides with the shortest possible vector ${\bf w}_\perp$, that is, if and only if
$\|{\bf w}\|^2=\|{\bm\ell}\|^{-2}$. This obviously happens whenever there is only one relevant saxion, simply because $\calh_{\bm\ell}$ becomes zero-dimensional.
More generically, $\|{\bf w}\|^2=\|{\bm\ell}\|^{-2}$ is satisfied in the asymptotic UV limit of EFT strings,
in which  all saxions in \eqref{flowWGC} asymptotically flow homogeneously as $s^i\simeq \sigma e^i[1+\calo(\sigma^{-1})]$. To see this, let us for instance consider a K\"ahler potential whose leading contribution is of the form $-\log P(s)$ with $P(s)$ a homogeneous function in all the saxions involved in the asymptotic string flow: $P(\lambda s)=\lambda^{n_{\bf e}} P(s)$. In this limit the K\"ahler potential is no-scale and then, as noted above,
\be
\|\ell\|^2=\frac12 n_{I}\, .
\ee
On the other hand, in this limit we also have that,  asymptotically
\be
\begin{aligned}
\calt_{\bf e}(\sigma)&=M^2_{\rm P}\langle {\bf e},{\bm\ell}(\sigma)\rangle=
\frac{M^2_{\rm P}}{2\sigma}\frac{s^i(\sigma)P_i(s(\sigma))}{2P(s(\sigma))}\left[1+\calo\left(\sigma^{-1}\right)\right]=\frac{n_{\bf e}M^2_{\rm P}}{2\sigma}\left[1+\calo\left(\sigma^{-1}\right)\right]\, .
\end{aligned}
\ee
Hence, from \eqref{QTstring} it follows that
\be\label{extTQ}
\frac{M^2_{\rm P}\calq^2_{\bf e}}{\calt^2_{\bf e}}\Big|_{\sigma\rightarrow \infty}= \frac{2}{n_{I}} =\frac{1}{\|{\bm\ell}\|^2}\, ,
\ee
which in particular reproduces \eqref{gaminex} when applied to \eqref{Kex}. This shows that, if we start from any EFT string defined at a cut-off scale $\Lambda$, the corresponding vector ${\bf w}(\Lambda)$  identifies a point in the polytope $\Delta\cap\calh_{{\bm\ell}(\Lambda)}$. If we take the limit $\Lambda\rightarrow\infty$, any such ${\bf w}(\Lambda)$ moves along this polytope and converges towards the shortest  vector ${\bf w}_\perp$.

On the other hand, if we  consider a BPS charge ${\bf e}$ which does {\em not} belong to $\cC^{\text{\tiny EFT}}_{\rm S}$, then the corresponding flow of ${\bm\ell}$ as one approaches the string enters the strong coupling regime and crosses the boundary of $\calp$  at a finite  $\sigma$ parameter and field distance. This means that the above geometrical description breaks down.\footnote{At this point, the string tension can stay finite or even vanish, obtaining a tensionless string at finite distance in field space. Since instanton corrections become important when this occurs, it is difficult to determine the fate of the charge to mass ratio of these strings. But they seem to be examples of non-extremal strings. A more detailed study of these strings will appear in \cite{Lanza:2021qsu}.} Hence, only EFT strings will lie over the polytope $\Delta\cap\calh_{{\bm\ell}(\Lambda)}$ along the entire RG flow and reach the UV attractor point \eqref{extTQ}, which suggests that only such EFT weakly coupled strings saturate the WGC bound.

\subsubsection{WGC for membranes}
\label{sec:WGCmem}

From \eqref{No-fo_Idm} it follows that the charge-to-tension ration of the EFT membranes is given by 
\beq
\label{QTmnoforce}
\frac{M_{\rm P}^2 \cQ^2_{\bf q}}{\cT_{\bf q}^2}=\left(\frac{\| \partial \cT_{\bf q}\|^2}{\cT_{\bf  q}^2}-\frac32\right) \, ,
\eeq
with $\cT_{\bf q}$ defined in \eqref{memT}.
Here, we have already restricted to those membranes which can be described in the EFT, i.e. with charges $q_a \in \Gamma_{\rm EFT}$. As happens with the strings, not any membrane satisfying the no-force condition \eqref{QTmnoforce} will automatically satisfy the WGC \eqref{WGC}. 

Extremal membranes are solutions of the form \eqref{DWmetr} that might potentially correspond to flat domain walls separating different flux vacua if the scalars happen to reach a minimum of the potential when flowing to the IR. Here, we can only study the classical backreaction of the fields near the brane and require that the scalar flow  \eqref{genflowa}-\eqref{genflowb}  is well behaved in the UV, but one should keep in mind that this is only a necessary but not sufficient condition. For an extremal domain wall solution to exist one would need to follow the flow away from the membrane and check that it does not break down before reaching a minimum for the scalar potential.\footnote{It is worthwhile to mention that, in principle, it might be possible to get extremal solutions which are not BPS if the potential \eqref{genmemV}  can be similarly written in terms of a fake superpotential such that there is a new $\tilde{\mathcal{Z}}= e^{\tilde K/2}\tilde W$ entering in \eqref{genflowa}-\eqref{genflowb} where $\tilde W$ is not the $\cN=1$ but a \emph{fake} superpotential. However, this would imply to consider membranes whose EFT tension is not given by $\cT=2e^{K/2}|f_a\Pi^a(t)|$ and, therefore, do not satisfy \eqref{No-fo_Idm}. Here we will neglect this possibility and consider only BPS extremal domain walls.}

Consider an elementary saxionic membrane charged under a single 3-form gauge field, with charge ${\bf q}$. As explained in \eqref{Tsigma}, these elementary membranes satisfy that
\beq
\partial_{i} \cT_{\bf q}=K_i\sigma_i \cT_{\bf q}\, ,
\eeq
with $\sigma_i$ some constant parameter which can be determined in the perturbative regime to be \eqref{sigma}. Let us first consider the case of a single saxionic field $s$ so that the K\"ahler potential takes the form \eqref{K1} in the perturbative expansion. By replacing this into \eqref{QTmnoforce}, we get 
\beq
\label{Msaxionic}
M_{\rm P}^2 \cQ^2_{\bf q}= \left(2\sigma^2 n -\frac32\right) \cT^2_{\bf q}\, ,
\eeq
This equation resembles saturating a WGC bound, but first one needs to check if there indeed exists an extremal solution to the RG flow equations with this charge and this value of $\gamma$. 

The RG flow of the scalars near the membrane was computed in the previous section and it is given by \eqref{lflow2}. 
Recall that EFT membranes have $\sigma>0$, implying that the saxion $s$ decreases when moving away from the membrane, leaving eventually the perturbative regime. Since these membranes are charged under a single 3-form gauge field whose gauge kinetic function behaves exponentially in terms of the canonically normalised saxion in the perturbative regime, we can apply the dilatonic formula \eqref{dilatonic} to compute the extremality factor, 
\beq
\label{Mext}
 \gamma^2= \left(\frac{|\alpha|^2}{4}-\frac32\right)\ .
\eeq
The dilatonic coupling $\alpha$ can be read from the scalar dependence of the  3-form  gauge kinetic function $T_{qq}$ corresponding to the charge {\bf q}. Using \eqref{growth}, we get that it goes as 
\beq
T_{qq}\sim s^{{r}}\, ,
\eeq
implying 
\beq
\alpha= \frac{\| \partial T_{qq}\|}{T_{qq}}=\sqrt{\frac{2}{n}}r=2\sqrt{2n}\,\sigma \, ,
\eeq
where we have used the field metric \eqref{K1} and replaced \eqref{sigma} in the last step. Hence, the extremality factor is given by
\beq
\gamma^2=2n \sigma^2-\frac32\, ,
\eeq
which precisely matches with the numerical factor in \eqref{Msaxionic}. Therefore, the EFT elementary membranes are extremal,  saturating the WGC bound in the UV regime. 

If the field metric is diagonal, as occurs to leading order in the perturbative expansion when following some non-degenerate string flow, we can still apply the dilatonic formula for multiple scalars and read the extremality factor from the scalar dependence of the gauge kinetic function. From the general structure of the three-form kinetic matrix \eqref{TAB}, upon using the asymptotic expressions for the charge \eqref{Qasymp} in the appendix  and \eqref{Ki} for the K\"ahler potential, one obtains 
\beq
\label{memalpha}
|\alpha|^2= \frac{\| \partial T_{qq}\|^2}{T_{qq}^2}=r_iK^{ij}r_j=8\sigma_i^2 (n_i-n_{i-1})\, ,
\eeq
with $n_0\equiv 0$, which in turn implies
\beq
\label{memgamma}
\gamma^2=\sum_{i=1}^I 2(n_i-n_{i-1}) \sigma_i^2-\frac32\, .
\eeq
On the other hand, the charge to mass ratio  of an elementary membrane \eqref{Msaxionic} generalised to multiple scalars with a diagonal field metric reads 
\beq
\label{Msaxionic2}
\frac{M_{\rm P}^2 \cQ^2_{\bf q}}{\cT^2_{\bf q}}=\left( 2 \sigma_iK_i K^{i\bar\jmath}K_{\bar\jmath} \sigma_j -\frac32\right)= \left(\sum_{i=1}^I 2(n_i-n_{i-1}) \sigma_i^2 -\frac32\right) \, ,
\eeq
so they still saturate the WGC. In the particular case in which all $\sigma=\sigma_i$ are equal, this reduces to $\gamma^2=2n_I\sigma^2-3/2$, with $n_I$ coinciding with the no-scale factor/homogeneity degree of $e^{-K}$.

In order to check whether membranes charged under several 3-form gauge fields saturate the WGC one needs first to determine the extremal region in the charge-to-tension ratio space, as explained around \eqref{CH}. The simple dilatonic formula \eqref{Mext} is not valid anymore, but we expect the extremality bound to form an ellipsoid in the charge-to-tension ratio plane. This is simply due to the fact that the tension is still linear in the charges but the membranes are generically non-mutually BPS, i.e. $\cT(q_1+q_2)\neq \cT(q_1) + \cT(q_2)$. This is analogous to the case of  particles and different from the case of strings of the previous section, where the extremality region was a polytope contained in a hyperplane. In fact, one can borrow the results of  \cite{Gendler:2020dfp} to show that there are only two non-degenerate directions and determine the principal radii of the ellipse in terms of the periods and the K\"ahler potential. For a diagonal gauge kinetic function, they can simply be computed in terms of the values of the elementary saxionic membranes, and there is again (as for the strings) a minimum value of the extremal factor $\gamma_{\rm min}$ corresponding to the minimal principal radius. The diagonal kinetic function is well justified in the asymptotic regimes of the moduli space, where the flux lattice splits into the nearly orthogonal subspaces $\Gamma_r$ as explained in section \ref{sec:EFTlattice}.
Since the computation is analogous to the case of particles\footnote{The main difference between this RG flow for the membranes, and the flow associated to black holes (particles) is that in the latter a minimum of the black hole potential $\partial \cQ=0$ can only occur at $\partial \mathcal{Z}=0$. However, for membranes, due to the minus sign in \eqref{QTmnoforce}, this minimum can also occur at $\partial \mathcal{Z}\neq 0$ representing a non-supersymmetric vacuum. We will comment a bit more on this and a possible correlation to superextremality of the domain walls in section \ref{sec:AdSremarks}.} in \cite{Gendler:2020dfp}, here we will only present an example to illustrate the key features.

But before discussing the example, a comment is in order. The scalar dependence that corrects the extremality bound by generating $\alpha\neq 0$ is in principle associated only to massless scalar fields. However, the scalars here are getting massive by the presence of fluxes or, equivalently, by the coupling to the 3-form gauge fields which induces a scalar potential. How is it possible then that they still contribute to the extremality bound in \eqref{Mext}? The resolution comes from noticing that the tension of the membranes is evaluated at the UV cut-off scale $\Lambda$ so the scalars are actually lighter ($m\ll \Lambda$) and therefore ``effectively'' massless. This also implies that, as we flow to the IR and hit some scalar mass threshold, the corresponding scalar should stop contributing to $\alpha$ and therefore lower the value of $\gamma$. Hence, we expect the extremality bound to be constant and independent of $\Lambda$ only while not crossing any mass threshold. The situation is analogous to the discussion about the RFC in section \ref{sec:RFC}, and it has some interesting consequences. If, by flowing to the IR, we reach some minimum of the potential at which all scalars are stabilised and integrated out but the membranes still belong to the spectrum of possible EFT operators, they should satisfy
\beq
M_{\rm P}^2\cQ_{{\bf q},{\rm IR}}^2= -\frac32 \cT_{{\bf q},{\rm IR}}^2\, ,
\eeq
in which $\alpha=0$.
Since $\cQ^2_{\bf q}=2V$, this is consistent with the well known fact that supersymmetric vacua with $W\neq 0$ only occur in AdS space; with negative vacuum energy. However, it gives some new non-trivial information: it is not possible to stabilise all moduli to get $\alpha=0$ at a de Sitter minimum (with positive vacuum energy $\cQ^2>0$). Of course, this result relies on assuming that there are no additional contributions to the scalar potential apart from those associated to the dynamical fluxes, which is not generically the case. As explained in Section~\ref{sec:EFTlattice}, the full scalar potential also involves the contribution from non-dynamical fluxes as well as non-perturbative corrections. But still, it is quite surprising that, according to this interpretation of the WGC applied to membranes, any EFT generating a positive contribution to the scalar potential coming from dynamical fluxes would be inconsistent with quantum gravity.

\subsection*{An example}

In order to exemplify the discussion, let us consider a simple case with two complex moduli, $t^1 = a^1 + \ii s$ and $t^2 = a^2 + \ii u$, described by the K\"ahler potential
\be
\label{memex_K}
K=-\log (s^3 u^4)\,,
\ee
which implies that $n_1 = 3$, $n_2 = 7$.
We include a superpotential which depends only on $t^1$ with effective nilpotency order (c.f. \eqref{Nd}) $d_1=3$ (so it has up to a cubic term). In the perturbative regime $s,u\rightarrow \infty$, the flux space splits into the subspaces $\Gamma_{\bf r}$ in \eqref{split-Vell} with $\bf{r}=(r_1,r_2)$. The tension of a membranes with charge $q_{\bf{r}}\in \Gamma_{\bf r}$ is given in \eqref{Tgeneral} as
\begin{equation}
\label{memex_Zq}
\mathcal{T}_{{\bf q_r}} =  2 \frac{q_r}{\left(\frac{r_1+3}{2}\right)!} s^{\frac{r_1}2} u^{\frac{r_2- r_1}2} \,,
\end{equation}
with $r_i = 2k_i -(n_i-n_{i-1}) $ and $k_i$ an integer. Since we did not turn on fluxes involving $t^2$ in the superpotential, we effectively set $k_2=0$ so that $r_2-r_1=-4$; while $r_1=2k_1-3$ with $k_1=(0,\dots,d_1)$ with $d_1=3$. Thus, the possible values for $\bf{r}$ are $(r_1,r_2)=\{(-3,-4),(-1,-4),(1,-4),(3,-4)\}$, yielding four types of elementary saxionic membranes  obeying \eqref{Tsigma} with
\beq
\sigma_1=-\frac{r_1}6\ ,\quad \sigma_2=\frac12 \ .
\eeq
The corresponding physical charges satisfy \eqref{Msaxionic2},
\be
\label{memex_Q2}
M_{\rm P}^2 \mathcal{Q}^2_{{\bf r}}= \frac16\left( r_1^2 + 3\right)\mathcal{T}^{\, 2}_{{\bf q}_{\bf r}} = \gamma^2_{\bf r} \mathcal{T}^{\, 2}_{{\bf q}_{\bf r}}\,.
\ee
Let us now consider composite membranes, whose charge is not the sole elementary ${\bf q}_r$ as above but some combinations thereof. For these membranes the central charge is generically given by
\be
\label{memex_Zqt}
\mathcal{Z}_{{\bf q}} = \sum\limits_{{\bf r}} \mathcal{Z}_{{\bf q}_{\bf r}} = \sum\limits_{r_1}  \frac{q_{r_1}}{\left(\frac{r_1+3}{2}\right)!} s^{-\frac{n_1}2} u^{-\frac{n_2-n_1}{2}} (t^1)^{\frac{n_1+r_1}{2}}\,.
\ee
It is important to notice that all the elementary central charges $\mathcal{Z}_{{\bf q}_{\bf r}}$ contributing to \eqref{memex_Zq}, assuming null axions, are actually misaligned, with their phases given by $\theta_{\bf r} = \frac{\pi}{4}(r_1+3)$.

\begin{center}
	\begin{figure}[ht]
		\centering
		\includegraphics[width=9cm]{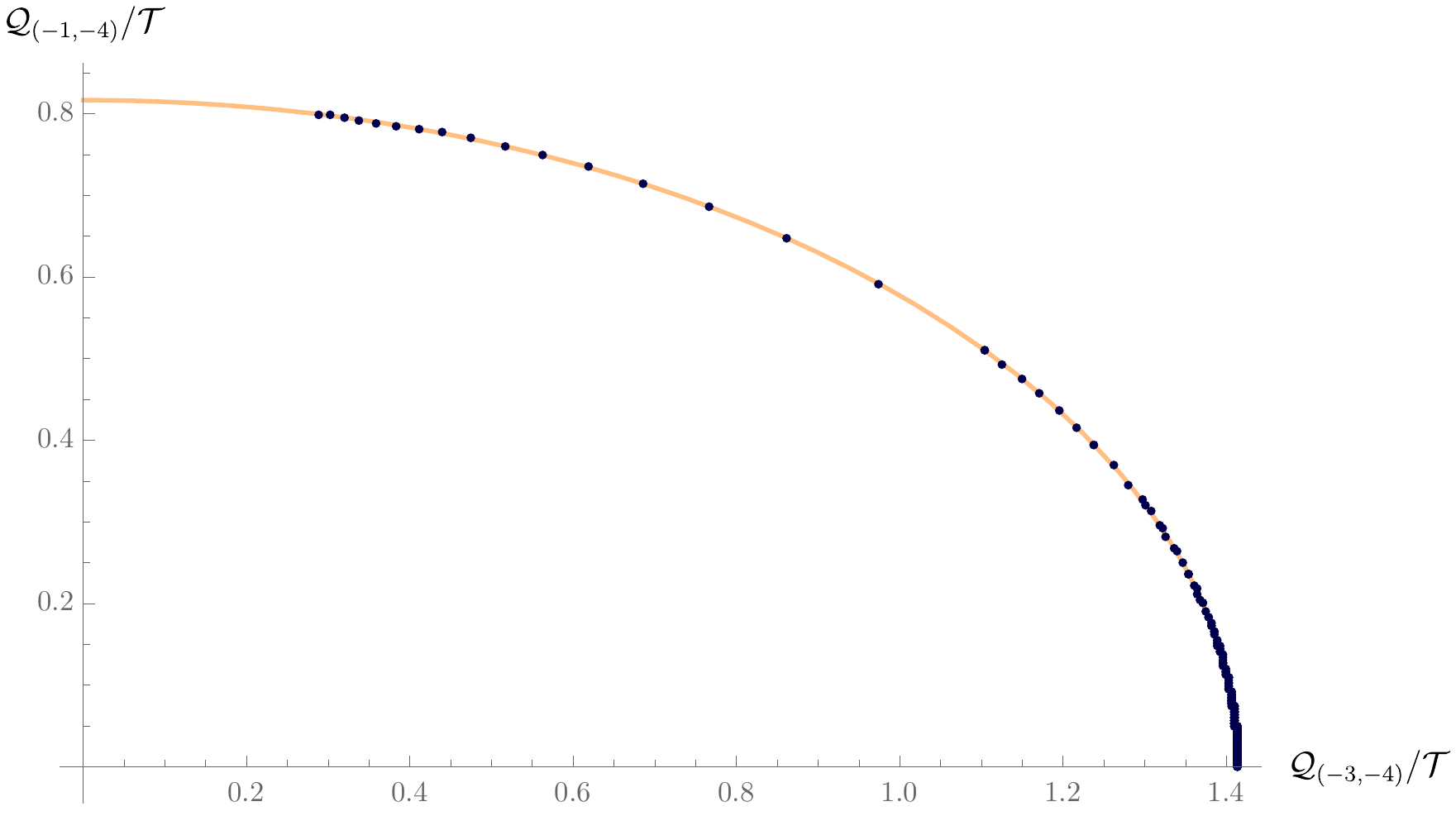}
		\caption{The charge-to-tension ratios of two non mutually BPS membranes. \label{Fig:QT_noBPSmem}}
	\end{figure}
\end{center}

This translates in the fact that two elementary membranes generically cannot simultaneously obey the supersymmetry preserving condition \eqref{DWproj}.\footnote{Notice that in \eqref{DWproj} $\theta$ is point dependent. Hence, generically, if two membranes are located at different points, sufficiently far apart, \eqref{DWproj}  may be realised for both of them. However, the EFT description is expected to break at long distances.} Thus, the tension of such a composite membrane is \emph{not} the sum of the tension of the single elementary membranes \eqref{memex_Zq}, but rather
\be
\mathcal{T}^2_{{\bf q}} = (\mathcal{T}_{{\bf q}_{(-3,-4)}} - \mathcal{T}_{{\bf q}_{(-1,-4)}})^2 +  (\mathcal{T}_{{\bf q}_{(1,-4)}} - \mathcal{T}_{{\bf q}_{(3,-4)}})^2\,.
\ee
On the other hand, the physical charge is given by
\be
\label{memex_Q2t}
 \mathcal{Q}^2 =  \sum\limits_{{\bf r}} \mathcal{Q}^2_{{\bf r}}\, ,
\ee
with $\mathcal{Q}^2_{{\bf r}}$ defined in \eqref{memex_Q2}. In such a case, it is less immediate to define the extremality factor.
As an example, consider a membrane with charges $q_0$, $q_1$, associated to the lightest elementary membranes, with tensions $\mathcal{T}_{{\bf q}_{(-3,-4)}}$ and $\mathcal{T}_{{\bf q}_{(-1,-4)}}$. We can introduce the charge-to-tension vectors 
\be
\label{memex_vQT}
\vec{z} = \frac{M_{\rm P}}{\mathcal{T}_{{\bf q}}} (Q_{(-3,-4)}, Q_{(-1,-4)})\,.
\ee
It can be seen that $\vec{z}$ draws an ellipsis, whose axes are specified by the extremality factors of the elementary membranes as follows
\begin{equation}
	\frac{1}{\gamma_0^2} \left(\frac{M_{\rm P} Q_{(-3,-4)}}{\mathcal{T}_{{\bf q}}}\right)^2 + \frac{1}{\gamma_1^2} \left(\frac{M_{\rm P} Q_{(-1,-4)}}{\mathcal{T}_{{\bf q}}}\right)^2 = 1\,,
\end{equation}
as depicted in Fig.~\ref{Fig:QT_noBPSmem}, and it is invariant under the saxionic flow. According to \eqref{memex_Q2}, $\gamma_{(-3,-4)}=\sqrt{2}$ and $\gamma_{(-1,-4)}=\sqrt{2/3}$, consistent with the figure.

\subsection{Distance Conjecture}
\label{sec:SDC}

The Distance Conjecture states that, at any infinite distance limit in field space, there is an infinite tower of states that becomes exponentially light in terms of the proper field distance as follows
\beq
m\sim m_0 \exp(-\lambda \Delta \phi)\, ,
\eeq
where $\lambda$ is some unspecified order one parameter and $\Delta\phi$ the geodesic field distance. A lot of work has been dedicated recently to check the conjecture in string theory and identify the tower of states that becomes light at the different asymptotic limits of supersymmetric moduli spaces \cite{Grimm:2018ohb,Grimm:2018cpv,Corvilain:2018lgw,Lee:2018urn,Lee:2019xtm,Lee:2019wij,Font:2019cxq,Baume:2019sry,Gendler:2020dfp,Joshi:2019nzi,Grimm:2019wtx,Marchesano:2019ifh,Xu:2020nlh}.

Since the vev of the moduli typically parametrise the couplings and masses of the EFT description, moving in the moduli space usually corresponds to varying the parameters of the EFT and exploring then the physics of different EFT's. However, it is also possible to explore large field variations within the same EFT, whenever there is some localised object which induces a non-trivial profile of the scalars over the non-compact space. Examples of the latter include bubbles of nothing, certain black holes and axionic strings, as studied in \cite{Klaewer:2016kiy,Draper:2019utz} (see also \cite{Draper:2019zbb,Buratti:2018xjt,Geng:2019bnn,Geng:2019zsx,Bonnefoy:2019nzv,Gendler:2020dfp}). This was dubbed as a Local Distance Conjecture in \cite{Draper:2019utz}. Interestingly, it seems that the EFT also breaks down when trying to engineer large field variations in this way, although it is not always associated to an infinite tower of states becoming exponentially light.

In this paper and in \cite{Lanza:2021qsu}, we are able to relate the usual Distance Conjecture with its local version by means of axionic strings.
As we have seen, the backreaction of the string creates a non-trivial profile for the saxions which can be interpreted as an RG flow of the brane couplings. As the cut-off $\Lambda$ increases so does the resolution of the profile, and tends towards \eqref{tsol}, where approaching the core of the string involves a trajectory $s^i\rightarrow \infty$ of infinite distance in field space. In other words, if we were able to trust the solution \eqref{tsol} with point-like precision within the EFT by sending $\Lambda \to \infty$ we would be probing infinite distances in field space!  However, this classical backreaction also implies that the tension of the string $\cT(\Lambda)$ decreases as $\Lambda$ increases, and the EFT breaks down when this tension becomes smaller than the cut-off $\Lambda$. This will occur at a scale $\Lambda_{\rm max}^2=\cT(\Lambda_{\rm max})$, which corresponds then to the maximum cut-off of the EFT consistent with the presence of the string. Since this profile also describes the physics of unperturbed vacua  when moving in the moduli space $\mathcal{M}$, $\Lambda_{\rm max}$ will act as a quantum gravity cut-off associated to an infinite tower of string modes becoming light. As we will show next, if the string satisfies the WGC, then this cut-off $\Lambda_{\rm max}$ decreases exponentially in terms of the proper field distance, as the SDC predicts.

Consider approaching the string from a reference scale $r_0$ to a distance from the core given by $r_{\rm max}=\Lambda^{-1}_{\rm max}$. This induces a saxionic flow from $s^i_0 = s^i(r_0)$ to $s^i(r_{\rm max})$ given by $s^i(r)=s^i_0+e^i\sigma(r)$ with $\sigma(r)=\frac{1}{2\pi}\log(r_0/r)$. The proper field distance up to $r_{\rm max}$, using \eqref{d*} and \eqref{monoT}, reads 
\be
\mathrm{d}_{\rm max}=\frac{1}{M_{\rm P}}\int_0^{\sigma_{\rm max}} \calq_{\bf e}\d\sigma=-\frac{1}{M_{\rm P}}\int_0^{\sigma_{\rm max}} \frac1{\calq_{\bf e}}\frac{\d\calt_{\bf e}}{\d\sigma}\d\sigma= \frac{1}{M_{\rm P}}\int^{\calt^0_{\bf e}}_{\calt^{\rm max}_{\bf e}} \frac1{\calq_{\bf e}}\d\calt_{\bf e}\, ,
\ee
where we have taken into account the fact that $\calt_{\bf e}^{\rm max}\equiv \calt_{\bf e}(\sigma_{\rm max})< \calt_{\bf e}^0\equiv \calt_{\bf e}(\sigma=0)$ since,  by \eqref{monoT},  $\calt_{\bf e}(\sigma)$ is always  decreasing along the  $\sigma$ flow. Then, by imposing the WGC bound $M_{\rm P}\calq_{\bf e}\geq \gamma\calt_{\bf e}$ with constant $\gamma$ we obtain
\be
\label{dmax}
\mathrm{d}_{\rm max}\leq  \frac{1}{\gamma}\int^{\calt^0_{\bf e}}_{\calt_{\bf e}^{\rm max}} \frac1{\calt_{\bf e}} \d\calt_{\bf e} = \frac1{\gamma} \log\frac{\calt^0_{\bf e}}{\calt_{\bf e}^{\rm max}}\, ,
\ee
and then
\beq
\label{Lmax}
\Lambda_{\rm max}^2=\cT_{\bf e}(\Lambda_{\rm max})< \cT_{\bf e}^0 \exp \left(-\gamma\  \mathrm{d}_{\rm max}\right)\, .
\eeq
Notice that this argument follows from the  WGC bound $M_{\rm P}\calq_{\bf e}\geq \gamma\calt_{\bf e}$ without any additional condition  on the K\"ahler potential (apart from the approximate  axial symmetry).

Hence, we obtain that the maximum value of the EFT cut-off $\Lambda_{\rm max}$ decreases exponentially with the proper field distance. This is precisely the SDC along paths generated by axionic string flows, with the tower of states 
corresponding to semiclassical states and excitation modes of the weakly coupled string\footnote{Thanks to the weakly coupled nature of the string, it is possible to argue in the semiclassical approximation for the presence of a continuum of new degrees of freedom when  $\mathcal{T}_{\bf e}\lesssim \Lambda^2$ \cite{Lanza:2021qsu}. We expect that the spectra of fluctuations would get discretised and yield the infinite tower upon quantisation, as it happens for critical strings. }. Therefore, in this case the SDC becomes simply a consequence of requiring that the EFT strings satisfy the WGC bound. Notice that the SDC factor $\lambda$ parametrising the decreasing rate is now fixed by the extremality factor $\gamma$ of the string. Thus, not only we recover the SDC from the WGC, but we also get information about the unspecified parameter $\lambda$, obtaining
\beq
\lambda=\frac{\gamma}{2}
\label{lg}
\eeq
so there are no more free parameters left in the conjecture! 

The above reasoning shows that the SDC along infinite distance paths generated by axionic string flows is a consequence of the WGC for strings. We may now invoke the Distant Axionic String Conjecture (DASC) stated in section \ref{sec:infinite}, namely that there is a one-to-one correspondence between EFT weakly coupled strings in $\cN=1$ theories and infinite field distance limits. We then see that the DASC and the WGC for strings imply the SDC, with a string associated to each infinite distance limit and the exponential decreasing rate of its tension providing the SDC factor. 
Moreover, it is compatible with the Emergent String Conjecture proposed in \cite{Lee:2019wij}, although there are some important differences. First of all, only in four space-time dimensional theories, the DASC and Emergent String Conjecture will feature the same object, as the key ingredient for the DASC is an object of codimension two, while the Emergent String Conjecture cares about the dimension of the state, referring to strings in any space-time dimension. Furthermore, we propose the ubiquitous presence of a string in 4d EFTs at any infinite distance limit, although the string modes are not necessarily the leading tower of states signalling the EFT breakdown. Contrary, the proposal in   \cite{Lee:2019wij} only requires the presence of a string in equi-dimensional infinite distance limits, a case in which it should generate the leading tower of light states. According to \cite{Lee:2019wij}, if the string modes are not the lightest ones, the leading tower should correspond to Kaluza--Klein states, while the DASC does not restrict their nature. The presence of additional towers of light states is a very important question, whose discussion and interplay with the string tension we postpone to \cite{Lanza:2021qsu}. In any event, we hasten to stress that even if the string does not provide the leading tower,  \eqref{lg} can still be used to provide a lower bound for the SDC factor, as the leading tower will necessarily have a bigger value of $\lambda$.

\begin{center}
	\begin{figure}[h]
		\centering
		\includegraphics[width=8.5cm]{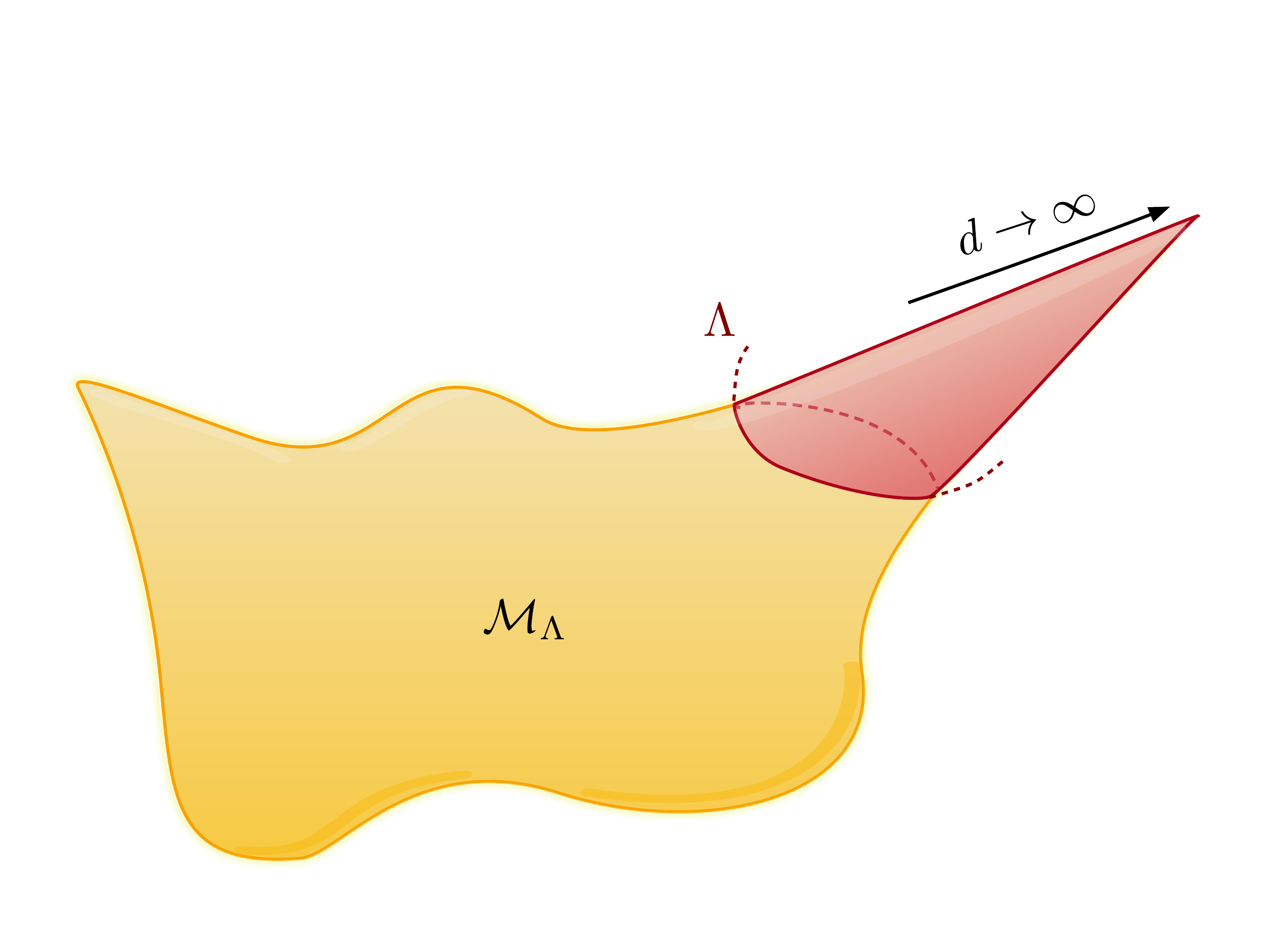}
		\caption{The moduli space $\mathcal{M}_\Lambda$ accessible within an EFT characterised by a cutoff $\Lambda$. \label{Fig:Asympstrmod}}   
	\end{figure}
\end{center}

Remarkably,  the correspondence between infinite field distance limits and string RG flows also provides an interesting realisation of the Emergence Proposal \cite{Harlow:2015lma,Heidenreich:2017sim,Grimm:2018ohb,Heidenreich:2018kpg}, for which the infinite field distance emerges from integrating out massive modes that are light with respect to the Planck mass. For a fixed value of the EFT cut-off $\Lambda$, the moduli space $\MM_\Lambda$  accessible by the EFT can be considered finite (see Fig. \ref{Fig:Asympstrmod}), since along the saxionic trajectories of infinite distance we will reach a point in which $\cT_{\bf e}(\Lambda) = \Lambda^2$ (denoted as $\Lambda_{\rm max}$ in \eqref{Lmax}) and the EFT will necessarily break down. As we lower $\Lambda$ and we integrate out 4d energetic modes, we will reach this point at a larger distance, and so $\MM_\Lambda$ will grow. In the limit $\Lambda \rightarrow 0$, $\MM_\Lambda$ will asymptote to the naive moduli space $\MM$, and points at infinite distance will emerge. Around these points, the string couplings and the kinetic terms for the saxion coupling to the string will be determined by the string RG flow towards the UV, which dictates the asymptotic behaviour of the system.

What can we say about the specific numerical value of $\gamma$? In the previous section, we argued that the EFT weakly coupled BPS strings are extremal, so that the WGC is saturated. 
The extremality factor is then completely determined in terms of the field metric or, in other words, in terms of the dilatonic factor of the gauge coupling of the 2-form gauge field. As derived in section~\ref{sec:WGCstr}, for elementary (non-degenerate) axionic strings of charge $e^i$ we have that
\beq
\label{L1}
\lambda= \frac{1}{\sqrt{2( n_i-n_{i-1})}}\, ,
\eeq
where $(n_i-n_{i-1})$ is the highest degree of $s^i$ entering in the K\"ahler potential (see \eqref{Ki}). 
As explained in the previous section, for strings charged under several gauge fields, $\gamma$ becomes field dependent as it is different for each direction in the charge-to-mass ratio $\vec{z}$-plane. However, the derivation \eqref{dmax} still applies if replacing $\gamma$ by the constant value $\gamma_{\rm min}$ defined in \eqref{gmin}, which corresponds to the minimal possible value of the extremality factor. We then get the following universal lower bound for the SDC factor,
\beq
\label{L2}
\lambda \geq \gamma_{\rm min}=\frac{1}{\sqrt{2n_I}}\, ,
\eeq
where recall that $n_I$ is the  no-scale/homogeneity degree factor of the K\"ahler potential. Deviations from no-scale could make $\gamma_{\rm min}$ non-constant, but we expect these deviations to become subleading when getting close enough to the string, as occurs when approaching asymptotic limits of the moduli space in string theory compactifications.
Notice that we could get a stronger bound if we allow that it is not the same string the one becoming light along the entire RG flow. For a given charge direction $\vec{z}$, there is always a constant $\gamma$, but fixing the charge direction $\vec{z}$ while varying the moduli implies that we also need to vary the quantised charges. Hence, there is a stronger bound corresponding to an intermediate value between \eqref{L1} and \eqref{L2} fixed by the extremal hyperplane, but it is realised by different strings at each point in moduli space. 

The relation between the SDC factor and black hole extremality has also been studied in \cite{Lee:2018spm,Lee:2019wij,Gendler:2020dfp}. Clearly, both quantities are related if the same tower that satisfies that WGC also satisfies the SDC, which seems to occur at any infinite field distance limit that has some  gauge coupling vanishing at infinite distance \cite{Gendler:2020dfp}. The particular relation \eqref{lg} can be derived in fact from imposing that the WGC coincides with the RFC, in the sense that extremal states feel no force among each other, as occurs for our weakly coupled strings. For the case of a tower of particles, the requirement of having a small gauge coupling is stronger than the original SDC, as there is a priori no reason to require the tower of states to be charged under some weakly coupled gauge field. However, as proposed in \cite{Gendler:2020dfp} this could indeed be a general feature if one allows the tower to be charged under some $p$-form gauge field (not necessarily with $p= 1$). The presence of a weakly coupled string becoming tensionless at every infinite field distance limit is indeed a confirmation of this expectation, as the string is charged under the 2-form gauge field dual to the axion, whose gauge coupling vanishes at infinite distance. Therefore, the SDC factor gets automatically fixed in terms of the extremality bound  associated to  strings. 
 Interestingly, the same numerical values \eqref{L1} and \eqref{L2} coincide with the bounds in \cite{Grimm:2018cpv,Gendler:2020dfp} for the case of towers of BPS particles in ${\cal N}=2$ Calabi-Yau string compactifications. This is because the bound only depends on the integers $n_i$ which characterise the type of singular limit in \cite{Grimm:2018cpv,Gendler:2020dfp}, regardless of whether we are considering particles in $\cN=2$ or strings in ${\cal N}=1$. The integers $n_i$ are upper bounded by the internal dimensions of the compactification, so that the maximum value is $n=3$ in a Calabi--Yau threefold, yielding $\lambda\geq 1/\sqrt{6}$ in the complex structure moduli space of Type IIB compactifications.

If we turn on fluxes, the string can become anomalous which forces it to be attached to membranes, as discussed in Section~\ref{sec:StrMem}. This also induces a scalar potential for the scalars, which might a priori obstruct to take the infinite distance limit and approach the string. However, the EFT membranes ending on the string induce an RG flow that makes the saxions grow as approaching the membrane, so there does not seem to be any obstruction to send $s^i\rightarrow \infty$. We will further discuss this point in section \ref{sec:dS}. This is encouraging, as it provides a realisation of the Distance Conjecture in a field space lifted by a scalar potential in $\cN=1$ theories and not just flat trajectories in a moduli space of extended supersymmetry. The tension of the string does not change in the presence of fluxes, so the lower bound of the SDC factor remains the same. Understanding the precise realisation of the Distance Conjecture in the presence of a scalar potential is crucial to apply the conjecture to constrain large field inflationary models, so we expect our analysis to be a starting point to address questions of phenomenological relevance.

One of the most exciting consequences of this analysis is that it might potentially provide a low energy explanation for the SDC. Even if the conjecture has been extensively tested in string compactifications, its underlying quantum gravity principle is still a mystery. But now we are translating the discussion on asymptotic limits in moduli spaces into the behaviour of the charge and tension of four dimensional strings that can be described in the low energy EFT. At the moment, it seems that the SDC holds at any asymptotic limit simply as a consequence of requiring the WGC to hold for codimension 2 objects, which is usually motivated by black hole physics. This uncovers another relation between the swampland conjectures, which are looking more and more as a web of interconnected conjectures instead of independent statements. It would be very interesting, though, to find the underlying quantum gravity explanation for the SDC just by thinking of the physical  properties of these strings, without mentioning the WGC, but we leave this exciting task for the future.

\subsection{de Sitter Conjecture\label{sec:dS}}

The advantage of rewriting the scalar potential in terms of the charge of membranes is that we can translate constraints on the scalar potential to constraints on the properties of the membranes. Vice versa, we can study the implications of constraints on the charge and tension of membranes (coming, for instance, from the WGC) to the behaviour of the scalar potential. From this perspective, the WGC and other conjectures about the behaviour of the potential (like the original de Sitter conjecture  \cite{Obied:2018sgi} or the AdS Distance Conjecture \cite{Lust:2019zwm}) are not so different and get interlinked by the properties of membranes. In this section we explore to what extent the WGC applied to membranes underlies the asymptotic version of the de Sitter conjecture, which amounts to the original de Sitter conjecture at asymptotic limits in field space \cite{Ooguri:2018wrx,Grimm:2019ixq}.\footnote{Interesting connections between the WGC applied to membranes and the AdS Distance Conjecture have been explored in \cite{Buratti:2020kda} where a discrete version of the WGC has been used to explain a potential scale separation in AdS.}

Consider some flux-induced scalar potential for chiral fields $t^i = a^i + \ii s^i$
\beq
\label{dS_V}
V=\frac12 T^{AB}(s,a) f_A f_B = \frac12 Z^{ab}(s) \rho_a(a) \rho_b(a) + f_a\Upsilon^a(s) + \hat V(s,a)\, ,
\eeq
where in the second equality we have explicitly separated the dynamical fluxes $f_a\in \Gamma_{\rm EFT}$ from the rest as in \eqref{potsplit} and introduced $\rho_a=e^{-N_i a^i} f_a$ as in \eqref{rho}. 

Let us first assume that only dynamical fluxes are turned on, so that the potential in \eqref{dS_V} reduces to
\beq
\label{dS_V0}
V_0\equiv  \frac12 Z^{ab}(s) \rho_a(a) \rho_b(a)\, .
\eeq

As explained in Section~\ref{sec:memscflow}, one may regard the potential $V_0$ as generated by a single membrane, whose physical charge is expressed in terms of the potential as in \eqref{genmemV}:
\beq
\label{dS_V0Q}
V_0=  \frac12 \cQ^2_{\bf f}\, ,
\eeq
where the quantised charges are given by the flux integers $q_a=f_a$. Such a generating membrane separates a fluxless configuration on one side, where trivially $V = 0$, from a flux generated potential \eqref{dS_V0} on the other side. Albeit this generating membrane satisfies the no-force condition \eqref{No-fo_Idm}, it does not necessarily satisfy the WGC. Therefore, imposing a WGC for the generating membrane provides constraints on the scalar potential, delivering interesting links with other Swampland Conjectures.

The original de Sitter conjecture \cite{Obied:2018sgi} constrains the slope of the potential to satisfy
\beq
\label{dS}
\|\del V_0\| \geq c\, V_0\,,
\eeq
where $\|\del V_0\| = (2 K^{i\bar\jmath}\partial_iV_0\partial_{\bar\jmath} V_0)^{1/2}$ and $c$ is, presumably, an order one factor. While this constraint is not expected to be valid at arbitrary points in field space (see \cite{Garg:2018reu,Ooguri:2018wrx} for refinements of the conjecture) it is still believed to apply in asymptotic limits \cite{Ooguri:2018wrx,Grimm:2019ixq}, which is the version of the conjecture that we will consider in the following. 

In light of the above discussion, for a generating membrane \eqref{dS} can be translated into a condition on its charge as
\beq
\label{triangle}
\|\del \cQ_{\bf f}^2\|\geq c\, \cQ_{\bf f}^2\, ,
\eeq
with $\|\del \cQ_{\bf f}^2\| = (2 K^{i\bar\jmath}\partial_i \cQ_{\bf f}^2\partial_{\bar\jmath} \cQ_{\bf f}^2)^{1/2}$.
Following \cite{Grimm:2019ixq}, using the Cauchy-Schwarz inequality one can show that
\be
\label{triangle2}
\|\del \cQ_{\bf f}^2\| \geq 2 \beta\, {\rm Re} (\kappa^i\partial_i \cQ_{\bf f}^2) \,, \ \text{ as long as }\ 2 K_{i\bar \jmath}\kappa^i\kappa^{\bar \jmath}\leq \beta^{-2}\, .
\ee
Hence, in order to satisfy the original dS conjecture \eqref{dS}, it is sufficient to show that 
\beq
\label{bound}
2 \beta\,  {\rm Re} (\kappa^i \partial_i \cQ^2_{{\bf f}}) \geq c \cQ^2_{{\bf f}}\, .
\eeq
which then provides a lower bound for the factor $c$ in \eqref{triangle}. 

Let us first consider that the membrane generating the potential is elementary saxionic, which implies that its tension satisfies $\partial_i\cT_{\bf f}=K_i\sigma_i\cT_{\bf f}$ (c.f.\eqref{Tsigma}). In this case, we found in section \ref{sec:WGCmem}, that the membrane saturates the extremality bound \eqref{Msaxionic}, and thus we have
\beq
\label{dQ}
M_{\rm P}^2 \partial_i \cQ_{\bf f}^2=\gamma^2 \partial_i(\cT_{\bf f}^2)=2\gamma^2 \cT_{\bf f}\partial_i\cT_{\bf f}=2 \gamma^2 K_i\sigma_i \cT_{\bf f}^2=2 M_{\rm P}^2 K_i\sigma_i  \cQ_{\bf f}^2\, ,
\eeq
with $\gamma$ being the extremality factor defined in \eqref{memgamma}. Summing over all saxions, this can be brought to the form
\beq
\label{dSform}
\sum_i \sigma_i \cQ^2+\sum_i \frac{s^i}{2(n_i-n_{i-1})}\partial_{s^i}\cQ^2=0\, ,
\eeq
where we have used \eqref{Ki}, assuming a diagonal field metric. Hence, \eqref{dSform} is precisely of the form \eqref{bound} with $\kappa^i=-\frac{\ii s^i}{2(n_i-n_{i-1})}$; and the second condition in \eqref{triangle2} is saturated with
$
\beta^{-2}=\sum_i\frac{1}{8(n_i-n_{i-1})}
$.
This implies that the original de Sitter conjecture is satisfied with 
\beq
\label{cbound}
c\geq (\sum_i\sigma_i)\beta=2\sqrt{2}(\sum_i\sigma_i)\left(\sum_i \frac{1}{n_i-n_{i-1}}\right)^{-1/2}\, ,
\eeq
providing a lower bound for the factor $c$.  In the case of diagonal metrics, one may canonically normalise the saxions such that \eqref{dSform} can be recasted in the same form than the no-go's in \cite{Hertzberg:2007wc,Andriot:2019wrs}, recovering the same bound \eqref{cbound}.\footnote{Alternatively, one could recast \eqref{dQ} in the form
\beq
\label{dSform2}
\sum_i \sigma_i (n_i - n_{i-1}) \cQ^2_{\bf f}+\frac12 \sum_i s^i\partial_{s^i}\cQ^2_{\bf f}=0
\eeq
which after identifying $\kappa^i=- 2\ii s^i$ and $\beta^{-2}=n_I$ yields the slightly stronger bound for $c$
\beq
c^2\geq 8\frac{(\sum_i (n_i- n_{i-1}) \sigma_i)^2}{\sum_i (n_i-n_{i-1})} =  8\frac{(\sum_i (n_i- n_{i-1}) \sigma_i)^2}{n_I}=2\frac{r_I^2}{n_I} \, \eeq
where we have used $\eqref{sigma}$ in the last step. The no-go then applies to any membrane with $\sum_i \sigma_i (n_i - n_{i-1})=-r_I>0$, which are precisely the membranes that become light along the string RG flow.
\label{footn_nogo}
}

To sum up, as long as $\sum_i\sigma_i$ is strictly positive, then $\cQ^2$ automatically satisfies \eqref{triangle} with positive $c$ and hence, its contribution to the scalar potential satisfies the original de Sitter conjecture. Interestingly, membranes with $\sum_i \sigma_i>0$ are only those which remain light  in the perturbative regime set by the RG flow of the strings\footnote{Strictly speaking, light membranes along string RG flow satisfy $\sum_i 2\sigma_i (n_i - n_{i-1})=-\sum (r_i-r_{i-1})=-r_I>0$. In general, this selects that same membranes than $\sum_i\sigma_i>0$, but if not, one can simply use the no-go in footnote \ref{footn_nogo} to ensure all lights membranes satisfy the de Sitter conjecture.}. Recall that all membranes with charges in $\Gamma_{\rm EFT}$ are of this type, so the contribution from EFT elementary membranes satisfies the no-go for de Sitter.

Given that we have determined that the sign of $c$ is positive for light membranes, we can actually compute the exact value for $c^2$ as follows 
\beq
||\partial \cQ_{\bf f}^2||=2 \sqrt{2} \sqrt{\sum_{i=1}^{I}  (n_i-n_{i-1})\sigma_i^2}\,  \cQ_{\bf f}^2 = |\alpha| \cQ_{\bf f}^2\, ,
\eeq
and $\alpha$ is precisely the dilatonic coupling obtained in \eqref{memalpha}. Therefore, the factor $c$ turns out to be given by the scalar contribution to the extremality factor:
\beq
\label{cdil}
c=|\alpha| = 2\sqrt{2} \sqrt{\sum_{i=1}^{I}  (n_i-n_{i-1})\sigma_i^2}\, .
\eeq
This is not a bound anymore but the exact result, and we find very interesting that indeed corresponds to the scalar contribution of the extremality factor. Notice that $\alpha$ was also playing an important role in the RG flow of the saxions in \eqref{incTmem} and determines the strong coupling scale at which the flow breaks down, $y_{\rm strong}=(\alpha^2 \Lambda_{\rm strong})^{-1}=(\alpha^2 \cT/M_{\rm P}^2)^{-1}$. Thus, the smaller the factor $c$ is, the further away we can move away from the membrane and the more we push $\Lambda_{\rm strong}$ to the IR.   

Notice that this result fails if some $n_i=n_{i-1}=0$, which in turn implies that the field metric is not diagonal to leading order in the perturbative regime (these are what we called degenerate flows, and they will be studied in more detail in \cite{Lanza:2021qsu}). This caveat is precisely the one appearing in \cite{Grimm:2019ixq}, where it was proved that the asymptotic potentials of F-theory flux compactifications to four dimensions satisfy the de Sitter conjecture for any two-moduli large field limit. The analytic proof of the no-go theorem in \cite{Grimm:2019ixq} only works as long as $n_i-n_{i-1}\neq 0$ $\forall i$,\footnote{When $n_i = n_{i-1}$ for some $i$, the singularity type does not increase in the enhancement chain, which typically occurs when there is some finite distance divisor involved, even if the final singularity is at infinite distance.}  while the rest of the cases were proven numerically and case-by-case using linear programming techniques.

It is worthwhile noticing that, in string compactifications, $c$ is both lower an upper bounded, as both $\sigma_i$ and $r_i$ are. The lowest possible value comes from setting $r_i-r_{i-1}=1$ in \eqref{sigma}, implying
\beq
\label{cmin}
c_{\rm min}=\alpha_{\rm min}\geq \sqrt{\sum_{i=1}^I\frac{2}{n_i-n_{i-1}}}\, ,
\eeq
taking into account that $n_i$ cannot exceed the internal dimension.\footnote{For Type II bulk fluxes in Calabi--Yau threefold compactifications, it becomes $c_{\rm min}\geq \sqrt{2/3}$. This value coincides with the TCC bound \cite{Bedroya:2019snp} (see also \cite{Andriot:2020lea}). However, since it depends on the internal dimension, it might get a bit smaller if coming from M-theory on $G_2$ or F-theory on CY$_4$.} We would like to stress that this is the same minimal value obtained for the exponentially decreasing mass rate $\lambda$ of the infinite tower of particles at different infinite field distance limits, either coming from string excitation modes as in section \ref{sec:SDC} (cf. eq.\eqref{L1}) in 4d $\cN=1$ compactifications or from BPS particle states in 4d $\cN=2$ setups, as in \cite{Grimm:2018cpv,Gendler:2020dfp}. In all these cases, the exponential rate both of the mass of the tower and of the potential is fixed by the extremality bound associated either to particles \cite{Gendler:2020dfp}, strings (section \ref{sec:WGCstr}) or membranes (section \ref{sec:WGCmem}). In practice, since the extremality factors are lower bounded by the maximum values of the degrees $n_i$ in the K\"ahler potential, one obtains
\beq
\label{mV}
\frac{||\partial m^2||}{m^2}|_{\rm min}\simeq \frac{||\partial V||}{V}|_{\rm min}\, ,
\eeq
which motivates the physical argument in \cite{Ooguri:2018wrx} for which the runaway potential emerges from integrating out the infinite tower of states at the infinite field distance limits. The correlation \eqref{mV} between the exponential mass rate of the tower and the potential was also conjectured to be general in \cite{Andriot:2020lea} although notice that here we are only pointing out that the minimal values that these quantities could take coincide; and not that for each type of asymptotic limit the actual exponential mass rate of the tower and that of the potential are equal. This latter stronger statement would require some correlation between the extremality bound for particles, strings and membranes, which we cannot guarantee, although it would be something interesting to explore in the future.

Up to now, we have only discussed the case of elementary membranes with charge in a single $q_r\in \Gamma_r$, but what about non-elementary saxionic membranes?
In the asymptotic regime, we can split the charge into a sum of elementary constituents since non-diagonal terms are polynomially suppressed as explained in Appendix~\ref{app:multimoduli}, obtaining
\beq
\label{QgT}
\cQ^2_{\bf f}\simeq \sum_r \cQ^2_{{\bf f}_r}=\frac{1}{M_{\rm P}^2}\sum_r\gamma_r^2 \cT^2_{{\bf f}_r}\, .
\eeq
 In the last step, we have used the extremality bound $\cQ_{{\bf f}_r}^2=\gamma_r^2 \cT_{{\bf f}_r}^2$ for the constituent membranes.
Furthermore, this also implies that
\beq
\partial_i \cQ^2_{{\bf f}}= \sum_r \partial_i(\cQ_{{\bf f}_r}^2) = \frac{2}{M_{\rm P}^2} \sum_r\gamma_r^2 \cT_{{\bf f}_r}\partial_i(\cT_{{\bf f}_r})\, .
\eeq
Therefore, employing the fact that the constituent membranes are elementary saxionic satisfying \eqref{Tsigma}, we get
\beq
\label{KdQ}
\partial_i \cQ^2_{{\bf f}}= \frac{2}{M_{\rm P}^2}K_i \sum_r\gamma_r^2 \sigma_i^{(r)} \cT_{{\bf f}_r}^2\, .
\eeq
The above result is again of the form \eqref{bound} only if
\beq
\label{Tbound}
\sum_r\gamma_r^2 \sigma_r \cT_{{\bf f}_r}^2\geq \sum_r\gamma_r^2 \cT^2_{{\bf f}_r}\, ,
\eeq
where $\sigma_r=\sum_i \sigma_i^{(r)}$. 
It is then clear that, if all the dilatonic couplings $\sigma_r$ are the same, we can factorise out $\sigma\equiv \sigma_r$ and immediately recover the de Sitter conjecture. If the more general case that includes non-elementary membranes, these couplings are generically different. Still, if all $\sigma_r>0$, we can get at least a bound by writing it in terms of the membrane of smallest charge to mass ratio as
\beq
2\sum_r\gamma_r^2 \sigma_r \cT_{{\bf f}_r}^2\geq 2\sigma_{\rm min}\sum_r\gamma_r^2 \cT^2_{{\bf f}_r}
\eeq
where $\sigma_{\rm min} = {\rm min} \{\sigma_r\} $. Using \eqref{QgT}, this implies
\beq
\sigma_{\rm min}\cQ^2_{{\bf f}}+\sum_i\frac{s^i}{2(n_i-n_{i-1})}\partial_{s^i} \cQ^2_{{\bf f}}\leq 0\, ,
\eeq
which reproduces the de Sitter conjecture with a lower bound for $c$ as in \eqref{cbound} upon replacing $\sum_i\sigma_i\rightarrow \sigma_{\rm min}$. Hence, also the contribution from non-elementary membranes satisfies the de Sitter conjecture as long as $\sigma_{\rm min}>0$. Recall again that if $\sigma_{\rm min}>0$, the membrane has a charge that belongs to $\Gamma_{\rm light}$, so all EFT membranes are of this type. Hence, the contribution from WGC-saturating BPS membranes to the scalar potential automatically satisfies the de Sitter conjecture and the bound \eqref{cmin} applies, regardless whether they are charged under one or several 3-form gauge fields so their charge belongs to several $\Gamma_r$ subspaces. In terms of the scalar potential, this implies that the contribution to the potential from the 3-form gauge fields vanishes asymptotically. However, this result only applies if the field metric can be approximated as a diagonal one at the perturbative limit, so the string flow is non-degenerate, otherwise more work is required. Examples in string  theory of fluxes whose contribution satifies $\sigma>0$ are RR fluxes in Type IIB and both RR and NS fluxes in IIA. Thus, the famous no-go for de Sitter in flux Type IIA compactifications at weak coupling and large volume \cite{Hertzberg:2007wc} would be a particular case of our story.

Our results fit well with the results of \cite{Grimm:2019ixq}, where a no-go theorem for de Sitter was derived for any asymptotic limit of the moduli space Calabi-Yau flux compactifications as long as the potential goes to zero in such a limit. On one hand, the no-go in \cite{Grimm:2019ixq} is more general as it includes contributions with both $\sigma>0$ and $\sigma<0$, but since the conclusion depends on the exact value of $\sigma_i$ (and whether \eqref{Tbound} is satisfied) then it was restricted to asymptotic limits of only two fields, where a classification of the possible perturbative potentials was performed.
On the other hand, the interpretation in terms of the charge and tension of membranes brings a new angle to the no-go theorem for de Sitter, since it becomes just a consequence of extremality. Hence, we can forget about the asymptotic geometry of the moduli space and simply state that the de Sitter conjecture is satisfied whenever the potential can be re-written as the charge of a light extremal membrane (i.e. a light membrane saturating the WGC). 

Needless to say, when adding other contributions like non-dynamical (heavy) fluxes or non-perturbative effects, the connection between the de Sitter conjecture and the WGC for membranes gets lost, as it involves membranes that cannot be described dynamically in the EFT (and that can have $\sigma<0$). Consequently, the potential generically develops some minima. It would be very interesting, though, to check if the presence of minima can somehow be related to superextremality or subextremality of the membranes, and whether it is only the latter that allows for the minima to have positive vacuum energy. At the moment, though, we will content ourselves with the observation that light extremal (WGC-saturating) membranes imply a contribution to the scalar potential in the form of a runaway that automatically satisfies the original de Sitter conjecture with an order one factor $c$ fixed by the extremality bound, and leave these other exciting questions for the future.

\subsection{Remarks on the AdS Instability Conjecture}
\label{sec:AdSremarks}

Membranes (codimension 1 objects) are featured in yet another Swampland Conjecture:  that any non-supersymmetric vacuum is ultimately unstable \cite{Ooguri:2016pdq,Freivogel:2016qwc}. It was mainly motivated by a sharpening of the Weak Gravity Conjecture stating that only BPS states can saturate the WGC. If the vacuum is supported by gauge fluxes on the internal dimensions, one can then apply the WGC to these fluxes (or equivalently to dual $(D-1)$-form gauge fields in $D$ space-time dimensions) and conclude the existence of a codimension-one brane satisfying $\cT\leq \cQ M_{\rm P}$. If the vacuum is non-supersymmetric, the sharpening of the WGC implies that this membrane needs to satisfy $\cT< \cQ M_{\rm P}$, mediating a bubble instability of the vacuum \cite{Maldacena:1998uz}. Hence, this is another example in which applying the WGC for membranes constraints the EFT flux-induced potentials  by not allowing the existence of completely stable non-susy vacua.

In the absence of scalars, the correlation between the WGC bound and vacuum instability is clear. However, the presence of scalars can complicate the story, since it is precisely in their presence where the differences between the extremality/WGC bound and the BPS bound become manifest \cite{Palti:2017elp,Lee:2018spm,Heidenreich:2019zkl,Gendler:2020dfp}. Furthermore, the scalar flow implies that the tension of the domain wall/bubble is different from the tension of the localised membrane. The former is the relevant quantity to determine vacua instability, while only the latter can be associated to a localised charged object. Although we do not have much to add to this discussion, we would like to discuss this conjecture in the context of our results regarding the RG flow of the tension and our EFT BPS `extremal' membranes.

In this paper, we have studied the physics of membranes that are electrically charged under 3-form gauge fields dual to dynamical fluxes, i.e.\ membranes whose transitions are consistent with the EFT cut-off $\Lambda$. Due to the strong scalar backreaction, we can only study the behaviour near the membrane (i.e.\ in the UV) since the scalars eventually leave the perturbative regime of control when flowing away to the IR and the EFT membrane description breaks down. Before this happens, we can define an effective scale-dependent tension for the domain wall $\cT^{\rm eff}(\Lambda)$ as in \eqref{effmemT} taking into account the  scalar backreaction. However, this is insufficient to determine the fate of the membrane in the IR. Our extremal membranes discussed in section~\ref{sec:memscflow}, which are BPS and saturate a WGC bound in the UV regime, are potential candidates to describe also extremal flat domain walls in the IR, in case the scalars succeed to flow to a minimum of the potential. But we cannot determine whether these domain walls will still saturate a WGC bound when evaluating the tension at the minimum of the potential, which is the relevant quantity to determine whether the vacuum is unstable. Interestingly, the AdS vacua instability is implying that the scalar flow have to be such that the domain wall still saturates the WGC bound when flowing to a supersymmetric minimum, while it becomes superextremal if flowing to a non-supersymmetric minimum. These constraints on the flow could in turn be translated to constraints on the continuation of the scalar potential away from the perturbative asymptotic regime. A way in which this different behaviour of the flow of the charge-to-tension ratio in SUSY and non-SUSY vacua could manifest is by means of violating the no-force identity \eqref{No-fo_Idm} when flowing to a non-susy vacuum at a lower cutoff scale. As discussed in section \ref{sec:RFC}, the membranes might become self-repulsive  when flowing to the IR upon integrating out some scalars as long as the vacuum is not supersymmetric. Otherwise, supersymmetry will force us to integrate out also the associated 3-form gauge fields such that the charge $\calq$ varies by the same amount and the object remains extremal. It would be interesting to further study this possibility and determine if  supersymmetry breaking necessarily forces the membrane to become self-repulsive and superextremal in the IR, which would further support the AdS Instability Conjecture.

\section{Conclusions}

In this work we have studied several aspects of $\half$BPS strings and membranes in 4d $\cN=1$ EFTs, serving a twofold purpose. On the one hand, intending to apply the different Swampland conjectures to the only BPS objects in 4d theories with low supersymmetry. On the other hand, aiming to unveil the role that low-codimension objects play with respect to quantum gravity constraints on EFTs.

A remarkable property of $\half$BPS strings and membranes is the off-shell identities \eqref{No-fo_Ids} and \eqref{No-fo_Idm}. In particular,  \eqref{No-fo_Ids} is obtained by assuming  approximate continuous shift symmetries of the 4d $\cN=1$ EFT. 
Such identities could be interpreted as no-force conditions, in analogy with the particle case. However, such an interpretation is more subtle for low-codimension objects, since generically they display a strong backreaction that takes us away from any perturbative regime of the theory, and may even involve scales above the EFT cut-off $\Lambda$. Nevertheless, one can still draw a physical interpretation of the said identities for {\em a subset} of strings and membranes. Indeed, if a string backreaction is such that an axionic symmetry is recovered for the metric in the vicinity of the string core, then  \eqref{No-fo_Ids} can indeed be interpreted as a no-force condition between two identical strings not too far from each other. This is a first example of a more general lesson. By understanding and classifying the different kinds of backreaction of strings and membranes, one may be able to extract valuable physical information of the 4d EFT. In fact, most of   the results of this paper stem from this observation.

To properly extract information from low-codimension objects we first require that they are seen as fundamental by the EFT, in the sense that they must be included as localised operators in the theory. In the case of 4d strings and membranes, this amount to require that their tensions satisfy \eqref{EFTregime}. One can check that membranes whose tensions lie in that range, characterised by a charge lattice $\Gamma_{\rm EFT}$, have a relatively mild backreaction, in the sense that up to a reasonable distance of order $\Lambda_{\rm strong}^{-1} > \Lambda^{-1}$ from their location we can still analyse the backreaction with EFT perturbative techniques. The case of strings is more subtle, and one must also require that a continue shift symmetry is recovered in the vicinity of the backreacted string metric. This subset of $\half$BPS fundamental strings is represented by the discrete cone of charges $\cC^{\text{\tiny EFT}}_{\rm S}$. 

In fact, to address the condition \eqref{EFTregime} one must first have a proper definition for the tension of 4d strings and membranes. Again, this is subtler than for codimension $> 2$ objects, whose tension and charge are typically measured at long distances from their core. Nevertheless, considering low-codimension fundamental objects as localised operators of the EFT gives a clear prescription on how to interpret their tension $\cT$ and charge $\cQ$. Indeed, following \cite{Goldberger:2001tn,Michel:2014lva} one can understand the backreaction of strings and membranes as a classical RG flow of their couplings, and the values entering in the EFT action as those evaluated at a cut-off scale $\Lambda$, namely at a distance $r_\Lambda \sim \Lambda^{-1}$ to each object. With this interpretation one can immediately identify fundamental membranes as relevant EFT operators by looking at their backreaction, while strings display a logarithmic profile corresponding to marginally relevant operators.

The backreaction profile of strings within $\cC^{\text{\tiny EFT}}_{\rm S}$ is particularly interesting, because when approaching their core one is driven to a regime in field space in which a continuous shift symmetry becomes exact. By general quantum gravity arguments one expects that such points with continuous global symmetries are located at infinite distance in the EFT moduli space, as confirmed in multiple string theory setups \cite{Lanza:2021qsu}. A careful analysis of these examples suggests that this may be a general feature, and that for every infinite distance perturbative limit there is a fundamental axionic string realising it by means of its backreaction. We have summarised this proposal as the {\em Distant Axionic String Conjecture} (DASC), whose details will be discussed in \cite{Lanza:2021qsu}. In this work we have focused on stressing the repercussion that this conjecture has on other Swampland proposals, like the Swampland Distance Conjecture \cite{Ooguri:2006in}.

Indeed, if the DASC is true, it opens the window to characterise the infinite distance paths of an EFT purely in terms of their physical objects of codimension 2. Also, since the physics behind this conjecture occurs at energies not far from the cut-off scale $\Lambda$, the statement is quite insensitive to IR physics. That is to say, it should not only apply to field space trajectories that are flat directions of the theory, but also to trajectories subject to a potential as long as the mass scales involved are below $\Lambda$, so that we can still consider these asymptotic regions to belong to the field space of the EFT. Even if the backreacted string solution of section \ref{sec:infinite} assumes a series of massless scalars, we have performed a first non-trivial test that suggests that our results should apply for massive scalars as well. Indeed, in the axionic string perturbative regimes one may generate a mass for a flowing scalar by means of a flux-induced superpotential. This will induce an anomaly on the string worldvolume, which can then be cured by a membrane ending on it. By the analysis of section \ref{s:asymptotic} we have seen that if the scalar has a mass below $\Lambda$, then $\Gamma_{\rm EFT}$ will contain an anomaly-cancelling membrane, such that one can always write a gauge-invariant string-membrane operator in the EFT. As such, the backreaction of the membrane will be mild enough such that the previous string flow should not be destabilised, although it would be interesting to corroborate this expectation by computing the explicit backreaction of the string-membrane system. More generally, each axionic string solution will split the full lattice of membrane charges $\Gamma_{\rm F}$ in terms of sublattices of super-Planckian  ($\Gamma_{\rm heavy}$) and sub-Planckian ($\Gamma_{\rm light}$) membrane tensions, the latter containing $\Gamma_{\rm EFT}$. While this picture was already present in  \cite{Lanza:2019xxg,Grimm:2019ixq}, in our current analysis we have been able to give precise formulas for such a splitting, in terms of the discrete data of the each string flow. 

The identification of string and membranes with their backreaction profile and the corresponding RG flow has important consequences when applying the Swampland Program philosophy to these objects. A clear example is given by the Weak Gravity Conjecture. In the original proposal \cite{ArkaniHamed:2006dz} the charge-to-mass ratio of a particle is evaluated using the IR asymptotic values. In the case of low-codimension objects, it is more natural to demand that the WGC is satisfied at the cut-off scale, where the couplings are defined. Then, as we vary the cut-off scale, we probe the whole RG flow generated by these objects, or in other words the backreaction of the physical objects. In practice, this amounts to impose the WGC at all scales, as in \eqref{WGC}. While with perturbative techniques we can only test scales above $\Lambda_{\rm strong}$, this already provides interesting implications for the EFT, since constraints on the UV physical charges and tensions of these objects translate to constraints on the axionic kinetic terms and scalar potentials of the EFT.

Indeed, one of our main results is that the WGC for strings in the UV region implies the Swampland Distance Conjecture. The RG flow of the scalars driven by the string  backreaction forces the string tension to asymptotically vanish $\Lambda\rightarrow\infty$. When the tension gets of order of the cut-off, namely $\cT_{\bf e}\sim \Lambda^2$, the EFT breaks down by the presence of an infinite tower of states corresponding to the string excitation modes. The field distance can be computed as the integral of the string charge, which only upon imposing the WGC, becomes logarithmic in the string tension. This implies that there is a maximal EFT cut-off which decreases exponentially in terms of the proper field distance, as stated by the Distance Conjecture. Interestingly, the exponential rate of the cut-off  (the SDC factor) is fixed by the extremality order one factor of the string, and the existence of a minimal charge-to-tension string ratio $\gamma_{\rm min}$ provides a lower bound for the SDC factor.  
Notice as well that the DASC can be seen as a different but complementary statement to the Emergent String Conjecture \cite{Lee:2019wij} applied to 4d EFTs. One difference is that the DASC always requires the presence of a weakly-coupled string at any type of infinite distance limit, although such a string does not necessarily yield the leading tower of states becoming light. Also, unlike the Emergent String Conjecture, the DASC does not specify whether other towers of massive modes that may be competing with the string vibrational modes form a KK tower or have a different nature. Finally, a potential generalisation of the DASC to dimensions higher than four would necessarily differ  from the Emergent String Conjecture substantially, as the DASC cares about the codimension and not the dimension of the object. Nevertheless, there is naively  no contradiction between both proposals in such higher dimensional cases, and it would be interesting to investigate their interplay in this more general context.

This correspondence between infinite field distance limits and string RG flows also provides a peculiar realisation of the Emergence proposal \cite{Harlow:2015lma,Grimm:2018cpv,Heidenreich:2018kpg,Palti:2019pca} for which the infinite field distance itself emerges from integrating out high energy modes. Recall that for a fixed cut-off $\Lambda$ there is a finite region $\MM_\Lambda$ in the moduli space accessible by the EFT before $\cT_{\rm str}$ reaches $\Lambda^2$. As we lower $\Lambda$ and we integrate out 4d high energy modes, we will reach this point at a larger distance, and so $\MM_\Lambda$ will grow. Only by sending $\Lambda\rightarrow 0$, infinite field distance points emerge.

Similarly, imposing the WGC for fundamental membranes translates into a particular behaviour for the potential. WGC-saturating membranes generate a runaway potential that realises the original de Sitter conjecture, with an order one factor fixed by the scalar dependence of the extremality bound for membranes. Although this only applies to the contribution to the scalar potential  from dynamical 3-form gauge fields, it provides an interesting physical interpretation of several no-go theorems for de Sitter at perturbative asymptotic limits \cite{Hertzberg:2007wc,Flauger:2008ad,Wrase:2010ew,Junghans:2018gdb,Roupec:2018mbn,Banlaki:2018ayh,Andriot:2019wrs,Grimm:2019ixq,Andriot:2020lea}. It would be interesting to study whether the presence of AdS or dS minima is somehow associated to superextremality or subextremality of the generating membranes.

 Four-dimensional $\cN=1$ theories are the starting point to answer questions of phenomenological interest, so these results open up a lot of future exciting directions. For instance, in this paper we have only studied in detail the RG flow of elementary membranes, which by themselves are not enough to generate a minimum of the potential. But if we were able to reach a minimum of the potential while staying in the perturbative regime, by extending the analysis to non-elementary membranes, we could ask whether the RG flow forces the backreacted membrane solution to become necessarily superextremal if the minimum is non-supersymmetric. The existence of this superextremal solution would describe a bubble instability, realising this way the AdS Instability Conjecture \cite{Ooguri:2016pdq}. 
On a different avenue, the realisation of the SDC in terms of the RG flow of  fundamental BPS strings provides a way to generalise the conjecture in a controlled way to field space trajectories  subject to a potential in 4d $\cN=1$ theories, as long as the mass scales involved are below the EFT cut-off. This can be the starting point to properly determine the maximal field range accessible in the EFT when constructing inflationary models and whether large field inflation is  in the Swampland or not. Since the extremality bound for the strings fixes the unknown parameter of the Distance Conjecture in this context, not only qualitative but quantitative bounds can be obtained, which are in fact consistent with the bounds already obtained in $\cN=2$ setups \cite{Grimm:2018cpv,Gendler:2020dfp}. This picture in terms of string RG flows might also shed some light into a low energy explanation for the SDC, as it becomes a consequence of the properties of fundamental EFT strings. Furthermore, the EFT membranes that need to be attached to the fundamental strings to cure anomalies induced by fluxes always generate a runaway potential along the string RG flow towards the perturbative regime. This provides indeed a correlation between the infinite tower of states and the runaway potentials, as proposed in \cite{Ooguri:2018wrx}. Although the lower numerical bounds for the exponential rate of the mass of the tower and the potential coincide in this setup, it would be interesting to explore if there is a more fundamental reason for it which relates the charge-to-tension ratio of strings and membranes.

\vspace{-.1cm}
 
\section*{Acknowledgements}

 \vspace{-.1cm}

We would like to thank David Andriot, Thomas Grimm, Alvaro Herr\'aez, Miguel Montero, Matthew Reece, Gary Shiu, Timo Weigand and Max Wiesner for useful comments and discussions. SL is supported by a fellowship of Angelo Della Riccia Foundation, Florence and a fellowship of Aldo Gini Foundation, Padova. FM is supported through the grants SEV-2016-0597 and PGC2018-095976-B-C21 from MCIU/AEI/FEDER, UE. LM is supported in part by the MIUR-PRIN contract 2017CC72MK\_003. IV is supported by Grant 602883 from the Simons Foundation.

\appendix

\section{RFC identities}
\label{app:RFC_Id}

Here we show that the $\cN=1$ supersymmetric structure of the EFTs forces BPS-strings and membranes to \emph{tautologically} saturate the RFC bound in \eqref{RFC}. We further illustrate that, macroscopically, this implies that identical supersymmetric strings or membranes do not exert any net force between each other.

\subsection{RFC identity for strings}
\label{app:RFC_Ids}

Consider a string electrically coupled to a set of gauge two-forms $\mathcal{B}_{2\,i}$ with charges $e^i$. The action that describes its dynamics is 
\be
\label{RFCs_stringS}
S_{\rm string}=-M^2_{\rm P}\, \int_\mathcal{S} \sqrt{-h}\, \mathcal{T}_{\bf e}+e^i \int_\mathcal{S} \mathcal{B}_{2\, i}\, ,
\ee
where $\mathcal{S}$ is the string worldsheet and the string tension $\mathcal{T}_{\bf e}$ may depend on the bulk fields. The inclusion of the action \eqref{RFCs_stringS} in a supersymmetric theory generically leads to the spontaneous breaking of \emph{all} the bulk supersymmetry over the string worldsheet. In order to preserve (partially) the bulk supersymmetry generators, further constraints need to be imposed on the possible dependence of $\mathcal{T}_{\bf e}$ on the bulk fields.

We focus on  the $\mathcal{N}=1$ theories considered in this paper -- the extension of the following calculations to non-supersymmetric settings is straightforward. We recall that in $\mathcal{N}=1$ supersymmetry the gauge two-forms $\mathcal{B}_{2\,i}$ are paired with  real scalar fields $\ell_i$, regarded as `dual saxions' in \eqref{dualfields}. These, along with their fermionic partners, build the linear superfields $L_i$ as an irreducible field representation of the super-Poincar\`e algebra. As shown in \cite{Lanza:2019xxg}, the bosonic components of the $\mathcal{N}=1$ bulk action describing the interactions among linear multiplets $L_i$ and gravity are
\be
\label{RFCs_dualaction}
S_{\rm bulk} = M^2_{\rm P}\int \frac{1}{2}R*1 -\frac12\int \cG^{ij}\left( M^2_{\rm P}\,\d\ell_i\wedge *\d\ell_j+\frac{1}{M^2_{\rm P}}\cH_{3\, i}\wedge * \cH_{3\, j} \right)\,.
\ee
Here $\cG^{ij}$ depends on the dual saxions $\ell_i$ and plays the role of both field metric for $\ell_i$ and gauge-kinetic function for the two-forms $\mathcal{B}_{2\,i}$. 

In order for the string with action \eqref{RFCs_stringS} to be able to maximally preserve $\mathcal{N}=1$ supersymmetry over its worldsheet once coupled to \eqref{RFCs_dualaction}, the tension of the string needs to depend on the dual saxions $\ell_i$ as
\be
\label{RFCs_T}
 \mathcal{T}_{\bf e} = M_{\rm P}^2 |e^i \ell_{i}| \, ,
\ee
with the same charges $e^i$ appearing in the two-form coupling in \eqref{RFCs_stringS}. Defining  the quadratic physical string charge as
\be
\label{RFCs_Q}
\cQ^2_{\bf e}=M_{\rm P}^2\, \cG_{ij}e^i e^j \, ,
\ee
we recognise that the string tension \eqref{RFCs_T} obeys the trivial identity
\be
\label{RFCs_TQ}
\cG_{ij} \partial_{\ell_i} \mathcal{T}_{\bf e} \partial_{\ell_j} \mathcal{T}_{\bf e}  = M_{\rm P}^2\cQ^2_{\bf e} \, ,
\ee
which is the no-force identity in \eqref{No-fo_RFCq}.

The relation \eqref{RFCs_TQ} has a neat macroscopic interpretation in terms of balance of forces between two identical strings. In order to see this, consider the coupling of the bulk action \eqref{RFCs_dualaction} to two identical supersymmetric strings as
\be
\label{RFCs_Stot}
S = S_{\rm bulk} + S^{(1)}_{\rm string} + S^{(2)}_{\rm string}\,,
\ee
where $S^{(i)}_{\rm string}$ is the action of the $i$th string as in \eqref{RFCs_stringS}, with the string tension as in \eqref{RFCs_T}.

In the static approximation, we may consider the worldsheets of the strings fixed in their transverse directions. Working in a `static gauge' such that the string worldsheets $\cS^1$ and $\cS^2$ stretch along the same spacetime directions $(t,x)$, the strings are considered `frozen' in the transverse directions ${\bf u} = (u,v)$ at the coordinates ${\bf u}^{(1)} =(u^{(1)},v^{(1)})$ and ${\bf u}^{(2)} =(u^{(2)},v^{(2)})$. In addition, motivated by the spacetime symmetries of the configuration, we may consider the bulk fields, namely the graviton, the dual saxions and gauge two-forms as depending only on the coordinates which are transverse to the string
\be
g_{\mu\nu} \equiv g_{\mu\nu}(u,v)\,,\quad \ell_i \equiv \ell_i(u,v)\,,\quad \cB_{\mu\nu\,i} = \cB_{\mu\nu\,i}(u,v)\,.
\ee
As a further simplification, given the geometry of the configuration, we use a gauge in which the only nontrivial component of $\cB_{\mu\nu\, i}$ is $\cB_{01\, i} \equiv \cB_i$. As usual in the background field method, we expand all these fields around some fixed background values $g_{\mu\nu}^0$, $\ell_i^0$, $\cB_i^0$ as
\be
g_{\mu\nu}(u,v)= g^0_{\mu\nu} + h_{\mu\nu}(u,v), \qquad \ell_{i}(u,v) = \ell_{ i}^0+ \hat\ell_{i}(u,v)\,, \qquad \cB_{i}(u,v) = \cB_{i}^0+ \hat\cB_{ i}(u,v)\,,
\ee
where $h_{\mu\nu}$, $\ell_{ i}$ and $\hat\cB_{ i}$ are understood as \emph{small perturbations} around the background values. As such, we may consider the background metric to be just the Minkowski metric $g_{\mu\nu}^0 = \eta_{\mu\nu}$. This perturbative regime is justified for the EFT strings considered in this paper and further  studied in \cite{Lanza:2021qsu},  
as long as the distance between the strings is not too large (i.e. below $\Lambda^{-1}_{\rm strong}$ as defined in section \ref{sec:stringflow}).

At a linearised level in the field variations, we regard the strings as \emph{external} sources for the bulk fields. Namely, we rewrite the string contributions to the action, as
\begin{equation}
\label{RFCs_SstrJ}
	 S^{(1)}_{\rm string} + S^{(2)}_{\rm string} = - \int \d^4 x\, \left(h^{\mu\nu} J_{\mu\nu}  + \hat\ell_i J_{\ell}^{i}  + \hat{\mathcal{B}}_{i} J_{\bf e}^i \right)+\ldots
\end{equation}
where
\begin{subequations}
\label{RFCs_J}
\begin{align}
	J_{\mu\nu}  &= - \frac{\delta}{\delta h^{\mu\nu}} (S_{{\rm string}}^{(1)}+  S_{{\rm string}}^{(2)}) = \frac{|e^j \ell_j^0|}2 \begin{pmatrix}
	1 & 0 & 0 & 0 \\0 & -1 & 0 & 0 \\ 0& 0& 0 & 0 \\ 0 & 0 & 0 & 0
	\end{pmatrix} \left[\delta^{(2)}\left(\mathcal{S}^1\right) + \delta^{(2)}\left(\mathcal{S}^2\right) \right]\,,
	\\
	J_{\ell}^i  &= -\frac{\delta}{\delta \hat\ell_i }  (S_{{\rm string}}^{(1)}+  S_{{\rm string}}^{(2)})= M_{\rm P}^2 e^{i} \left[\delta^{(2)}\left(\mathcal{S}^1\right) + \delta^{(2)}\left(\mathcal{S}^2\right) \right]\,,
	\\
	J_{\bf e}^i  &= \frac{\delta}{\delta \hat\cB_i }  (S_{{\rm string}}^{(1)}+  S_{{\rm string}}^{(2)})= e^{i}\left[\delta^{(2)}\left(\mathcal{S}^1\right) + \delta^{(2)}\left(\mathcal{S}^2\right) \right]\, ,
\end{align}
\end{subequations}
which denote the classical sources for the (perturbations of the) graviton, dual saxions and two-forms, respectively. In \eqref{RFCs_SstrJ} we have also chosen $\calb^0_i=\ell^0_i$.

Collecting the field variations as $\psi^{\cA} \equiv (h^{\mu\nu}, \hat\ell_i, \hat\cB_i^{\mu\nu})$ and the sources as $J_{\cA} \equiv (J^{\mu\nu},J_{\ell}^i,J_{\bf e}^i)$, the action \eqref{RFCs_Stot} may be written as
\begin{equation}
	S =  \int \d t\, \d x\, \d^2 {\bf u}  \left(   \int \d^2 {\bf u}' \frac12 \psi^{\cA} ({\bf u}) Q_{\cA,\cB}({\bf u};{\bf u}') \psi^{\cB} ({\bf u}')   - J_{\cA}({\bf u})\psi^{\cA} ({\bf u})  \right)\,.
\end{equation}
Then, by integrating out the dynamical fields $\psi^{\cA}$ at the classical level, we get
\begin{equation}
\label{RFCs_SJJ}
S = \frac{\ii}2 \int \d t\, \d x\, \d^2 {\bf u}  \int \d^2 {\bf u}' J_{\cA} ({\bf u}) \Delta^{\cA,\cB}({\bf u};{\bf u}') J_{\cB} ({\bf u}')  \,,
\end{equation}
where $\Delta^{\cA,\cB} = \ii (Q_{\cA,\cB})^{-1}$. 

\begin{center}
	\begin{figure}
		\centering
		\includegraphics{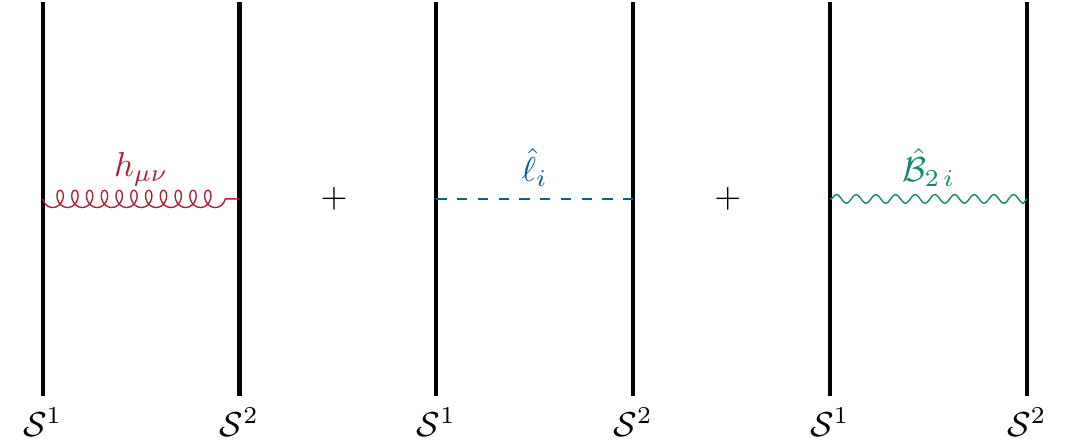}
		\caption{The mutual interactions of two parallel strings mediated by the graviton $h_{\mu\nu}$, dual saxions $\hat\ell_i$ and gauge two-forms $\hat\cB_{2\,i}$.    \label{Fig:FT_intstrings}}
	\end{figure}
\end{center}

The potential per unit area of the configuration is defined as
\begin{equation}
\label{RFCs_Vdef}
\int \d t\, \d x\, V  \equiv - S \,.
\end{equation}
In the present case, it splits as
\begin{equation}
\label{RFCs_Vst}
V = V_{\rm gr} + V_{\ell} + V_{\cB}\,,
\end{equation}
entailing the contributions of gravity, the dual saxions and the gauge two-forms. These may be computed from the general formula \eqref{RFCs_SJJ} as
\begin{subequations}
\label{RFCs_Aexp}
\begin{align}
V_{\rm gr} &= - \frac{\ii}2 \int \d^2 {\bf u} \int \d^2 {\bf u}'\, J_{\mu\nu} ({\bf u}) J_{\rho\sigma} ({\bf u}') \Delta^{\mu\nu,\rho\sigma} ({\bf u};{\bf u}')\,,
\\
V_{\ell} &= - \frac{\ii}2 \int \d^2 {\bf u} \int \d^2 {\bf u}'\, J_{\ell}^i ({\bf u}) J_{\ell}^j({\bf u}') \Delta^{\ell}_{ij} ({\bf u};{\bf u}')\,,
\\
V_{\cB} &= - \frac{\ii}2 \int \d^2 {\bf u} \int \d^2 {\bf u}'\, J_{\cB}^i ({\bf u}) J_{\cB}^j({\bf u}') \Delta^{\cB}_{ij} ({\bf u};{\bf u}')\,,
\end{align}
\end{subequations}
where the graviton, dual saxions and two-form propagators are respectively given by
\begin{subequations}
	\label{RFCs_Prop}
	\begin{align}
 \Delta^{\mu\nu, \rho\sigma}  ({\bf u};{\bf u}')  &= ( \eta^{\mu\rho}\eta^{\nu\sigma}+\eta^{\mu\sigma}\eta^{\nu\rho} - \eta^{\mu\nu}\eta^{\rho\sigma}) \Delta  ({\bf u};{\bf u}')\,,
 \\
 \Delta_{i j}^{\ell}  ({\bf u};{\bf u}') &= \mathcal{G}_{ij}(\ell^0) \Delta  ({\bf u};{\bf u}')\,,
 \\
 \Delta_{i j}^{\cB}  ({\bf u};{\bf u}') &= -M_{\rm P}^4 \mathcal{G}_{ij}(\ell^0) \Delta  ({\bf u};{\bf u}')\,.
\end{align}
\end{subequations}
where $\Delta  ({\bf u};{\bf u}')$ is the  two-dimensional scalar propagator, in the coordinate representation, along the spacetime directions transverse to  the strings.\footnote{This is the usual field theory propagator for a scalar field in the $n$-spacelike dimensions. It is defined via
	\be 
	\int \d^n u \, Q({\bf x}, {\bf u})  \Delta ({\bf u},{\bf x}')  = \ii \delta^n({\bf x} - {\bf x}') \;,
	\ee
	where $Q({\bf x}, {\bf x'})$ is the quadratic kinetic operator
	$Q({\bf x}, {\bf x'})= M_{\rm P}^2 \delta^{(n)}({\bf x} - {\bf x}')  \del_{\bf x} \cdot \del_{\bf x}$.
}
The full potential of the two-string configuration is then obtained by plugging \eqref{RFCs_J} and \eqref{RFCs_Prop} in \eqref{RFCs_Vst}.  However, it is important to remark that \eqref{RFCs_Vst} includes both contributions from the self-energy of both strings and interactions between them. The former  lead to classical divergences  which need to be  regularised, so that the string tension acquires a logarithmic cut-off dependence, as discussed in section \ref{sec:stringflow}.\footnote{Furthermore, because of supersymmetry, the renormalised tension at a scale $\mu<\Lambda$ can be identified with the `bare' one given by the values of the running dual saxions $\ell_i(r)$ at a distance $r=\mu_0^{-1}$, as in   \cite{Dabholkar:1990yf,Buonanno:1998kx}.} Hence, to first order in our  perturbative approximation, in order to compute the effect of the forces between the two strings it is enough to consider  the mutual exchange of gravitons, dual saxions and two-forms as depicted in Fig~\ref{Fig:FT_intstrings}. We then get
\begin{equation}
\label{RFCs_Acomp}
	V_{\rm gr} = 0\,,\qquad V_{\ell} = - V_{\cB} = - {\ii} M_{\rm P}^4 \mathcal{G}_{ij}(\ell^0) e^i e^j \Delta  ({\bf u};{\bf u}')\, .
\end{equation}
where $\Delta  ({\bf u};{\bf u}') = \frac{\ii}{4\pi} \log |{\bf u}-{\bf u}'| \equiv \frac{\ii}{4\pi} \log r$. Therefore we recognise that the static potential \eqref{RFCs_Vst} is zero. This expresses the balance of forces between the strings:
\begin{equation}
    F_{\rm tot}= \frac{1}{4\pi r} \left(f_{\ell} + f_{\cB}\right) = 0\,\qquad f_{\ell} = - f_{\cB} = - M_{\rm P}^4\mathcal{G}_{ij}(\ell^0) e^i e^j
\end{equation}
More precisely, the strings do not exert any gravitational forces between themselves, while the electric two-form mediated repulsion is exactly compensated by the attraction due to the mediation of the dual-saxions. 

\subsection{RFC identity for membranes}
\label{app:RFC_Idm}

In full generality, the action which describes the interaction between a membrane and the bulk fields may be written as
\be
\label{RFCm_memS}
S_{\rm mem}=-M^2_{\rm P}\, \int_\cW \sqrt{-h}\, \mathcal{T}_{\bf q}+q_a \int_\cW C_3^a\, .
\ee
Here $\cW$ is the membrane worldvolume and the membrane tension $\mathcal{T}_{\bf q}$ is allowed to depend on the bulk scalar fields, which we here assume to be a set of $n$ complex scalar fields $\phi^\alpha$. Furthermore, the membrane is assumed to be electrically charged under a set of gauge three-forms $C_3^a$ with charges $q_a$.

Coupling a membrane to a bulk theory requires the presence of the kinetic terms for the gauge three-forms in the bulk. In a fully-fledged supersymmetric theory, gauge three-forms need to be accommodated within a proper multiplet, irreducible representation of the supersymmetry algebra.  In $\cN =1$ supergravity, it was shown in \cite{Farakos:2017jme,Bandos:2018gjp,Lanza:2019xxg} that a set of gauge three-forms $C_3^a$ may be embedded within chiral multiplets $\Phi^\alpha$, in replacement of the usual scalar auxiliary fields. The gauge three-forms therefore share the same multiplets as the $n$ scalar fields $\phi^\alpha$ and, as illustrated in \cite{Lanza:2019xxg}, their number is capped off by $2(n+1)$. The action which describes the interactions between the bosonic components of these redefined chiral mutiplets is \cite{Lanza:2019xxg}
\be
\label{RFCm_dualaction}
S_{\rm bulk} = M^2_{\rm P}\int_{X_4} \left(\frac{1}{2}R*1  - K_{\alpha \bar\beta} \d \phi^\alpha \wedge * \d \bar \phi^{\bar\beta} \right) - \frac12 \int_{X_4} T_{ab} F_4^a *F_4^b + \int_{\partial{X_4}}  T_{ab} * F_4^a C_3^b\, ,
\ee
with the three-form kinetic matrix \eqref{TAB}. 

Owing to the fact that gauge three-forms do not carry propagating degrees of freedom, they can be most readily integrated out from the action, delivering
\be
\label{RFCm_actionos}
S_{\rm bulk} = M^2_{\rm P}\int_{X_4} \left(\frac{1}{2}R*1  - K_{\alpha \bar\beta} \d \phi^\alpha \wedge * \d \bar \phi^{\bar\beta} \right) -  \int_{X_4} V*1\,,
\ee
where $V$ is the usual Cremmer et al. potential
\begin{equation}
\label{RFCm_V}
V = e^K \left( K^{\alpha\bar\beta} D_\alpha W \bar D_{\bar\beta} \bar W - 3 |W|^2\right) = \frac12 T^{ab} f_a f_b\,,
\end{equation}
expressed in terms of the superpotential
\begin{equation}
\label{RFCm_W}
	W_{\bf f} = {\bf{f}} \cdot {\bf{\Pi}}(\phi) = f_a \Pi^a(\phi)\,.
\end{equation}

The explicit presence of the three-forms in the action \eqref{RFCm_dualaction} allows for coupling a membrane with the action \eqref{RFCm_T}. However, as also noticed for strings, the general \eqref{RFCm_T} would lead to the spontaneous breaking of all the bulk supersymmetry generators over the membrane worldvolume. In order to preserve supersymmetry partially, we need to constrain \eqref{RFCm_T}. In \cite{Bandos:2018gjp} it was shown that, in order to preserve the maximal $\mathcal{N}=1$ supersymmetry over the membrane worldvolume, one needs to choose the membrane tension
\be
\label{RFCm_Tmem}
\cT_{\bf q}= 2 M^3_{\rm P} e^K |q_a \Pi^a (\phi)|\,,
\ee
where $\Pi^a(\phi)$ are the same periods that appear in the superpotential \eqref{RFCm_W}.

As for supersymmetric strings, also supersymmetric membranes specified by \eqref{RFCm_memS} with \eqref{RFCm_Tmem} tautologically satisfy the no-force identity in \eqref{No-fo_RFCm} off-shell, albeit less trivially. In order to show this, let us first rewrite the membrane tension as
\be
\label{RFCm_T}
\cT_{\bf q}(\phi) = 2 e^{\frac K2} | W_{\bf q} (\phi)| \, ,
\ee
where we have introduced the charge-dependent superpotential\footnote{Notice that this quantity describes the discontinuity of the superpotential on the membrane worldvolume:
	\be
	\label{RFCm_DeltaWb}
	W_{\bf q} = W_{\bf f+q} - W_{\bf f}\,.
	\ee
}
\be
\label{RFCm_DeltaW}
W_{\bf q}  (\phi)\equiv {\bf{q}} \cdot {\bf{\Pi}}(\phi) = q_a \Pi^a(\phi)\,.
\ee
The holomorphicity of $W_{\bf q}$ implies that
\be
\del_{i} \overline{ W}_{\bf q} =  0 \qquad \Rightarrow \qquad \del_\alpha |W_{\bf q}| = \ii  |W_{\bf q}| \del_\alpha \theta
\ee
with $\theta = \arg W_{\bf q}$. On the other hand
\be
\del_\alpha | W_{\bf q} |  = \del_\alpha \left( e^{-\ii\theta}  W_{\bf q} \right) = -\ii  | W_{\bf q}| \del_\alpha \theta + e^{-\ii\theta} \del_\alpha W_{\bf q} \qquad \Rightarrow \qquad \del_\alpha \theta = -\frac\ii2 e^{-\ii \theta}   \frac{\del_\alpha W_{\bf q}}{|W_{\bf q}|}
\ee
and then
\be
\del_\alpha |W_{\bf q}|=\frac12 e^{-\ii \theta} \del_\alpha W_{\bf q}\,.
\ee
Using this identity, it is immediate to show that
\be
\label{RFCm_TDW}
\del_\alpha \cT_{\bf q}(\phi) =  2 \del_\alpha( e^{\frac K2} | W_{\bf q}| ) =  e^{\frac K2 - \ii \theta} D_\alpha (W_{\bf q})\, ,
\ee
with $D_i = \partial_i + K_i$ the covariant derivative associated to the holomorphic line bundle for which the superpotential is a section with weights $(1,0)$.\footnote{We choose the weights such that, under a K\"ahler transformation $K \to K + f + \bar f$, a quantity $Q$ transforms as $Q e^{-p f - q \bar f}$. The associated holomorphic and anti-holomorphic covariant derivative are $D^{p}_i Q =  (\del_{i} + p K_i) Q$, $\bar D^{q}_{\bar \imath} Q =  (\del_{\bar\imath} + q K_{\bar\imath}) Q$. The superpotential transforms as a section of a holomorphic line bundle $W \to W e^{-f}$; instead, $\mathcal{Z} = e^{\frac K2} W$, has weights $(-\frac12,\frac12)$, not being a holomorphic section. } Thus, we may rewrite
\be
\label{RFCm_TensidB}
\|\del\cT_{\bf q}\|^2 \equiv 2 K^{\alpha\bar\beta } \del_{\alpha} \cT_{\bf q} \bar\del_{\bar\beta} \cT_{\bf q} = 2 e^K  K^{\alpha\bar\beta } D_\alpha W_{\bf q} \bar D_{\bar\beta} \overline W_{\bf q}\, ,
\ee
which, combined with \eqref{RFCm_T}, gives
\be
\begin{split}
	&\|\del\cT_{\bf q}\|^2 - \frac32 \cT_{\bf q}^2 = 2 e^K \left(K^{\alpha\bar\beta} D_\alpha W_{\bf q} \bar D_{\bar\beta} \overline W_{\bf q} - 3 | W_{\bf q}|^2 \right)\,.
\end{split}
\ee
Recalling the structure of the superpotential \eqref{RFCm_DeltaW}, this quantity may be easily recast as
\be
\begin{split}
\label{RFCm_id}
	&\|\del\cT_{\bf q}\|^2 - \frac32 \cT_{\bf q}^2 = 2 e^K q_a q_b \left[K^{\alpha\bar \beta} D_\alpha \Pi^a \bar{D}_{\bar\beta} \bar\Pi^b - 3 \Pi^a \bar\Pi^b \right] = 2 M^2_{\rm P} V_{\bf q}(\phi) \, ,
\end{split}
\ee
where $V_{\bf{q}}$ is defined as in \eqref{RFCm_V} with the replacement $f_a \to q_a$. If we introduce the quadratic membrane charge
\be
\cQ_{\bf q}^2 \equiv 2 V_{\bf q}(\phi) \,,
\ee
we recognise that \eqref{RFCm_id} does coincide with the equality in \eqref{No-fo_RFCm}.

Along the lines of the previous section, the forces between two identical membranes may be computed starting from the potential of a configuration with two parallel membranes, in the non-relativistic approximation (see \cite{Herraez:2020tih} for a more general computation). We therefore couple the bulk action to two identical membranes with actions \eqref{RFCm_memS} with \eqref{RFCm_Tmem}:
\be
S = S_{\rm bulk} + S^{(1)}_{\rm mem} + S^{(2)}_{\rm mem}\,,
\ee
where $S^{(i)}_{\rm mem}$ is the action of the $i$th membrane spanning the worldvolume $\mathcal{W}^i$. For convenience, we here recast the bulk action in terms of a real basis of fields 
\be
\label{RFCm_dualactionb}
S_{\rm bulk} = M^2_{\rm P}\int_{X_4} \left(\frac{1}{2}R*1  - \frac12 \cG_{\alpha\beta}\d \phi^\alpha \wedge * \d \phi^{\beta} \right) - \frac12 \int_{X_4} T_{ab} F_4^a *F_4^b + \int_{\partial {X_4}}  T_{ab} * F_4^a C_3^b
\ee
where $\phi^\alpha$ denote all chiral fields, including possible axionic ones. 

As discussed in section \eqref{sec:EFTmem}, the a membrane backreaction does not lead to UV divergences, while growths to strong coupling as one approaches an IR scale of order $\Lambda_{\rm strong}=\calt_{\bf q}/M^2_{\rm P}$. Then, it can be treated in a perturbative regimes at within distances of order $E^{-1}$, with $E\gg \Lambda_{\rm strong}$. Notice that  in \eqref{RFCm_dualactionb} we have neglected possible additional  terms appearing in \eqref{dualF4lagr}.  This approximation assumes  that such corrections are  irrelevant at the energy scales $E$ such that $\Lambda_{\rm strong}\ll E \lesssim \Lambda$, where the perturbative regime is reliable.

We then consider two membrane configurations at a transversal distance $d\ll \Lambda^{-1}_{\rm strong}$, which can be considered as straight and parallel at this length scale.  They both stretch along the same spacetime directions $(x^0,x^1,x^2)$ and they are located at the positions $y = -\frac d2$ and $y = \frac d2$ along the transverse coordinate. In analogy with the  string case, all the fields are assumed to depend only on the transverse coordinate $y\equiv x^3$
\be
g_{\mu\nu} \equiv g_{\mu\nu}(y)\,,\quad C^a_{\mu\nu\rho} \equiv C^a_{\mu\nu\rho}(y)\,,\quad \phi^\alpha = \phi^\alpha(y)\, ,
\ee
and, additionally, we  will consider a gauge in which $C^a_{012} \equiv C^a$ is the only nontrivial components of the gauge three-forms $C^a_3$. In the limit $d\ll \Lambda^{-1}_{\rm strong}$, since we are interesting in the mutual interaction between the membranes, both $C^a$ and $\phi^\alpha$  be considered as small perturbations around some background values $\phi^\alpha_0$, $C^a_0$. The background metric around the membranes is taken to be Minkowski, with $h_{\mu\nu}$ representing small perturbations thereof. 

As for the strings, we can regard these membrane configurations as classical sources and perturbatively compute the effective potential describing their mutual interaction due to the exchange of bulk fields. We then expand the fields as
\be
g_{\mu\nu}(y) = \eta_{\mu\nu} + h_{\mu\nu}(y)\,,\quad C^a(y) \equiv C^a_0+ \hat C^a(y)\,,\quad \phi^\alpha(y) = \phi^\alpha_0 + \hat\phi^\alpha(y)\, ,
\ee
where $h_{\mu\nu}(y)$, $\hat C^a(y)$ and $\hat\phi^\alpha(y)$ are regarded as small perturbations around the background values $\eta_{\mu\nu}$, $\hat C^a_0$ and $\hat\phi^\alpha_0$. Then, the membranes couple to the field perturbations $h_{\mu\nu}$, $\hat\phi^\alpha$, $\hat C^a$ via the background currents $J_{\mu\nu}$, $J_\alpha^\phi$, $J^{\bf q}_a$. These are computed expanding the membrane action retaining only the linearised terms as 
\begin{equation}
S^{(1)}_{\rm mem} + S^{(2)}_{\rm mem} = - \int \d^4 x\, \left(h^{\mu\nu} J_{\mu\nu}  + \hat\phi^\alpha  J_\alpha^\phi  + \hat C^a J^{\bf q}_a \right)+\ldots
\end{equation}

which leads to the correspondence
\begin{subequations}
	\label{RFCm_J}
	\begin{align}
	J_{\mu\nu}  &= - \frac{\delta}{\delta h^{\mu\nu}} (S_{{\rm mem}}^{(1)}+  S_{{\rm mem}}^{(2)}) = \frac{\mathcal{T}_{\bf q}^0}2 \begin{pmatrix}
	1 & 0 & 0 & 0 \\0 & -1 & 0 & 0 \\ 0& 0& -1 & 0 \\ 0 & 0 & 0 & 0
	\end{pmatrix} \left[\delta\left(y+\frac d2\right) +  \delta\left(y-\frac d2\right)\right]\,,
	\\
	J_\alpha^\phi  &= -\frac{\delta}{\delta \hat\phi^\alpha}  (S_{{\rm mem}}^{(1)}+  S_{{\rm mem}}^{(2)})= (\del_{\phi^\alpha} \mathcal{T}_{\bf q})^0 \left[\delta\left(y+\frac d2\right) +  \delta\left(y-\frac d2\right)\right]\,,
	\\
	J^{\bf q}_a  &= \frac{\delta}{\delta \hat C^a}  (S_{{\rm mem}}^{(1)}+  S_{{\rm mem}}^{(2)})= q_a \left[\delta\left(y+\frac d2\right) + \delta\left(y-\frac d2\right)\right]\,.
	\end{align}
\end{subequations}
Here $\mathcal{T}_{\bf q}^0 \equiv 2 M^3_{\rm P} e^K |q_a \Pi^a (\phi_0)|$, $(\del_{\phi^\alpha} \mathcal{T}_{\bf q})^0 \equiv (\del_{\phi^\alpha} \mathcal{T}_{\bf q})|_{\phi^\alpha_0}$ and $T^{ab} (\phi_0)$ depend on the background values of the scalar fields $\phi^\alpha_0$.

\begin{center}
	\begin{figure}
		\centering
		\includegraphics{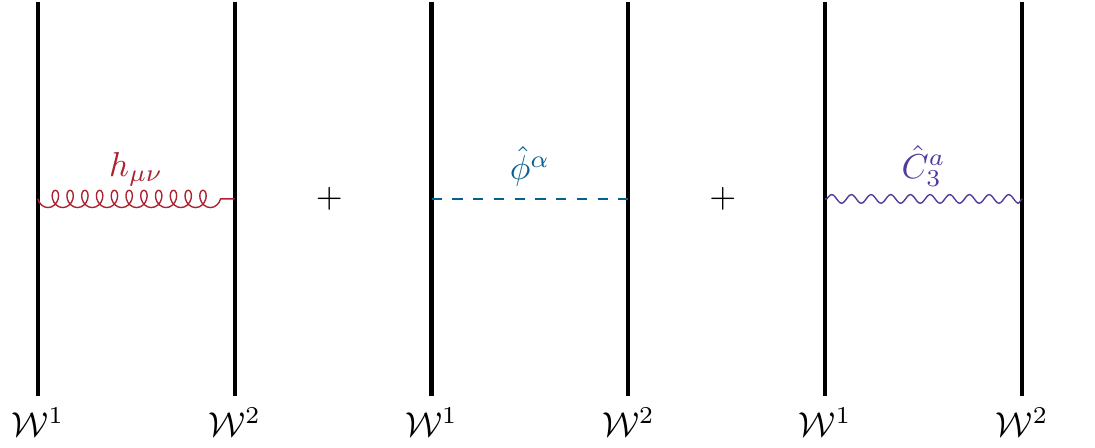}
		\caption{The mutual interactions between two parallel membranes mediated by the graviton $h_{\mu\nu}$, scalar fields $\hat\phi^\alpha$ and gauge three-forms $\hat C_3^a$.  \label{Fig:FT_intmem}}
	\end{figure}
\end{center}

In analogy to our computation of mutual forces between strings in the previous section, we may define the potential per unit area as
\begin{equation}
\label{RFCm_Vdef}
\int \d^3 x\, V \equiv - S = - \frac{\ii}2 \int \d^3 x\, \d y  \int \d y' J_\cA (y) \Delta^{\cA,\cB}(y;y') J_\cB (y')  \,.
\end{equation}
where $\d^3 x = \d x^0\,\d x^1\, \d x^2$ and we have collected the classical sources as $J_\cA = (J_{\mu\nu},J_\alpha^\phi,J^{\bf q}_a)$ and introduced the one-dimensional propagators $\Delta^{\cA,\cB}(y;y')$. We can then split the potential
\begin{equation}
V = V_{\rm gr} + V_{\phi} + V_{\bf q}\, ,
\end{equation}
singling out the contribution from gravity, scalar fields and gauge three-forms as depicted in Fig~\ref{Fig:FT_intmem}. Explicitly, these are given by
\begin{subequations}
	\label{RFCm_Vexp}
	\begin{align}
	V_{\rm gr} &= - \frac{\ii}2 \int \d y \int \d y'\, J_{\mu\nu} (y) J_{\rho\sigma} (y') \Delta^{\mu\nu,\rho\sigma} (y;y')\,,
	\\
	V_{\phi} &= - \frac{\ii}2 \int \d y \int \d y'\, J_{\alpha}^\phi (y) J_{\beta}^\phi (y')\Delta^{\alpha\beta}_\phi(y;y')\,,
	\\
	V_{C} &= - \frac{\ii}2 \int \d y \int \d y'\, J_{a}^{\bf q} (y) J_{b}^{\bf q} (y')\Delta^{ab}_C (y;y')\,.
	\end{align}
\end{subequations}
The graviton, scalar and three-form propagators obtained from the action \eqref{RFCm_dualactionb} are
\begin{subequations}
	\label{RFCm_Prop}
	\begin{align}
	\Delta^{\mu\nu, \rho\sigma}  (y;y')  &= ( \eta^{\mu\rho}\eta^{\nu\sigma}+\eta^{\mu\sigma}\eta^{\nu\rho} - \eta^{\mu\nu}\eta^{\rho\sigma}) \Delta (y;y')\,,
	\\
	\Delta^{\alpha\beta}_\phi  (y;y') &= \mathcal{G}^{\alpha\beta}(\phi^0) \Delta  (y;y')\,,
	\\
	\Delta^{ab}_C  (y;y') &= -M_{\rm P}^2 T^{ab}(\phi^0) \Delta  (y;y')\,,
	\end{align}
\end{subequations}
where $\Delta  (y,y')$ is the one-dimensional scalar propagator
\begin{equation}
	\label{ForceG_Sc_G}
	\Delta (y,y') = \frac{\ii}{2M_{\rm P}^2} |y-y'|\,.
\end{equation}
In order to obtain the mutual forces between the membranes, it is sufficient to focus on the interacting part in \eqref{RFCm_Vdef} only, neglecting the self-energy contributions. Combining \eqref{RFCm_J}, \eqref{RFCm_Prop} and \eqref{RFCm_Vexp}, we get that the following contributions to potential of the two-membrane configuration
\begin{equation}
\begin{split}
V_{\rm gr} &=  -  \frac{3 }{4 M^2_{\rm P}} \cT_{\bf q}^2(\phi_0) d\,,
\\
V_{\phi} &=  \frac{1}{2 M^2_{\rm P}} \cG^{\alpha\beta}(\phi_0) \del_\alpha \cT_{\bf q}(\phi_0)\del_\beta \cT_{\bf q}(\phi_0) d\,,
\\
V_{\bf q} &= -  \frac12 q_a q_b T^{ab} (\phi_0) d\,.
\end{split}
\end{equation}
These correspond to the forces per unit area
\begin{subequations}
\label{RFCm_forces}
\begin{align}
F_{\rm gr} &= \frac{3 }{4 M^2_{\rm P}} \cT_{\bf q}^2(\phi_0)\,, \label{RFCm_fgr}
\\
F_{\phi} &= - \frac{1}{2 M^2_{\rm P}} \cG^{\alpha\beta}(\phi_0) \del_\alpha \cT_{\bf q}(\phi_0)\del_\beta \cT_{\bf q}(\phi_0)\,, \label{RFCm_fphi}
\\
F_{\bf q} &=  \frac12 q_a q_b T^{ab} (\phi_0)\,. \label{RFCm_fel}
\end{align}
\end{subequations}
We emphasise that \eqref{RFCm_fgr} expresses that the gravitational force between membranes is \emph{repulsive} rather than attractive. This unusual feature of membrane dynamics was already prefigured in \cite{Vilenkin:1981zs,Garriga:2003gv}. Furthermore, we also notice that, since the matrix $T^{ab} (\phi_0)$ is generically not positive definite, the electric forces between two identical membranes may be either attractive or repulsive according to the chosen charges $q_a$.

Finally, the condition for balance of forces per unit area
\begin{equation}
\label{RFCm_bal}
	F_{\rm gr} + F_{\phi}  + F_{\bf q} = 0\, ,
\end{equation}
coincides with \eqref{RFCm_id} upon employing \eqref{RFCm_forces}. We also remark that, while \eqref{RFCm_id} is an off-shell identity, the relation \eqref{RFCm_bal} does depend on the background values of the scalar fields $\phi^a_0$.

\section{Flux lattice for multi-moduli limits in string compactifications}
\label{app:multimoduli}

Consider some asymptotic limit in field space given by sending several scalars
\beq
t^i\rightarrow \ii\infty \quad i=1,\dots,I
\eeq
with $t^i=a^i+\ii s^i$.
Each of these limits corresponds to a different perturbative regime from the EFT perspective.

The perturbative superpotential $W=\langle f,\Pi\rangle$ (c.f. \eqref{compsup}) can be written in terms of some period vectors  $\Pi$ which undergo a monodromy transformation \eqref{monper} by encircling the singular divisors $\cD_i=\{ e^{2\pi i t^i}=0\}$ (i.e. by shifting the axions $a^i\rightarrow a^i+1$). If the monodromy transformation is of infinite order, one can define a non-trivial nilpotent matrix as $N^i=\log T^i$.  They form a commuting set of matrices that further satisfy $\langle \cdot,N_{i}\cdot\rangle=-\langle N_{i}\cdot,\cdot\rangle$.

The Nilpotent Orbit Theorem states that the periods can be approximated as the following nilpotent orbit $\Pi_{\rm nil}$ up to exponentially suppressed corrections,
\beq
\Pi=\Pi_{\rm nil}+\mathcal{O}(e^{2\pi it^i})=e^{t^iN_i}\Pi_0(\chi)+\mathcal{O}(e^{2\pi it^i})
\eeq
where $\chi$ are the rest of the scalars not sent to the large field limit. The information encoded in $N_i$ and $\Pi_0$ can be associated to a set of  commuting $sl(2,\cC)$ triples $(N_i^-,N_i^+,Y_i)$ which captures the asymptotic behaviour of the periods. All these data determines a limiting mixed Hodge structure as described in \cite{Grimm:2019ixq}. To define these triples, one needs to divide the space into different growth sectors of the scalars
   \begin{equation}\label{growthsector}
    \mathcal{R}_{12 \cdots I} = \Big\{t^j = \phi^j + i s^j \Big| \frac{s^1}{s^2} > \gamma , \ldots, \frac{s^{I - 1}}{s^I} > \gamma , s^I > \gamma, \phi^j < \delta \Big\}\ ,
\end{equation}
with $\gamma>0$ a positive constant, such that there is a commuting $sl(2,\cC)$ triples for each sector (i.e., for each ordering of the scalars).

The $sl(2)$ orbit theorem then states that the nilpotent orbit can be further approximated by the $sl(2)$ orbit $\Pi_{sl(2)}$ up to polynomial corrections of order $\mathcal{O}(s^{i+1}/s^i)$, which become subleading in the growth sector \eqref{growthsector},
\beq
\label{Psl2}
\Pi_{nil}=e^{-a^iN_i}p(s)\Pi_{sl(2)}=s_1^{\frac{r_1}2}s_2^{\frac{r_2-r_1}2}\dots s_I^{\frac{r_I-r_{I-1}}2}p(s)e^{a^iN_i}\tilde\Pi_0+\dots
\eeq
where $p(s)$ includes the polynomial suppressed corrections. This way, we can single out the leading dependence on the saxions for any path within a given growth sector \eqref{growthsector} even for multi-moduli limits, which is a highly non-trivial result due to all the potential path dependence issues.

The  $sl(2,\cC)$ triples can be used to split the total flux vector space into eigenspaces of $Y_i$ as follows,
  \beq \label{split-Vellap}
\Gamma_{\rm F} = \bigoplus_{\bf r} \Gamma_{\bf r}\ ,\qquad  \textbf{r}=(r_1, \ldots, r_I)\ ,
\eeq
where $\textbf{r}=(r_1,r_2,\dots r_I)$ are integers representing the eigenvalues of $Y_{(i)}=Y_1+\dots +Y_i$\footnote{These $\Gamma_r$ vector spaces are the same as the $V_\ell$ defined in \cite{Grimm:2019ixq} upon identifying $r\rightarrow \ell_i-D$.}. 
Hence,
\beq
f_r\in\Gamma_r \ \Leftrightarrow\  Y_{(i)}f_r=r\, f_r
\eeq
When the fluxes belong to the primitive cohomology group $H^D_p(X_D,\mathcal{R})$, these eigenvalues are bounded by $-D\leq r_i \leq D$ where $D$ is the weight of the Hodge structure. 
The vector spaces $\Gamma_r$
satisfy  $\text{dim}\,  \Gamma_{\bf r} =   \text{dim}\,  \Gamma_{- \bf r}$ and are pairwise orthogonal as follows
 \beq
   \langle f_{\bf r}, f_{\bf r'} \rangle =  0 \quad \text{unless} \quad \textbf{r}+\textbf{r}' = 0\ ,
\eeq
due to the fact that $\langle \cdot,Y_{(i)}\cdot\rangle=-\langle Y_{(i)}\cdot,\cdot\rangle$. 

 Furthermore, if we restrict ourselves to a specific growth sector \eqref{growthsector}, the leading behaviour of the Hodge norm is completely determined by the discrete data associated to the asymptotic limit as follows
  \begin{equation} \label{general-norm-growth}
 \langle f,*f \rangle \ \sim\  \sum_{\bf r} \Big( \frac{s^1}{s^2}\Big)^{r_{1}} \cdots \Big(\frac{s^{\nmod-1}}{s^\nmod} \Big)^{r_{I-1}}
 (s^\nmod)^{r_{I}}\, \langle \rho_{\bf r}(f,\phi),*\rho_{\bf r}(f,\phi) \rangle_\infty +\dots\ 
\end{equation}
where $\rho(f, \phi) = e^{-\phi^i N_i} f $ and  $\langle\cdot,*\cdot\rangle_\infty$ is the asymptotic hodge norm defined at the singular boundary, $t^i=\ii\infty$. Hence, $\langle\cdot,*\cdot\rangle_\infty$ can still depend on the rest of the scalars $\chi$ but not on the saxions $s^i$, so it gives a finite result. The subspaces $\Gamma_r$ are orthogonal with respect to this asymptotic Hodge norm as follows,
 \beq
   \langle f_{\bf r}, *f_{\bf r'} \rangle_{\infty} =  0 \quad \text{unless} \quad \textbf{r}=\textbf{r}'\ .
\eeq
This implies the following exact direct sum decomposition on the split \eqref{split-Vellap},
\beq
 \langle f,*f \rangle_\infty =\sum_{\bf r} \langle f_{\bf r},*f_{\bf r} \rangle_\infty
 \eeq
 Contrary, the direct sum decomposition in \eqref{general-norm-growth} is only a good approximation away from the singularity if we are restricted to a growth sector \eqref{growthsector} with $\gamma\gg 1$, as it neglects polynomial suppressed corrections.
Keeping only this leading term in \eqref{general-norm-growth} (and therefore neglecting any non-diagonal term in the hodge norm) was denoted as the strict asymptotic limit in \cite{Grimm:2019ixq}, which implies that we are neglecting both exponentially suppressed non-perturbative corrections $\cO(e^{2\pi i t^j})$ as well as perturbative polynomial corrections $s^{i+1}/s^i$ which are subleading due to our restriction to the growth sector \eqref{growthsector}.

The asymptotic behaviour of the hodge norm allows us to further split the flux space as
\beq
   \Gamma=  \Gamma_{\rm light} \oplus  \Gamma_{\rm heavy} \oplus \Gamma_{\rm rest}\ , 
   \label{ap:GLHR}
\eeq
where we define 
\bea
     \Gamma_{\rm light} &=&  \bigoplus_{\bf r}  \Gamma_{ \bf r}\ , \qquad \text{ with } \ \textbf{r} =  \{ r_1,\ldots,r_{I-1} \leq 0 , r_I <0 \}\ , \\
      \Gamma_{\rm heavy} &=&  \bigoplus_{\bf r}  \Gamma_{ \bf r}\ , \qquad   \text{ with } \ \textbf{r} =  \{ r_1,\ldots,r_{I-1} \geq 0 , r_I >0 \}\ . 
\eea
The hodge norm of fluxes in $\Gamma_{\rm light}$ (i.e. their contribution to the scalar potential) goes to zero in the asymptotic limit for any path in the growth sector \eqref{growthsector}; while those in $\Gamma_{\rm heavy}$ diverge. Fluxes in $\Gamma_{\rm rest}$ can vanish or diverge asymptotically depending on the path. If we follow the path determined by the RG string flows, some fluxes in $\Gamma_{\rm rest}$ can be allocated into light or heavy, being the light ones those with $\sum_i (r_i-r_{i-1})=r_I <0$.

Notice that $\langle \Gamma_{\rm light},\Gamma_{\rm light}\rangle =\langle \Gamma_{\rm heavy},\Gamma_{\rm heavy}\rangle=0$
while $\langle \Gamma_{\rm light},\Gamma_{\rm heavy}\rangle\neq 0$. In typical string theory examples \cite{Grimm:2019ixq}, this bilinear product can be identified with the tadpole condition in \eqref{tadpole}. Hence, by restricting to $\Gamma_{\rm light}$ we are also guaranteeing that the fluxes appear only linearly in the tadpole, so they can belong to $\Gamma_{\rm EFT}$.

The perturbative expansion of the K\"ahler potential in the asymptotic limit reads
\beq
\label{KP}
K=-\log(P(s)+\mathcal{O}(e^{2\pi i t^i}))
\eeq
where $P(s)$ is a polynomial of the saxion whose coefficients can also depends on the rest of the scalars $\chi$. If the K\"ahler potential can be written as $K=-\log\langle\Pi^\dagger,\Pi\rangle$, \footnote{This occurs at the complex structure moduli space of Calabi-Yau manifolds, but also in the K\"ahler moduli space of $CY_3$ due to mirror symmetry.} this perturbation expansion is a consequence of plugging the nilpotent orbit for the periods \eqref{nil}, so that the total degree of $P(s)$ is $d_I$. The leading term of the polynomial in a given growth sector \eqref{growthsector} reads
\beq
\label{KP1}
K=-\log (s_1^{d_1}s_2^{d_2-d_1}\dots s_{I}^{d_{I}-d_{I-1}}+\dots)
\eeq
where $d_i$ is defined as 
\beq
\label{Nid}
N^{d_i}_{(i)}\Pi_0\neq 0\quad ,\ N^{d_i+1}_{(i)}\Pi_0=0
\eeq
with $N_{(i)}=\sum_{j=1}^i N_j$. Keeping only this leading term is at the same approximation level as keeping only the leading term in \eqref{general-norm-growth}. For the non-degenerate string flows defined in the main text, it is enough to only keep these leading terms. However, the perturbative correction become important when considering the degenerate flows, because then some $d_i-d_{i-1}=0$ which makes necessary to consider the next term in the perturbative expansion.

If the K\"ahler potential cannot be written in terms of the periods as above, it still makes sense to consider the perturbative expansion in \eqref{KP} and \eqref{KP1}. However, the degrees of the polynomial do not need to coincide anymore with the effective nilpotency orders in \eqref{Nid}. For this reason, to keep the discussion as general as possible, we have kept the degrees in $K$ as free integers in the main draft, which requires that we need to replace $d_i\rightarrow n_i$ in \eqref{KP1} and $r_i\rightarrow \hat r_i=r_i-(n_i-d_i)$ in \eqref{general-norm-growth}. However, the orthogonality conditions are still determined in terms of $r_i$, which implies that some light fluxes might appear quadratically in the tadpole, so $\Gamma_{\rm EFT}\subset \Gamma_{\rm light}$.

For convenience, we can also write explicitly the asymptotic behaviour of the tension and charge of a membrane. Plugging \eqref{Psl2} and \eqref{general-norm-growth} into \eqref{memT}, we get
\beq
\label{Tasymp}
\calt_{{\bf q}_r}\simeq M^3_{\rm P}T_0(\chi,\bar\chi)\rho_{\bf  r}(q,a^i) s_1^{\frac{\hat r_1}2}s_2^{\frac{\hat r_2-\hat r_1}2}\dots s_n^{\frac{\hat r_n-\hat r_{n-1}}2} \, ,
\eeq
 \beq \label{Qasymp}
 \cQ_{\textbf{q}_r}
 \ \simeq\ M^4_{\rm P} \cQ_0(\chi)\,  |\rho_{r}(q,a)|^2\, s_1^{\hat r_1}s_2^{\hat r_2-\hat r_1}\dots s_n^{\hat r_n-\hat r_{n-1}} \ .
 \eeq


\providecommand{\href}[2]{#2}\begingroup\raggedright\endgroup

\end{document}